\documentclass[conference]{IEEEtran}
\IEEEoverridecommandlockouts 
\usepackage{graphicx}
\usepackage{epsfig}
\usepackage{epstopdf}
\usepackage{multirow}
\usepackage{caption}
\usepackage{subcaption}
\usepackage{cite}

\usepackage[utf8]{inputenc}
\usepackage[english]{babel}
\usepackage{amssymb,amsmath,amsthm}
\usepackage{comment}
\newtheorem{theorem}{Theorem}

\usepackage{mathtools} 

\def\BibTeX{{\rm B\kern-.05em{\sc i\kern-.025em b}\kern-.08em
    T\kern-.1667em\lower.7ex\hbox{E}\kern-.125emX}}

\begin{document}

\title{Breath to Pair (B2P): Respiration-Based Pairing Protocol for Wearable Devices}



\author{
  \IEEEauthorblockN{Jafar Pourbemany, Ye Zhu}
\IEEEauthorblockA{\textit{Department of Electrical Engineering and Computer Science} \\
\textit{Cleveland State University}\\
Cleveland, USA \\
pourbemany@ieee.org, y.zhu61@csuohio.edu}
\and
\IEEEauthorblockN{Riccardo Bettati}
\IEEEauthorblockA{\textit{Department of Computer Science and Engineering} \\
\textit{Texas A\&M University}\\
Texas, USA \\
bettati@tamu.edu}
}
\maketitle
\begin{abstract}
  We propose Breath to Pair (B2P), a protocol for pairing and shared-key generation for wearable devices that leverages the wearer's respiration activity to ensure that the devices are part of the same body-area network. We assume that the devices exploit different types of sensors to extract and process the respiration signal. We illustrate B2P for the case of two devices that use respiratory inductance plethysmography (RIP) and accelerometer sensors, respectively. Allowing for different types of sensors in pairing allows us to include wearable devices that use a variety of different sensors. In practice, this form of sensor variety creates a number of challenges that limit the ability of the shared-key establishment algorithm to generate matching keys. The two main obstacles are the lack of synchronization across the devices and the need for correct noise-induced mismatches between the generated key bit-strings.
  
  B2P addresses the synchronization challenge by utilizing Change Point Detection (CPD) to detect abrupt changes in the respiration signal and consider their occurrences as synchronizing points. Any potential mismatches are handled by optimal quantization and encoding of the respiration signal in order to maximize the error correction rate and minimize the message overheads. Extensive evaluation on a dataset collected from 30 volunteers demonstrates that our protocol can generate a secure 256-bit key every 2.85 seconds (around one breathing cycle). Particular attention is given to secure B2P against device impersonation attacks.
   
\end{abstract}



%
  \section{Introduction}\label{sec:intro}
  Wearable devices have become remarkably popular in many fields, such as health care, health management, the workplace, education, and scientific research. Nowadays, there are hundreds of different products in the market, including smartwatches, smart wristbands, smart glasses, smart jewelry, smart straps, smart clothes, smart belts, smart shoes, smart gloves, skin patches, and even implanted medical devices (IMD) \cite{Guk2019, Vhaduri2019, chong2014survey, Seneviratne2017, Bianchi2016, JohnDian2020, Poongodi2020, Pantelopoulos2010}, and they are utilized in a variety of applications \cite{JohnDian2020}. For example, they can monitor vital signs like heart rate \cite{jayanth2017wearable,majumder2019energy, brezulianu2019iot}, respiratory rate \cite{milici2016wireless, shah2019cloud, mahbub2017low, naranjo2018smart}, body temperature \cite{wan2018wearable, yoshida2018development}, blood pressure \cite{lamonaca2019overview}, blood oxygen \cite{murali2018pulse}, and blood glucose \cite{sargunam2019iot}, to detect disorders.
  
  Wearable devices can collect a wide range of information about human activities and behaviors. They usually sense, analyze, and store data based on the needs of their applications. In many cases, they need to share data with a base station or other wearable devices. Data shared between wearable devices is usually sensitive. Examples are health data or commands to adjust an implanted medical device. Therefore, the communication between wearables should be protected by encryption, and the devices need to have a shared secure key for encryption to ensure that an attacker cannot compromise the process.
  
  Traditional pairing techniques either need user intervention or implicitly rely on device proximity. A common approach that is based on user intervention is Bluetooth pairing, which requires selecting a target device from the list of available devices with/without an additional PIN requirement. PIN-based methods need interaction with some sort of input device or display, which may be either inconvenient or impossible in many wearables \cite{zeng2017wearia, unar2014review}. Furthermore, PIN-based pairing is vulnerable to observation attacks such as shoulder surfing \cite{unar2014review}. Public key cryptography (PKC), on the other hand, cannot be used to create a secure key on wearable devices because it requires a public key infrastructure (PKI) \cite{shim2015survey}. In addition, expensive computing methods such as PKC are not suitable for resource-limited IoT devices. Various techniques have been proposed to mitigate these limitations that take advantage of common features in the different wearable devices.
  
  Proximity-based pairing, such as NFC or RFID, requires a tightly controlled transmission range in order to avoid relay attacks. Pairing based on observation of a biometric activity is a natural choice because wearable devices are attached to the same body. Both behavioral biometrics (e.g., gait) and physiological biometrics (e.g., heart rate and muscle activity) are used for wearable device pairing and authentication. In recent years, various biometric-based pairing techniques have been provided that exploit a variety of sensors such as accelerometers, gyroscopes, magnetometers, ECG sensors, PPG sensors, EMG sensors, and piezo sensors, among others. 
  
  \textbf{Contributions:}
  To the best of our knowledge, this paper is the first study to use the breathing signal to generate a shared key and so securely pair wearable devices. One difficulty with using respiration as the underlying biometric activity for pairing is the variety of how different devices observe the respiration activity.
  We designed and implemented a key generation protocol called Breath-to-Pair (B2P), which uses the observation of the respiration activity to generate the same keys for two wearable devices that use two different types of sensors. We use RIP and accelerometer in this paper. 

  According to our knowledge, the study is also the first attempt to pair devices with different sensing mechanisms.  Because of the fundamental differences in the sensing mechanism, the pairing task is challenging: (1) The sampling rates used by the devices can be different due to the differences in the sensing mechanisms. (2) The differences in the sensing mechanism can cause big differences in the properties of signals collected from the wearable devices, such as signal amplitude. (3) The dynamics of the signals collected from the wearable devices can be very different because of the differences in the sensing mechanisms (for example, a change in respiration such as tachypnea can bring very different changes in respiration signals collected with different sensing mechanism).  (4) Synchronization between signals collected by different devices (which is necessary for the pairing) is difficult because of the differences in the sampling rates, signal properties, and dynamics.
  
  The rest of the paper is organized as follows. Section 2 provides a background of the respiration mechanism and the techniques to observe respiration signals. Section 3 discusses the trust model and the adversary model. In Section 4, we motivate and describe the different stages of the B2P protocol. In Section 5, we explain the experiment setup and performance metrics. Then, we present the best results and investigate the impact of different factors on the B2P performance and the resistance to attacks. The discussion and related work are provided in Section 6 and Section 7. Finally, Section 8 concludes the paper.
  
  \section{Background}\label{sec: Background}
  \subsection{Respiration Mechanism} 
\begin{figure} 
    \centering
    \subfloat[\label{fig:res_mechanism}]{
    \includegraphics[width=.5\columnwidth]{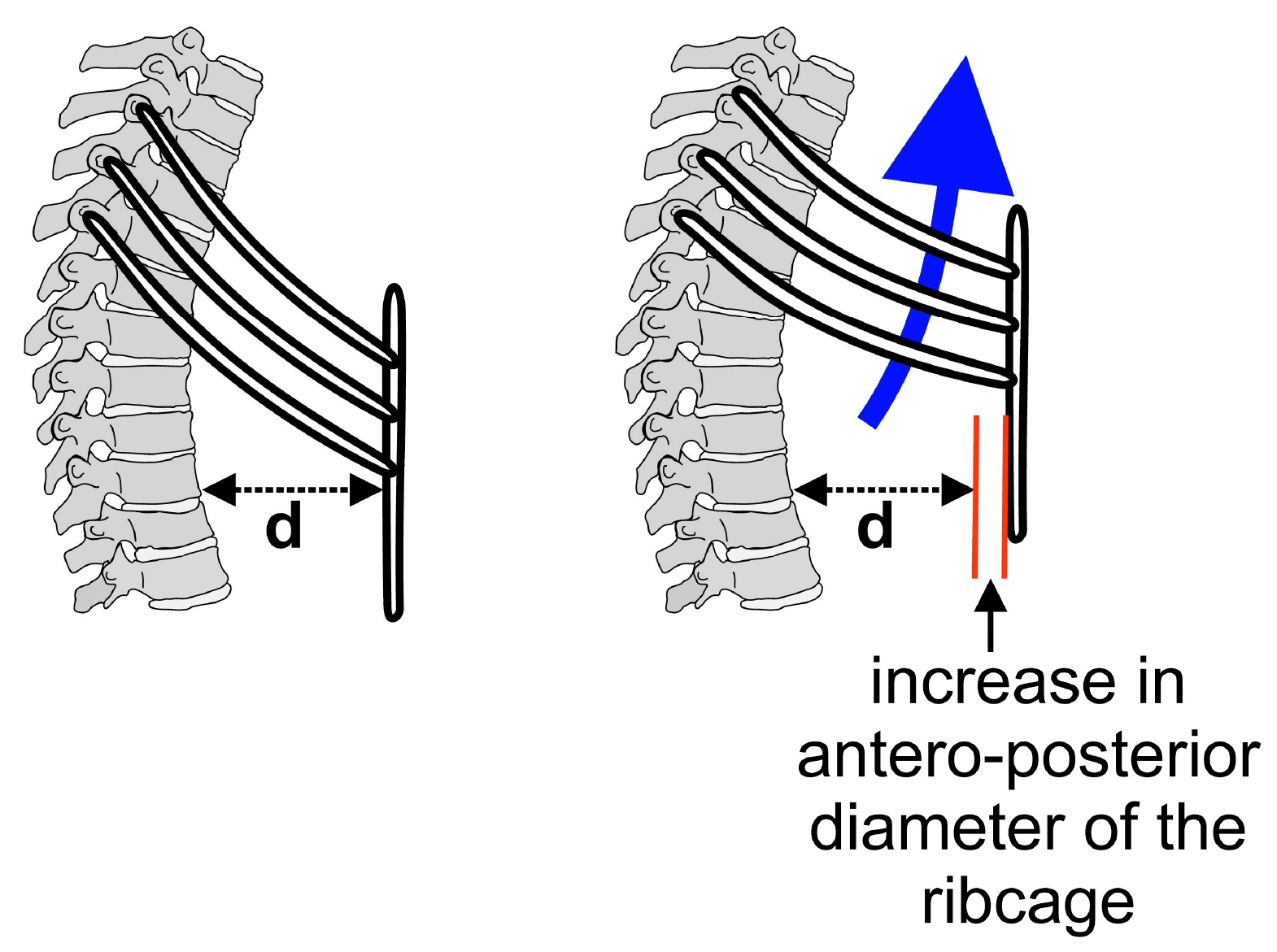}}
    \hfill
    \subfloat[\label{fig:Rip_Bands}]{
    \includegraphics[width=.4\columnwidth]{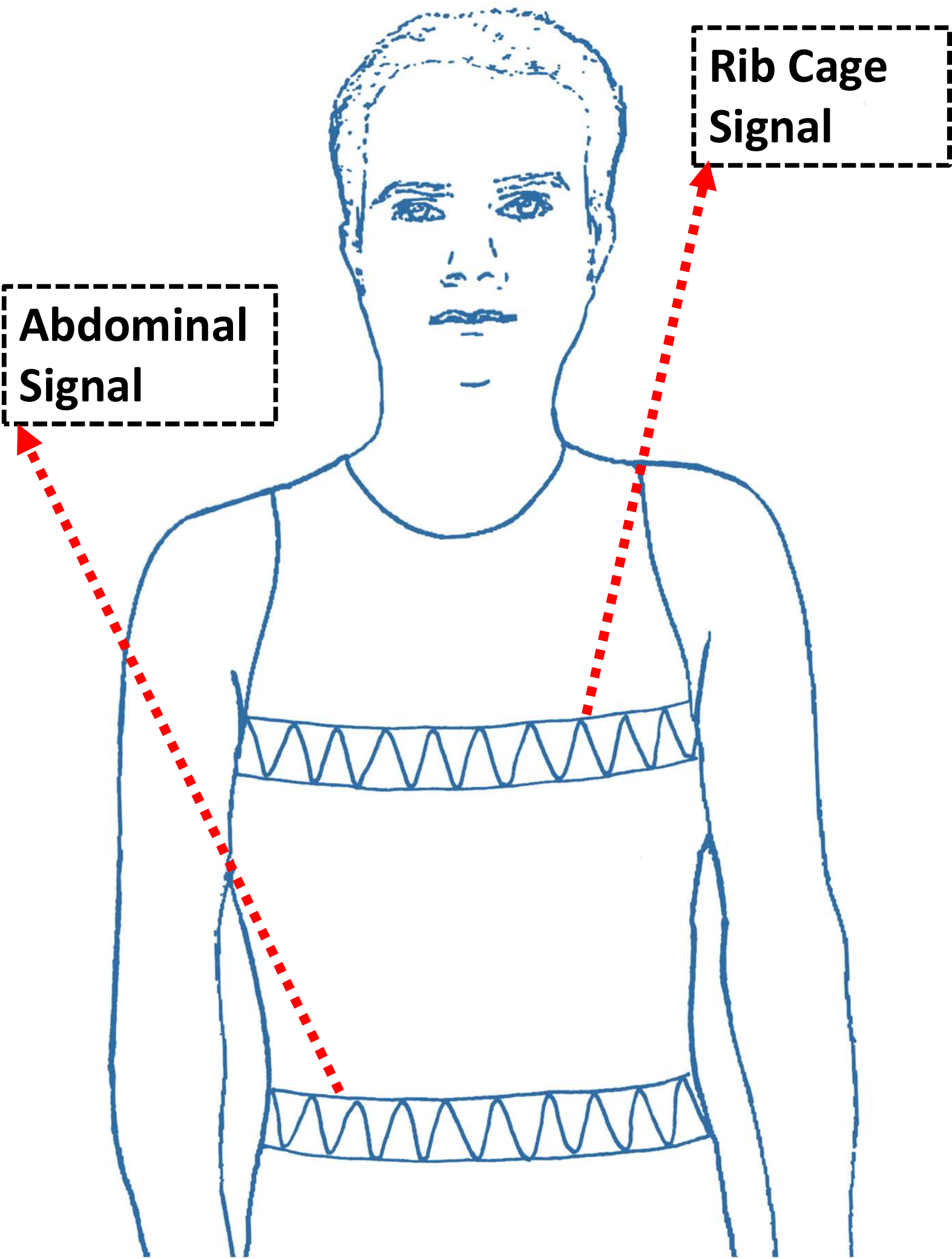}}

    \caption{(a) Movement of the Rib Cage in Breathing \cite{wikiRespiratory}, (b) RIP Bands at Chest and Abdomen}
    \label{fig:res_mechanism_RIP}
    \vspace{-0.2in}
  \end{figure}

  A respiration cycle consists of inhalation and exhalation in which the lungs will be inflated or deflated via three muscles: diaphragm, sternocleidomastoid (SCM), rib muscles (intercostal). During inhalation, the external intercostal contracts to bring out the rib cage, the SCM contracts to pull the rib cage upwards, and the diaphragm contracts and goes down to expand the thoracic cavity and increase the air volume in the lungs. During exhalation, the diaphragm relaxes and goes up, and the internal intercostal and abdominal muscles contract to make the lungs deflate back. The movement of the rib cage is shown in Figure~\ref{fig:res_mechanism}.
  The normal breathing rate for a healthy adult human is 12-18 breaths per minute at rest.
  This rate varies based on age. For example, in infants, the average resting respiratory rate varies from 25 to 40 breaths per minute, while for seniors over 80 years old, the breathing rate is between 10 and 30 breath cycles per minute.
  \subsection{Respiration Sensing Mechanisms} 
  Continuous measurement of respiratory rate (RR) is necessary for healthcare and sports applications. A traditional approach to measuring respiratory rate and tidal volume (the volume of air inhaled and exhaled with each breath) measures the airflow using a flow meter. There are other approaches that use air features, such as air temperature, air humidity, air components, and respiratory sounds \cite{massaroni2019contact}.
  In addition to airflow detection, RR can be observed using chest and abdominal movement and other signals.
  
  Several respiratory measurement techniques have been developed based on detecting chest and abdominal movement. Respiratory inductance plethysmography (RIP) is a common method to measure pulmonary ventilation. RIP consists of two belts of sinusoid wire coils positioned around the thorax and abdomen, as shown in Figure~\ref{fig:Rip_Bands}. The changes in the anteroposterior diameter of the abdomen and the rib cage during respiration change the coils' self-inductance and oscillation frequency. The latter is converted to a digital signal, the amplitude of which is proportional to the respiratory volume \cite{clarenbach2005monitoring}. A number of recent studies \cite{Preejith2017, Vertens2015, JafariTadi2014} show the use of an accelerometer to detect the respiration signal based on the movement of the chest and abdominal wall. The main challenge of this method is the removal of motion artifacts.
  
  Many other contact-based and noncontact-based techniques exist to measure respiration parameters. These techniques use a wide variety of sensors, such as anemometers, thermistors, thermocouples, pyroelectric detectors, fiber optic sensors, capacitive sensors, resistive sensors, microphones, nanocrystal and nanoparticles sensors, and many others \cite{Vanegas2020}.
  
  Given the wide variety of sensor mechanisms to observe respiration, it is unlikely in practice that two wearable devices use the same signal to observe respiration. This leads to the difficulties described above when using respiration for pairing.
  In this paper, we develop a pairing protocol that addresses these difficulties, and we demonstrate how it enables effective pairing devices that use different sensor technologies (RIP and accelerometers in our case) to measure respiration. We expected that the protocol design developed for the example of RIP and accelerometer can be directly applied to other combinations of sensing techniques as well.
  \begin{figure} []
    \centering
    \includegraphics[width=.65\columnwidth]{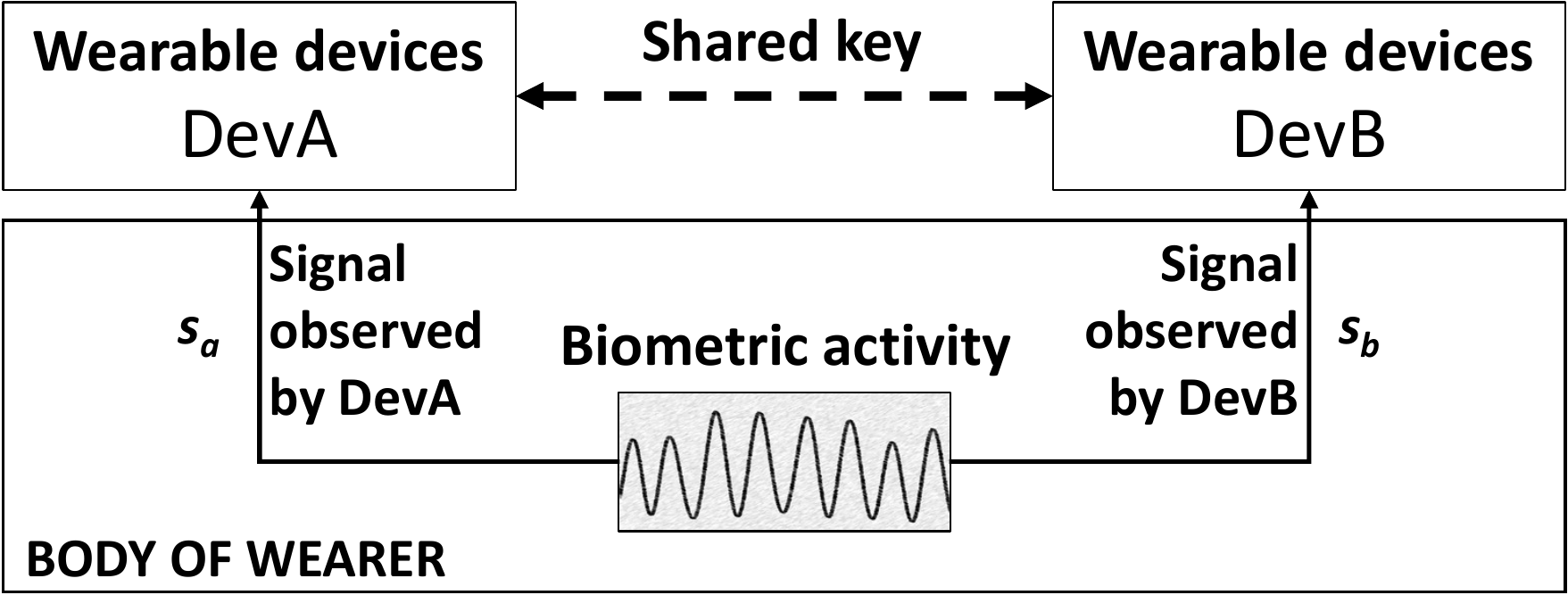}
    \caption{System Model of the Biometric-based Key Generation}
    \label{fig:system_model}
     \vspace{-.2in}
  \end{figure}
  \section{System Models} 
  \subsection{System Model}  
  The goal of pairing two wearable devices $DevA$ and $DevB$ is two-fold: 
    1) Ensure that both $DevA$ and $DevB$ are "worn" by the same individual.
    2) Establish a shared key using information that is shared only by the two devices.
  An approach that addresses both requirements is to use the biometric activity $S$ that can be accessed by the two devices but not by devices that are not connected to the body of the wearer (see Figure~\ref{fig:system_model}). In this paper, we choose respiration to be the observed biometric activity. The challenge with using biometric activity is that the sensors at a device do not observe $S$ directly, but rather infer $S$ from physical measurements of features that are accessible to the device (in the example of respiration, these features are ventilation for RIP and chest movement for accelerometer sensors). For a device $DevX$ the observed signal, call it $s_x$, for activity $S$ can be affected as follows: 

    1) The signal $s_x$ can be noisy. Noise can be due to measurement errors, or it can be caused by external disturbances. For example, an accelerometer may capture motions that are not caused by breathing.
    2) The Signal $s_x$ may be delayed. This may be due to time discretization or other data collection delays.
    3) The signal $s_x$ may have inconsistent specifications, such as different sampling rate or different amplitude range. 

    As a result, the two signals $s_a$ and $s_b$, observed by $DevA$ and $DevB$, are not equal. They may even be difficult to correlate at a granularity fine enough to allow for the reliable and low-latency establishment of shared keys.

  We make the following assumptions for the pairing: (1) The wearable devices are attached to the same body, and the pairing is based on the breathing activity of the subject. (2) The two devices collect respiration signals with two different mechanisms, in our case, a RIP sensor and an accelerometer. (3) Since the two devices are collecting respiration signals with different mechanisms, the signals collected by the two devices are not synchronized. Also, different sampling rates are used by the two devices for data collection. (4) The two devices do not share any prior information before the pairing. 
  \subsection{Threat Model} 
  We assume a powerful adversary who knows the details of the B2P protocol. Due to the broadcast nature of the wireless communication between wearable devices,  the adversary is capable of eavesdropping on the communication between legitimate wearable devices. Through eavesdropping, the adversary can access the content of the exchanged packets between the two devices before their successful pairing, as these packets are not encrypted. Although the adversary cannot attach a wearable device to the body to pair with a legitimate device, the adversary can possibly obtain average respiration rate and dynamic respiration rate by remotely observing the victim's breathing activities, typically using a camera, and then process the observed the collected data either manually, or using automated tools \cite{Jafar_HR_RR}. She can then generate an approximation of the breathing activity based on the average respiration rate. With the knowledge of the pairing protocol, the access to the content of packets exchanged between legitimate wearable devices, and access to the biometric activity (respiration in our case), the adversary is in a position to launch impersonation attacks, which distinguish themselves primarily by how the biometric activity is observed. With the generated signal, the adversary attempts to pair with a legitimate device through the B2P protocol.
  \section{Protocol Design}
For a biomedical-process-based pairing protocol to support a variety of devices, it must address the challenges described above: (1) the features collected by the different devices give rise to signals with different sample frequencies, amplitudes, and noise characteristics (measurement noise vs. disturbances). (2) The signals collected by the different devices are likely not sufficiently synchronized for the devices to use them as shared data needed for the key exchange. (3) In order to minimize its own quantization error, each device optimizes its quantization levels. When devices have different quantization levels, they interpret the signal differently, which leads to errors in the pairing process. (4) Finally, even if all challenges above are addressed, errors may remain across the collected signals, which again leads to errors in the pairing.
In the following, we describe how we address each of these challenges. At the end of this section, we describe how this naturally gives rise to the design, architecture, and finally to an implementation of the B2P pairing protocol. 
  \subsection{Signals have Different Characteristics} 
  We handle the problem of signals being collected at different frequencies, having different amplitudes, and different noise characteristics as follows:
   
  \noindent{\bf Different Sample Frequencies:} 
  Suppose that the two signals $s_a$ and $s_b$ are collected with sample rates denoted $f_a$ and $f_b$, respectively. If, say, $f_a > f_b$, then we need to resample the signal $s_a$ with the sampling rate $f_b$. In other words, we need to down-sample the signal generated by the higher sampling rate. For downsampling, we can use skipping (where the sample closest to a sampling point is kept and the rest is skipped), or averaging (where the new sample is interpolated from the closest samples). 
  
  \noindent{\bf Different or Unequal Signal Amplitudes:} This can be easily handled by normalization, such that the signal lies in the range of $[-1, 1]$.
  
  \noindent{\bf Different Noise Characteristics:} All biometric sensors suffer from measurement errors, which can be easily eliminated by filtering high frequency components. In addition, the collected data contains noise that is generated by external disturbances. In our example, accelerometers pick up motion that is unrelated to breathing. Similarly, RIP sensors may get disturbed by a poor fit of the vest \cite{bentsen2016electromagnetic}. 
  
  Since humans respiration rate is within a range from 6 to 30 per minute \cite{sanyal2018algorithms},  a low-pass filter with the cut-off frequency of 0.5
  Hz and a band-pass filter with cut-off frequencies of 0.1-0.5 Hz can be used to filter out both measurement noise and disturbances. In order to decrease the difference between the two devices,
  we also applied a Savitzky-Golay filter to eliminate the effect of
  possible noise in the range of the band-pass filter’s cut-off frequencies
  and smoothen the respiration signal without changing its features.
  
  \subsection{Lack of Synchronization}\label{sec:CPD}
  Keeping the signals synchronized is important to pairing performance. The synchronization is  especially critical for pairing with signals collected with different sensing mechanisms. In such cases, the mode transitions in the biometric activity (inhalation vs. exhalation in the case of breathing) may be detected at different points in time by the different sensing mechanisms.
  One straightforward method for synchronization is to synchronize signals according to obvious features such as peaks and valleys. 
  In practice, this is quite unreliable, mostly because of noise caused by disturbances, such as noise caused by motion to RIP sensors \cite{pmid23201991}. Instead, we propose a synchronization based on both the obvious features as well as statistical features with the change point detection method described below.  
  
  A Change Point (CP) is a point at which some statistical features of a signal, like mean, standard deviation (STD), root mean square (RMS), etc., changes abruptly. In this study, we apply a change point detection, and we compare the three main statistical features, i.e., mean, STD, and RMS, in terms of the pairing performance.
  
  Due to the motion artifacts, the accelerometer data is noisier than RIP data. Hence, some sudden change may happen in the noisy parts of the accelerometer signal, which is not a CP in the RIP signal. To address this problem and find the common CPs precisely, we first find respiration cycles based on signal peaks and valleys, and then we find the biggest CPs within each peak-valley. We compared multiple CPs per peak-valley in Section~\ref{sec: experiment}.
  \subsection{Lack of Uniform Quantization Levels} \label{sec:quantization}
  Quantization converts continuous infinite values to a smaller set of discrete finite values that can be represented by a fixed number of bits. On the one hand, quantization can possibly remove differences in the biomedical signals that are caused by differences in sensing mechanisms and sensing noise by converting close values of two signals into the same discrete value that can be represented by the same set of bits.  On the other hand, quantization can also cause a loss of information. For example, a small fluctuation in a respiration signal caused by a small change in breathing may be detected by one device but not by another because they do not use the same quantization levels. The number of quantization levels largely determines the amount of differences that the quantizer will remove by quantizing the different values into the same quantization level. So a smaller number of quantization levels leads to larger intervals between two successive quantization levels and then possibly removal of more differences.  On the other hand, the number of quantization levels determines the number of bits needed to represent each quantization level. So a larger number of quantization levels means more bits can be generated from each sample. In other words, more information about the respiration dynamics can be captured. So the parameter can affect the pairing performance greatly.
  
\begin{figure} []
    \centering
    \subfloat[\label{fig:quant_level}]{%
    \includegraphics[width=.45\columnwidth]{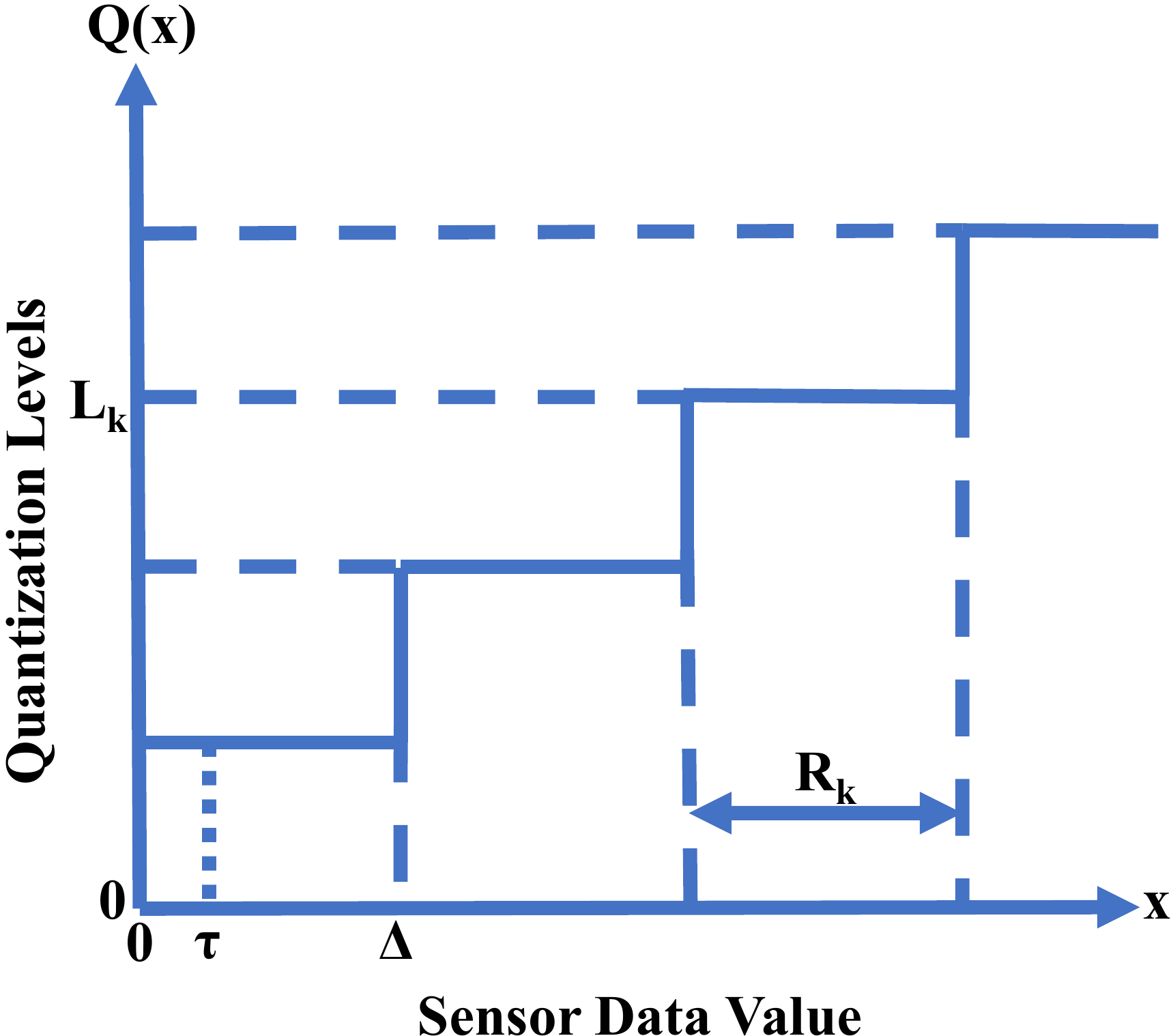}}
    \hfill
    \subfloat[\label{fig:res_distrobution}]{%
    \includegraphics[width=.5\columnwidth]{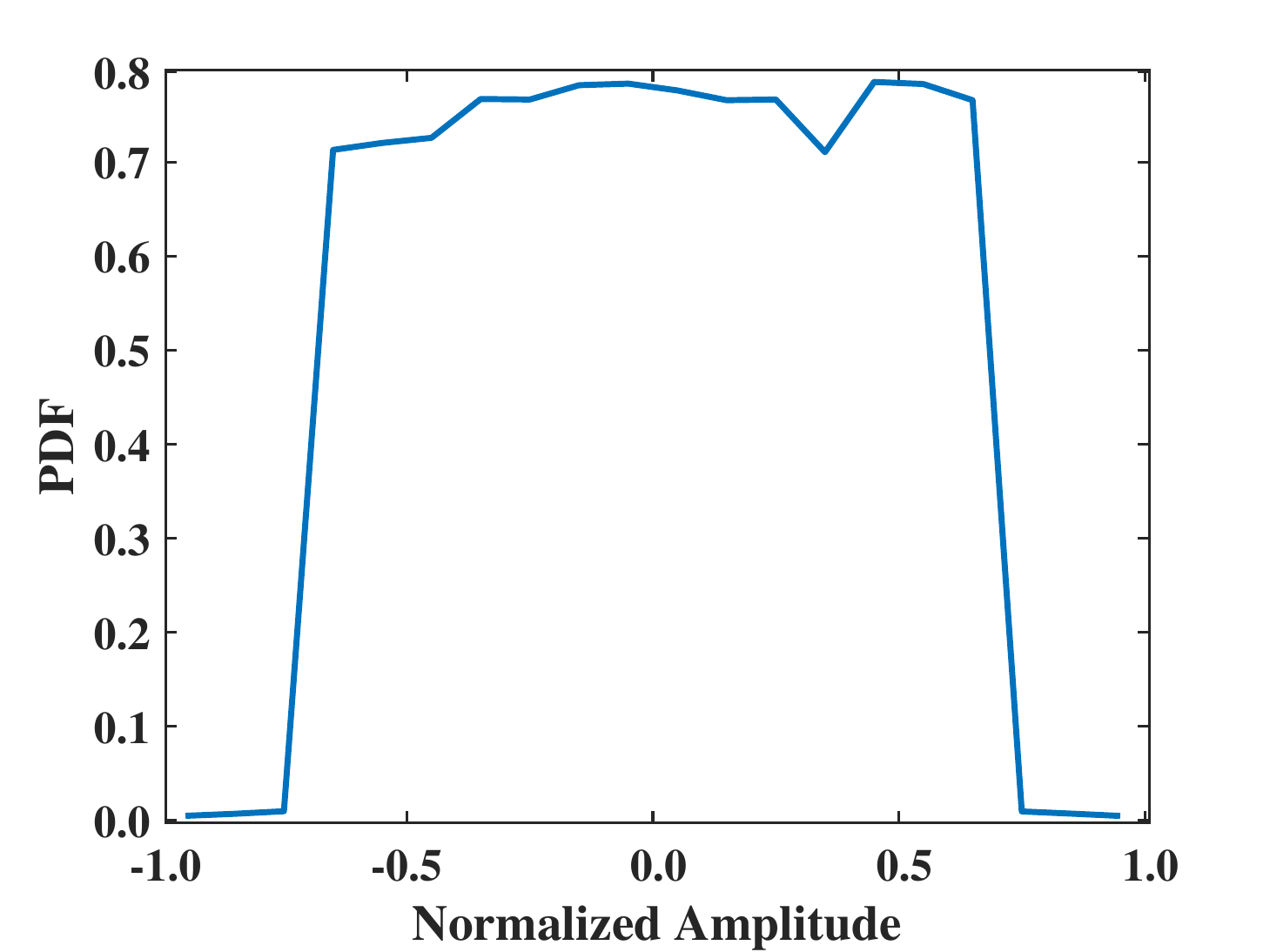}}
    \caption{(a) Quantization Levels {\em vs.} Sensor Data Value, (b) Breathing Signal's Distribution}
    \label{}
    \vspace{-.3in}
  \end{figure}

  \noindent{\bf Optimal Number of Quantization Levels:} Traditional quantizers are optimized to minimize the quantization error, defined as the difference between actual values in the x-axis and the corresponding quantization levels as shown in Figure \ref{fig:quant_level}. For pairing with different types of sensors, we need to maximize the number of bits that can be agreed upon by two sensors of different types on each sample, which is also the ultimate goal for pairing.
  
\begin{figure} [htbp]
    \centering
  \subfloat[\label{res_acc_diff}]{%
       \includegraphics[width=.5\linewidth]{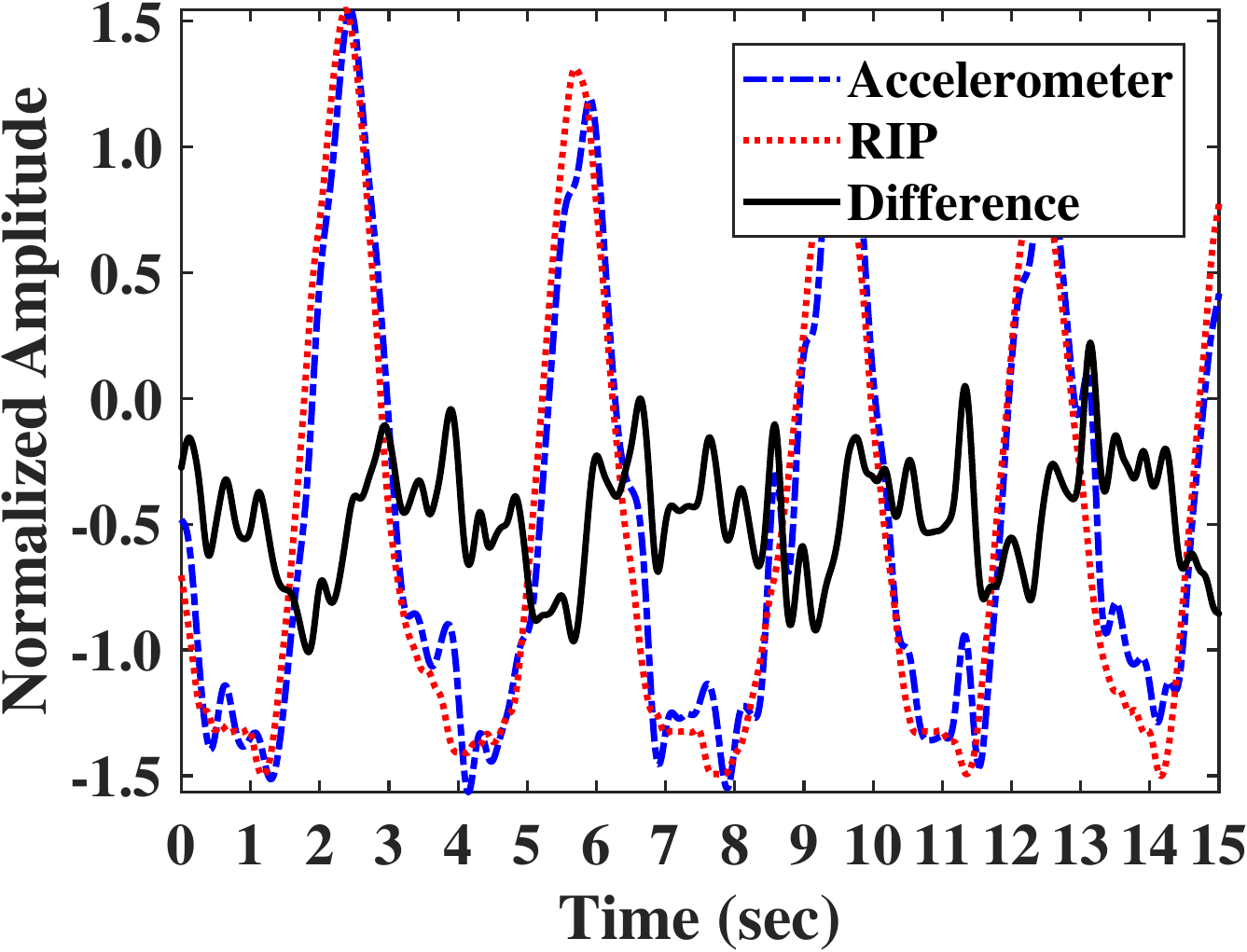}}
    \hfill
  \subfloat[\label{diff_pdf}]{%
        \includegraphics[width=.5\linewidth]{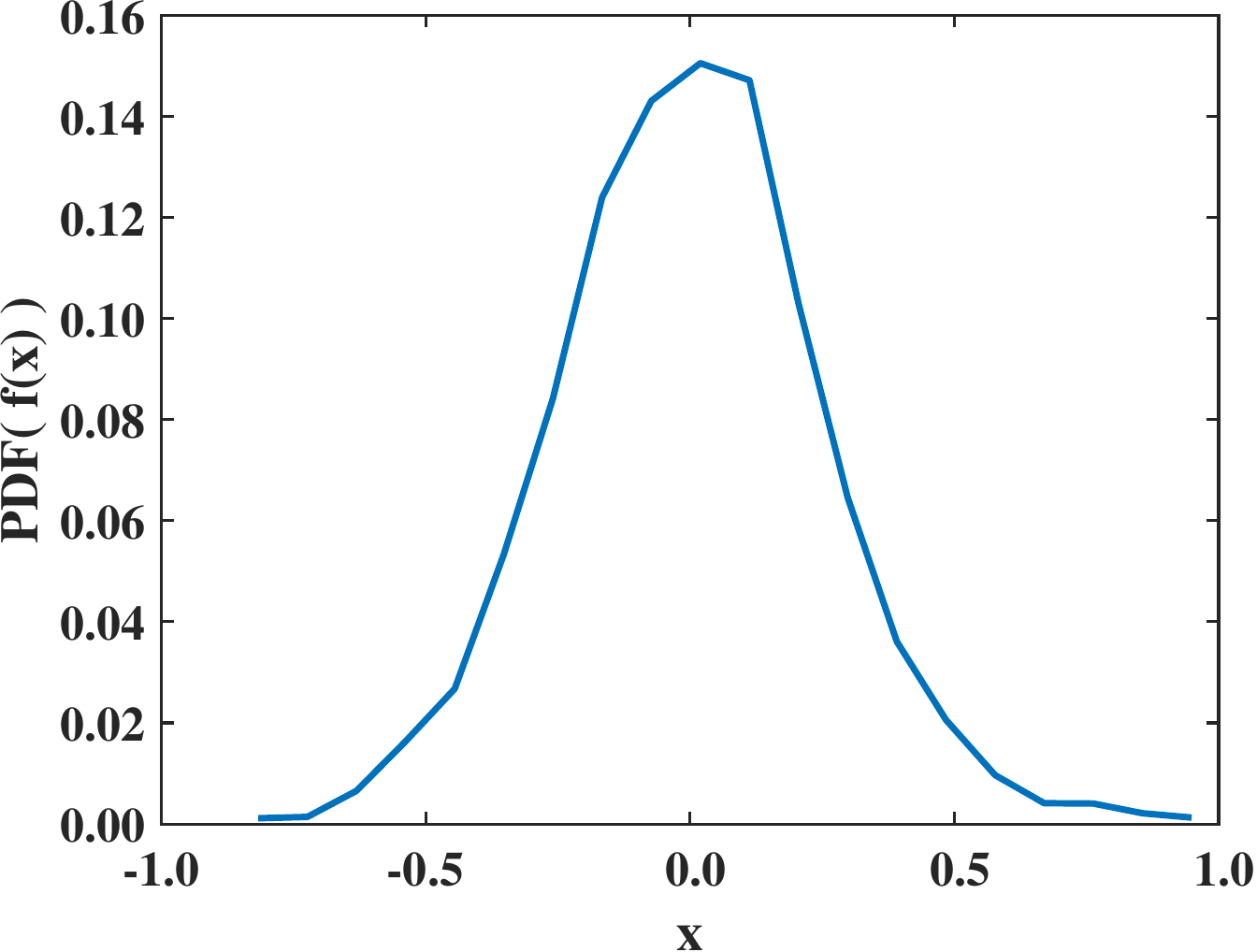}}
    \hfill  
    \caption{(a) Respiration Signals Measured by RIP and Accelerometer Sensor (b) Probability Density Function of Difference Between the Two Signals}
    \label{fig:res_acc_diff}
    \vspace{-.2in}
  \end{figure}

  In this paper, we focus on pairing between RIP-based and accelerometer based sensors. We collected RIP sensor data and accelerometer data from 30 participants.  A typical example of respiration signals   collected by a RIP sensor and an accelerometer is shown in Figure \ref{fig:res_acc_diff}(a). The difference between the two respiration signals is also shown in Figure \ref{fig:res_acc_diff}(a). \ref{fig:res_acc_diff}(b) shows that the difference can be modeled as a Gaussian distribution. With the modeling on the difference, we can derive  the following theorem.
  
  \begin{theorem}
  \label{thm:optimal}
  With the differences between two respiration signals
modeled as a Gaussian distribution with the mean $\mu$ and
standard deviation $\sigma$, and the signal values modeled with a uniform distribution in the range $[-\frac{d}{2}, \frac{d}{2} ]$, the optimal 
number of bits per sample, denoted as $b$, that can maximize the bit agreement on a shared $K$-bit secure key by the two devices collecting the two respiration signals satisfies the following equation:  
   \begin{displaymath}
    \begin{array}{l}
      {b\,\left(-\frac{\lambda_2 }{\lambda_3 }+\frac{2\,{\mathrm{e}}^{-\frac{{\lambda_1 }^2 }{2\,\sigma^2 }}}{2^{2\,b} \,d\,\sqrt{2\, \pi \, \sigma^2}}\right)}\,{{\left(-\frac{\lambda_2 }{\lambda_3 }\right)}}^{\frac{2\,K}{b}-1} +\mathrm{ln}\left(-\frac{\lambda_2 }{\lambda_3 }\right)=0~~,\\
      \end{array}\\
    \end{displaymath}
    where $\lambda_1 =\mu -\frac{2}{2^b }$, $\lambda_2 =\mathrm{erf}\left(\frac{\lambda_1 }{\sqrt{2\,\sigma^2}}\right)-\mathrm{erf}\left(\frac{\mu }{\sqrt{2\,\sigma^2}}\right)$, and $\lambda_3 =2^b \,d$. 
  \end{theorem}
  In the Theorem~\ref{thm:optimal}, the function $\mathrm{erf(.)}$ denotes the Gauss error function which is defined as $\mathrm{erf}(z)=\frac{2}{\sqrt{\pi}}\int_{0}^{z}e^{-x^2}dx$.
  The theorem is on $b$, the number of bits per sample. Usually, the  number of quantization levels,  denoted as $q$, can be determined as $q=2^b$. 
  
  The proof of Theorem \ref{thm:optimal} is straightforward. With the modeling on the differences, we can define the objective function as the product of $b$ and the probability of containing the differences between the signals within the same quantization interval, for all the bits in the key $K$. The optimization can be solved with  derivatives. 

  Our data collected from the 30 participants also show that the normalized signal values collected from the sensors can be approximated as a uniform distribution as shown in Figure \ref{fig:res_distrobution}. So, according to Theorem \ref{thm:optimal}, the optimal number of quantization levels is four, and correspondingly the number of bits per sample is two.
  
  \noindent{\bf Lloyd-Max Quantization:}
  Given the optimal parameter on the number of quantization levels, we further relax the constraint on the uniform quantizer by allowing quantization intervals of  different lengths. The Lloyd-Max quantizer \cite{lloyd1982least,max1960quantizing} is known for minimizing the quantization error by optimizing the quantization intervals. It finds the best values for quantization intervals and their corresponding quantization levels by considering the boundaries of the quantization intervals as the midpoints of the corresponding quantized values, $a_{i}=\frac{1}{2}(L_{i}+L_{i+1}), i=1, ..., M-1$, and the quantized values as the centroid of the quantization intervals, 
  \begin{displaymath}
  L_{i}=\frac{\int_{a_{i-1}}^{a_{i}}{xf_{X}(x)dx} }{\int_{a_{i-1}}^{a_{i}}{f_{X}(x)dx} }, i=0, ..., M-1, 
  \end{displaymath}
  where $f_{X}(x)$ is the probability density function (PDF) of the signal $X$. In this paper, we assume that the algorithm knows the PDF of signals in advance. The Lloyd-Max quantizer essentially allocates more quantization intervals to those ranges of values with a larger probability of occurrence. We hypothesize that the optimization by Lloyd-Max quantizer can also maximize the number of bits
  that can be agreed on by two sensors, as more quantization intervals, essentially more  bits, are allocated to the ranges with high probabilities of occurrence. Therefore, using the Lloyd-Max quantization, all the samples can be used during the key generation phase, and we can use the change points for synchronization purposes. 
  \subsection{Remaining Errors} 
  Both down-sampling and normalization, as well as quantization, can remove differences in biometric signals collected with different sensing mechanisms. But there can still remain some differences not removed by these techniques. These remaining differences can cause mismatches between bits generated by the quantizers of both devices. So, it is necessary to use some form of error correction to remove bit mismatches.  
   However, using a plain error correction code (ECC) not only is vulnerable to data leakage but also adds overhead to  the protocol and can decrease entropy. In this study, we modified the BCH error correction code to address these issues and correct the mismatches. We exploit the BCH method both as an encryptor and as an ECC without leakage vulnerability in the error correction stage.
  
   \noindent{\bf BCH-based Error Correction Technique:}\label{sec:ModifiedBCH}
  Assume a bit-string of length $K$. A BCH code can encode and decode it using a codeword of length $N=2^{m} -1$ ($m\geq 3, m\in \mathbb{N} $) such that it can correct up to $t$ errors where $t<2^{m-1}$ and the number of check bits is $N-K\leq mt$ \cite{han2010bch}. If we directly use this method for correcting mismatches, for a bit-string with length $K=128$, according to the table of the valid BCH pairs $(N,K)$ in \cite{han2010bch}, we need at least a codeword of length $N=255$ ($N=2^{m} -1$, $m=8$). Indeed, it needs 127 $(N-K=255-128)$ check bits. Hence, it can correct up to $t=16$ ($N-K\leq mt, N=255, K=128, m=8$) mismatched bits in a codeword. The drawback of using plain BCH code is that the overhead of check bits is too high for resource-constraint devices like wearables. 
  
\begin{figure*} [ht]
    \centering
    \includegraphics[width=5.6 in]{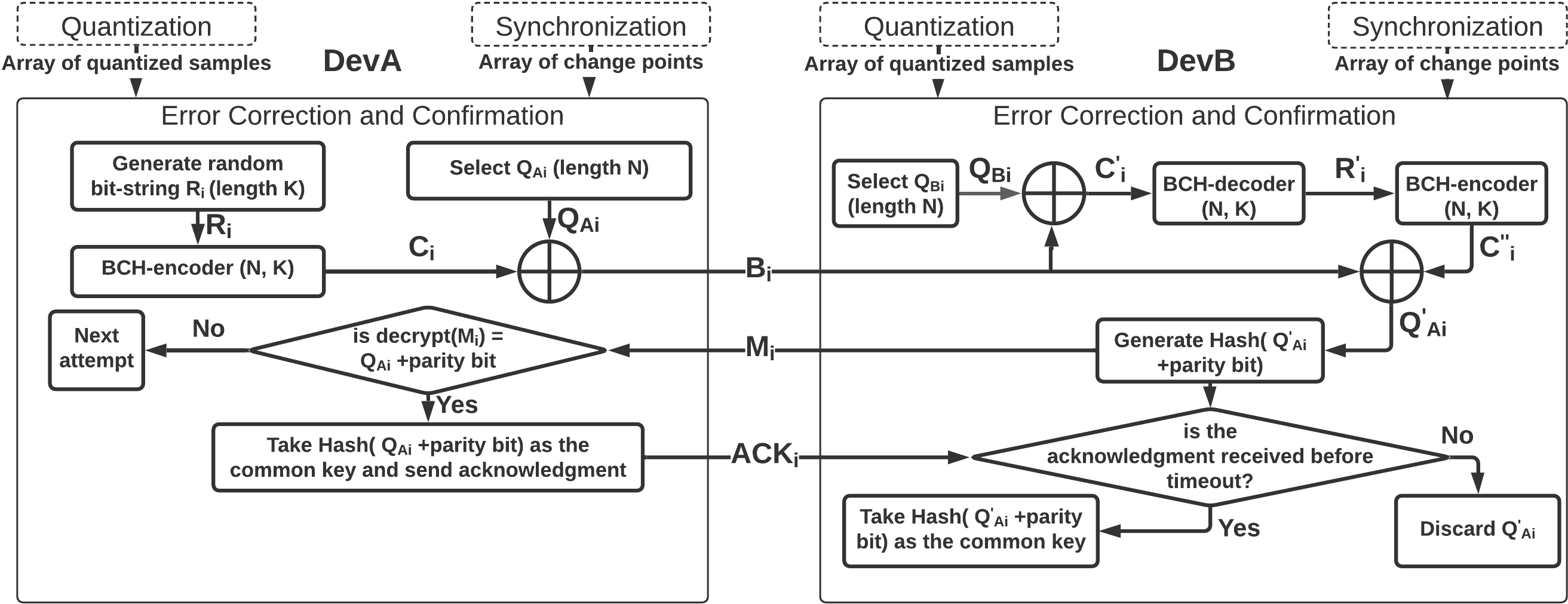}
    \caption{Steps for BCH Error Correction and Confirmation}
    \label{fig:bch_steps}
  \end{figure*}

  To eliminate the overhead, we modified the BCH method such that a BCH code of length $N$ can carry a message of length $N$.
  Our solution (illustrated in Figure~\ref{fig:bch_steps}) works as follows: For the pairing based on the $i$th CP, $DevA$ generates a random bit-string $R_{i}$, of length $K$, and uses a BCH-encoder to encode it into codeword $C_{i}$, of length $N$. Then $DevA$ selects $Q_{Ai}$ which is $N$ bits of quantized bits around the $i$th CP, and generates block $B_{i}$ by performing an exclusive or (XOR) operation on $C_{i}$ and $Q_{Ai}$. It then sends $B_{i}$ to $DevB$ over an unencrypted channel. When $DevB$ receives $B_{i}$, it selects its $Q_{Bi}$ and performs XOR on $B_{i}$ and $Q_{Bi}$ to extract $C'_{i}$. Then it uses the BCH-decoder to extract $R'_{i}$. As long as the number of mismatch bits between $Q_{Ai}$ and $Q_{Bi}$ is less than t, we have $R'_{i}=R_{i}$. Afterward, $DevB$ uses the BCH-encoder to encode $R'_{i}$ to $C''_{i}$ and does XOR on $C''_{i}$ and $B_{i}$ to derive $Q'_{Ai}$. Finally, $DevB$ uses the derived key $Q'_{Ai}$ to encrypt a confirmation message $M_{i}$ and sends it to $DevA$. 
  
   Upon receiving $M_{i}$, if $DevA$ was successful in decrypting $M_{i}$, it means $Q'_{Ai}=Q_{Ai}$, therefore, $DevA$ uses $Q_{Ai}$ as the common secure key and sends an acknowledgment message to $DevB$. If $DevA$ fails to decrypt $M_{i}$, it will continue with the next pairing attempt. This procedure will be continued until a common key is found or a timeout occurs. Since $N=2^{m} -1$ is an odd number (e.g., 127 or 255), we can use a parity bit for $Q_{Ai}$ at the final stage. In addition, for further amplification, a hashing algorithm like SHA2-256 can be applied to the generated key. Wearable devices can run the B2P protocol regularly to generate a new secure key.
  
   \noindent{\bf Security Analysis of BCH Error Correction and Confirmation:}
  We note that the adversary can receive the $B_{i}$ on the public channel. She is not able, however, to derive $Q_{Ai}$ from $B_{i}$ because she has no knowledge about $R_{i}$ and the quantized bits in $DevA$. Even if the adversary were to mimic the real user's respiration, it would be hard to achieve a common key because she cannot precisely synchronize her signal with $DevA$. 
\begin{figure} []
  \centering
  \includegraphics[width=.5\columnwidth]{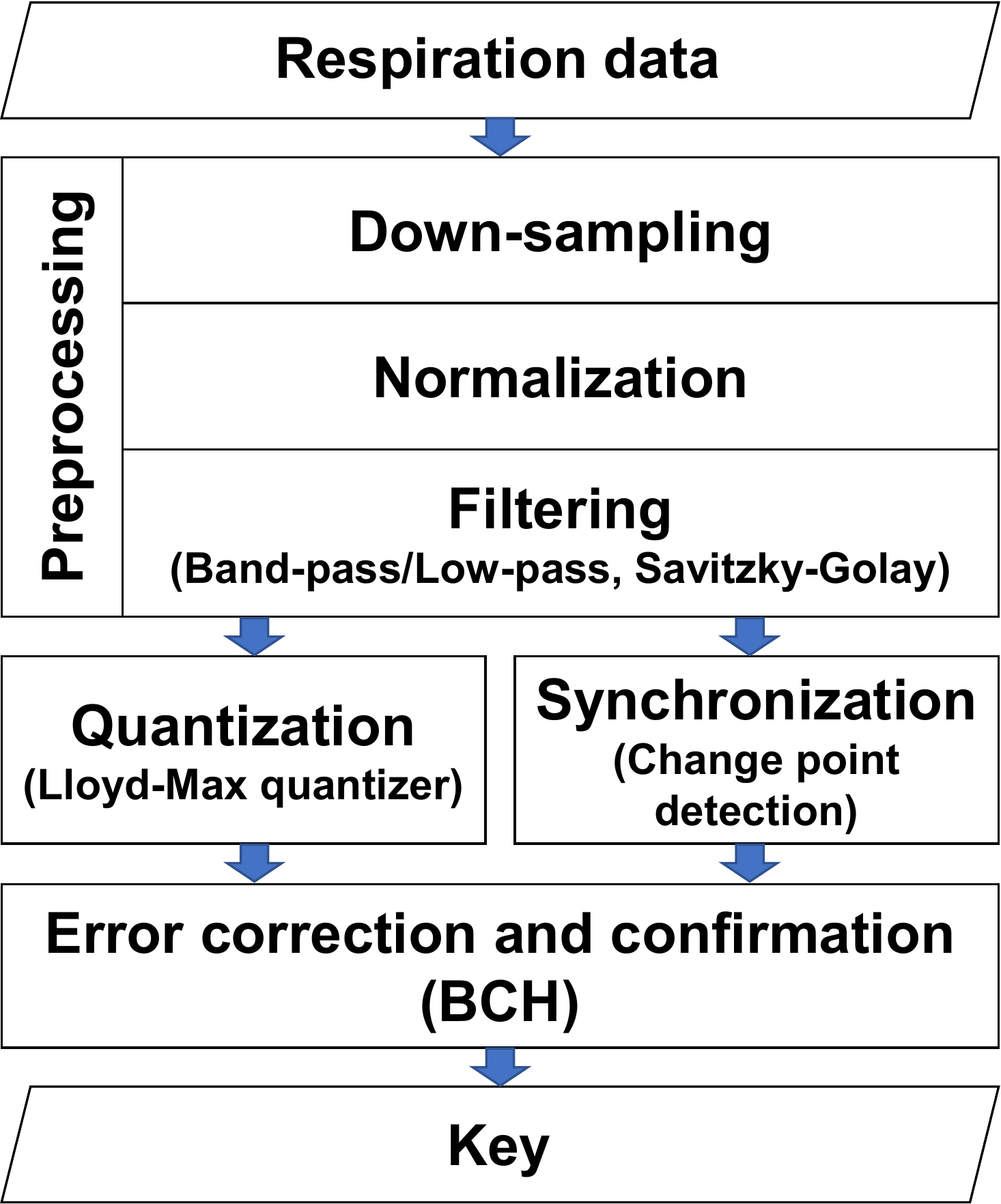}
  \caption{Protocol Overview}
  \label{fig:method_steps}
  \vspace{-0.2in}
\end{figure}
\subsection{Design and Implementation of the B2P Protocol}
The description of the challenges faced by B2P naturally
gives rise to the an architecture of the protocol, which is described
in Figure~\ref{fig:method_steps}.
The protocol consists of four main modules. The first
module, called {\em Signal Preprocessing}, is designed to handle the differences in the data collected by different sensing mechanisms and the
filtering of the noise outside the respiration frequency range. It
implements the necessary downsampling of the data, the normalization
of the amplitudes and the filtering. 

The {\em Quantization}  module is designed to 
tolerate the differences (primarily different quantization levels)
caused or applied by the differences in sensing
mechanisms.

The {\em Synchronization} module uses change point
detection (CPD) methods to detect changes that can be used to
synchronize signals collected by the different sensing mechanisms.

The {\em Error Correction and Confirmation} module is designed to correct bit
errors caused by differences that can not be removed by the
Quantization module. It uses a variation of the Bose Chaudhuri
Hocquenghem (BCH) code do do that. This module also contains the
mechanics of the shared-key generation, which we described in
Section~\ref{sec:ModifiedBCH}. The output of the protocol
implementation is the shared key.
  \subsection{ Impersonation Attack Mitigation}
Since breathing activities can be easily observed by an adversary, impersonation attacks can be launched with the information obtained from the observations to pair with legitimate devices. The impersonation attacks can be especially effective in compromising the pairing between devices with different sensing mechanisms.  For example, as described above, filtering techniques  are used to remove differences caused by different sensing mechanisms for better agreement on key bits. But the filtering can also smoothen out sudden changes that can be captured by the respiration sensors but may not be observable by the adversary. So instead of relying on the sudden changes that can not be observed by the adversary to mitigate impersonation attacks, we propose a method to mitigate the attacks by segmenting the signals. 

The segmenting method divides an output bit string from the quantizer, with respect to a change point, into $n_{seg}$ segments. Instead of using bits from successive segments for pairing, the method may skip neighboring segments and use only bits from selected segments for pairing. The rationale behind the method is as follows: The smoothing effect brought by the filtering can make the next few samples of a sensor signal predictable if the adversary can estimate the value of one sample with reasonable accuracy. But the predictability can reduce greatly if the samples are relatively distant from the sample that the adversary may estimate accurately.  So the best segment choice should contain segments that are distant from each other.
In our implementation, when $n_{seg}=10$ and five segments are needed for pairing, we choose segments with index  ${2, 5, 6, 7, 10}$. This selection has proven to be effective in our experiments.
  \section{Evaluation}\label{sec: experiment}
\begin{figure} [ht]
    \centering
    \includegraphics[width=.5\columnwidth]{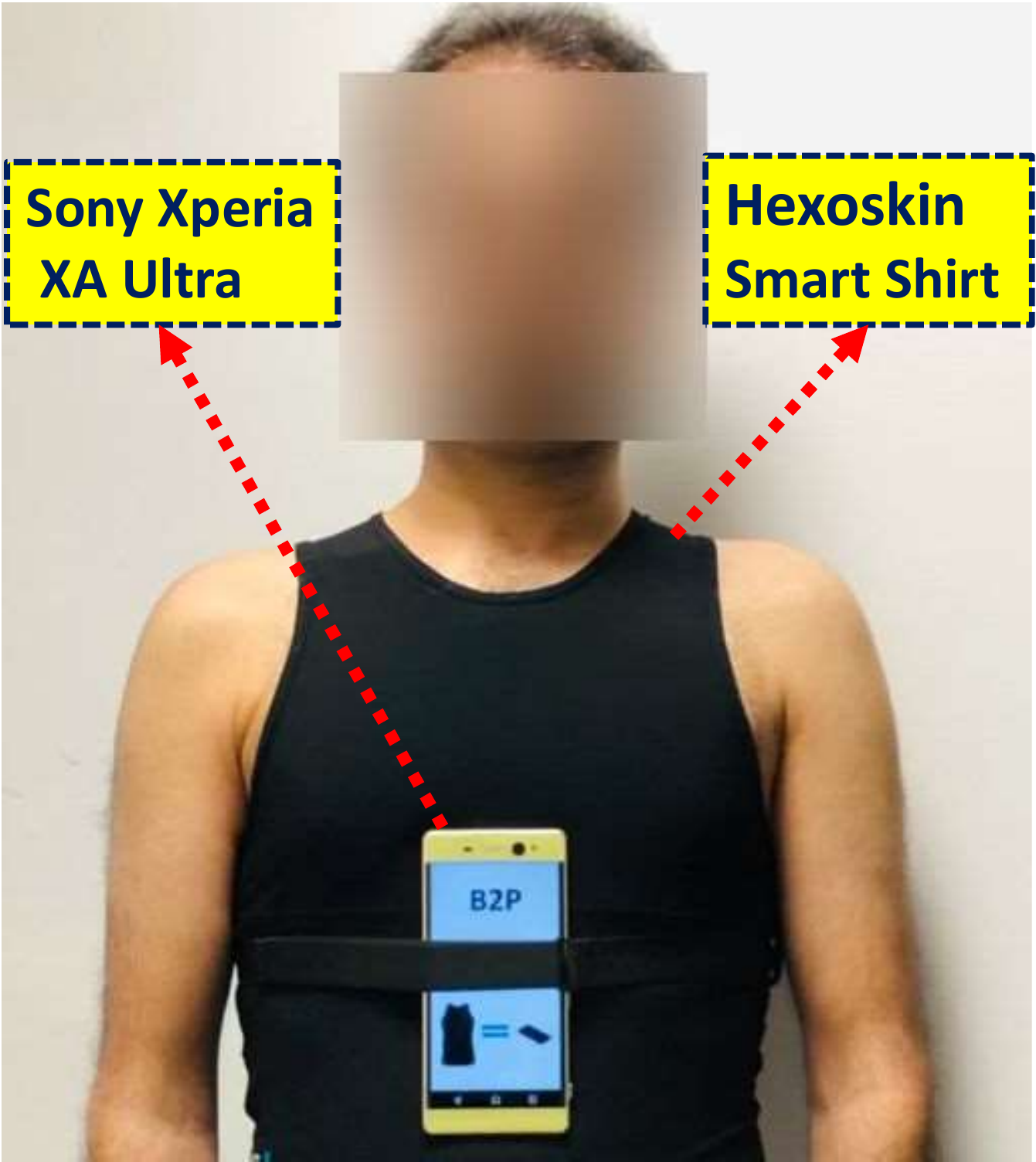}
    \caption{Devices Used in the Experiment }
    \label{fig: experiment_setup}
    \vspace{-.14in}
  \end{figure}

  In this section, we evaluate the performance of the B2P pairing protocol. The devices used for the performance evaluation are a Hexoskin smart shirt \cite{Hexoskin} and two smartphones. More specifically, we use the RIP sensor on the Hexoskin smart shirt and the accelerometers on the smartphones to collect respiration signals for pairing. The specifications of the sensors used in pairing are presented in Table \ref{devices_spec}. As shown in Table \ref{devices_spec}, the RIP sensor and the accelerometers sense the respiration activities in very different ways.  
  \begin{table*}[]
    \centering
    \caption{Sensors Specifications}
    \begin{tabular}{|l|l|l|l|l|l|}
    \hline
    \textbf{Sensor} & \textbf{Device} & \textbf{Sensing Mechanism} & \textbf{Sampling Rate (Hz)} & \textbf{Resolution (bit)} & \textbf{Unit} \\ \hline
    RIP & Hexoskin Smart Kit & Measuring pulmonary ventilation & 128 & 16 & mL \\ \hline
    \multirow{2}{*}{3-axis Accelerometer} & Sony Xperia XA Ultra & \multirow{2}{*}{Measuring acceleration forces} & \multirow{2}{*}{100} & \multirow{2}{*}{13} & \multirow{2}{*}{G} \\ \cline{2-2}
     & HTC One E9 Plus &  &  &  &  \\ \hline
    \end{tabular}
    \label{devices_spec}
    \end{table*}
  After receiving approval from the Institutional Review Board (IRB) in our university, we collected data from 30 participants, including 19 males and 11 females, with a mean age of 30.14 years (between 20 and 43). The experiment setup is shown in Figure \ref{fig: experiment_setup}. The  participants were asked to wear the smart shirt, attach the smartphone to their chest, and sit on a chair. The experiment duration was about 5 minutes per participant.
  \subsection{Performance Metrics}
  We evaluate the pairing performance using the following metrics:
  
  \noindent{\em Key Generation Rate:} This metric is defined as the number of keys generated per second. Practical systems often use keys with a length of 128 or 256 bit. We, therefore, focus on these two key lengths in these evaluations. In general, a higher key generation rate is desired. 
  
  \noindent{\em Entropy of Generated Keys:} This metric is defined as a measurement of ``surprise'' contained in the generated keys by measuring the entropy \cite{klir2013uncertainty}. Given a randomly generated key $W$, its entropy can be computed as $H(W)=-P(w_0)\log_2 P(w_0)-P(w_1)\log_2 P(w_1)$, where $P(w_0)$ and $P(w_1)$ are the probability of all zeros and ones in $W$, respectively.
  The ideal entropy for generated keys is 1.
  
  \noindent{\em False-Positive Ratio:} In this paper, we define the false-positive ratio as the percentage of pairing attempts that incorrectly generate common keys among all the expected unsuccessful attempts. The metric indicates the likelihood of the adversary successfully pairing with a legitimate device by the adversary. The FPR can be computed as:$FPR=\frac{FP}{EN}$, where $FP$ is the number of incorrectly generated keys, and $EN$ is the number of expected unsuccessful attempts.
  
  \noindent{\em False Negative Ratio:} This metric is defined as the percentage of incorrectly unsuccessful pairing attempts among all the expected successful pairing attempts. It indicates the probability that two legitimate devices attached to the same body can not pair successfully. FNR is computed as $FNR=\frac{FN}{EP}$, where $FN$ is the number of incorrectly missed keys, and $EP$ is the number of expected successful attempts.
  
  \noindent{\em Bit Agreement Ratio (BAR):} We define this metric as the percentage of matched bits in bit strings generated by two devices.  We use BAR to evaluate B2P's resistance to impersonation attacks, such as those described in Section \ref{subsec:attacks}. Since a successful pairing attempt needs a 100\% bit agreement ratio on key bits, the BAR indicates how close  an adversary is  to pair with a legitimate device. 
  
  Table~\ref{tbl:B2P Performance} shows all the performance metrics used in this study and the best B2P results.
  \begin{table}[]
    \centering
    \caption{Achievable Key Generation Rates}
    \begin{tabular}{|l|l|l|}
    \hline
    \textbf{Performance   Metric} & \textbf{128-bit Key} & \textbf{256-bit Key} \\ \hline
    Key Generation Rate (key/sec) & 0.3526 & 0.3508 \\ \hline
    Entropy & 0.99 & 0.97 \\ \hline
    False Positive Ratio (\%) & 0.064 & 0.028 \\ \hline
    False Negetive Ratio (\%) & 0.0003 & 0.0007 \\ \hline
    Bit Agreement Ratio (\%) & 64 & 62 \\ \hline
    \end{tabular}
    \label{tbl:B2P Performance}
    \end{table}
  \subsection{Impact of Parameters and Methods Choices}\label{subsec: impact_of_param}
\begin{figure} [htbp]
    \centering
    \subfloat[\label{fig: code_length}]{%
       \includegraphics[width=0.5\linewidth]{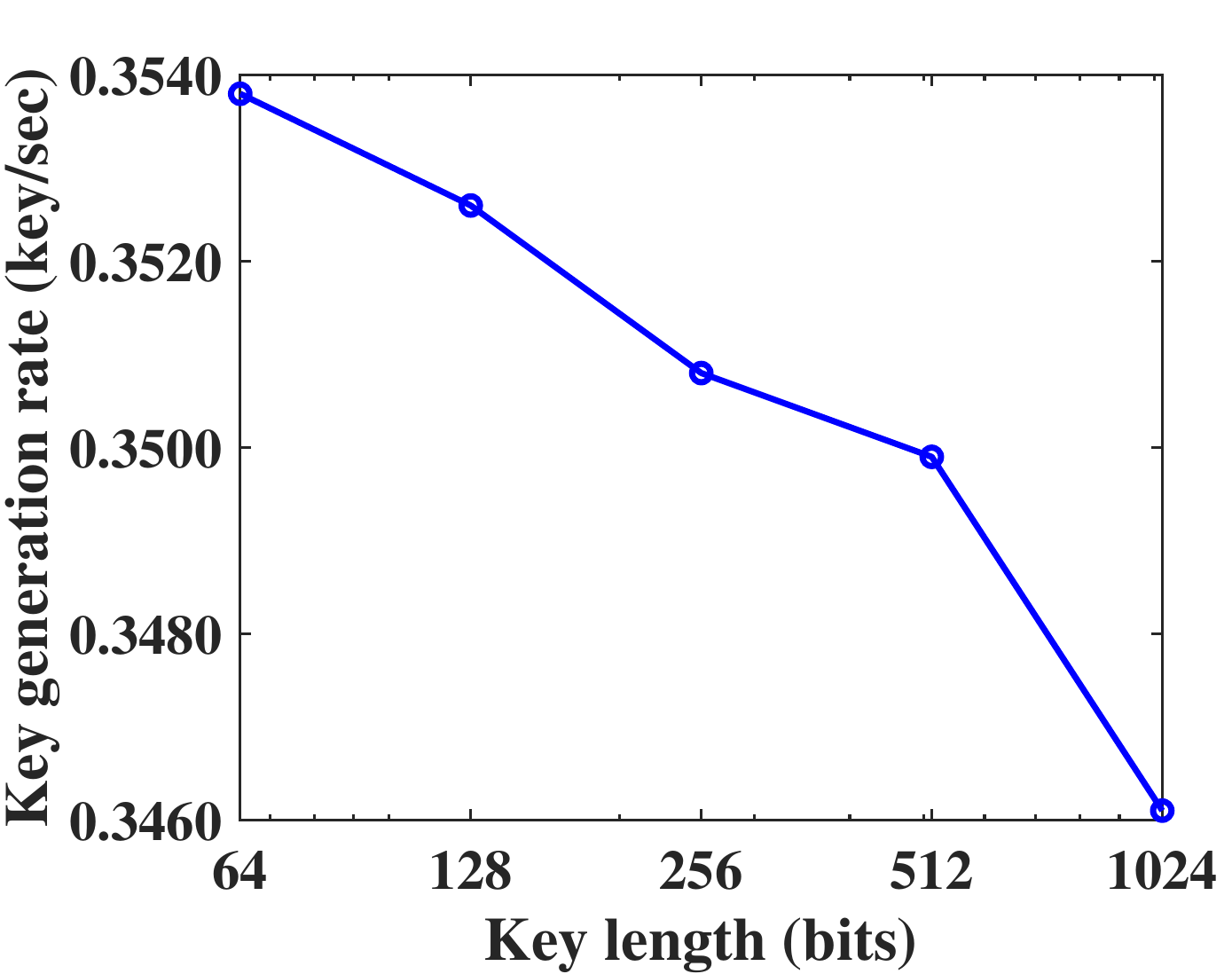}}
       \hfill
       \subfloat[\label{fig: down-sampling effect}]{%
       \includegraphics[width=0.48\linewidth]{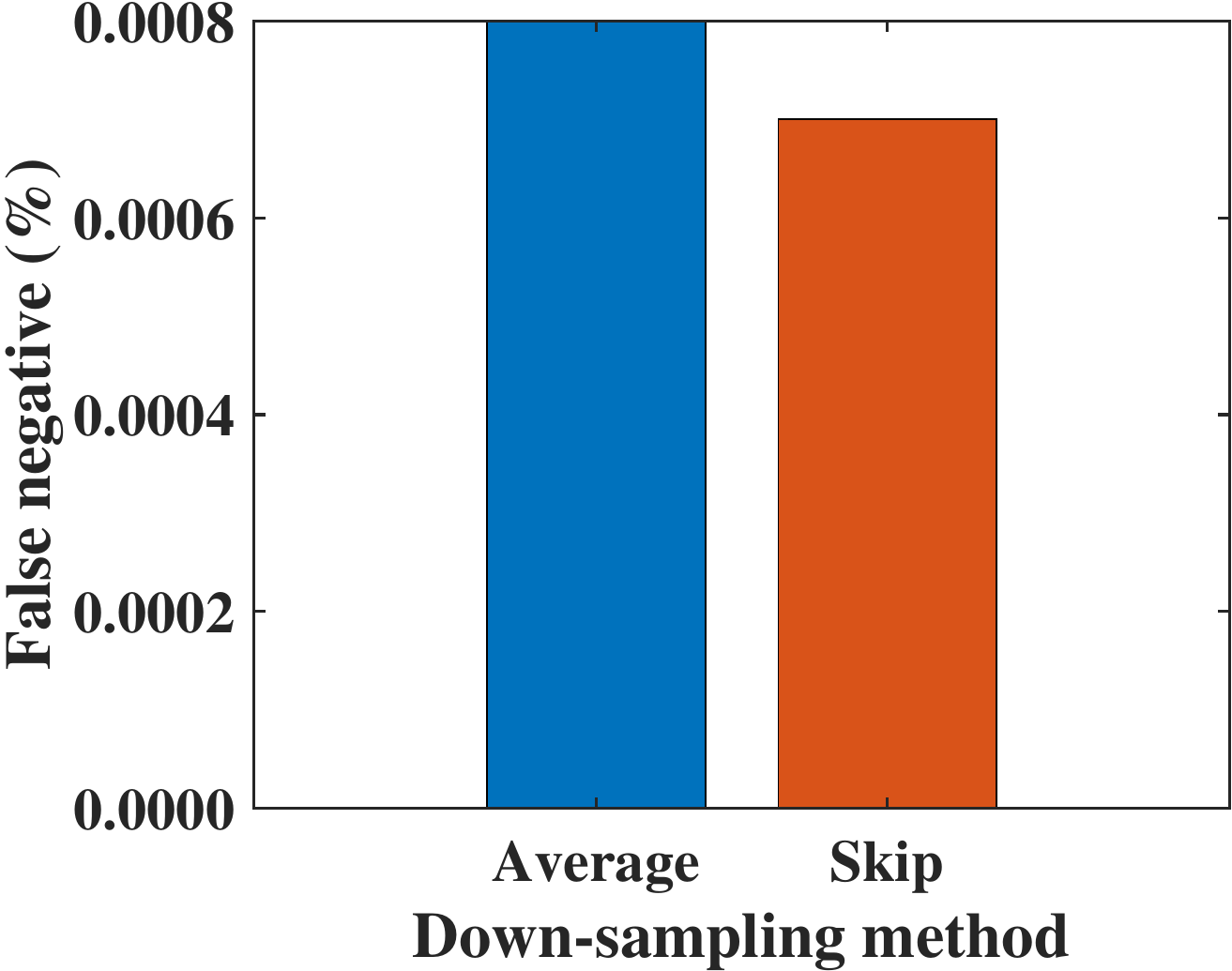}}
  \caption{(a) Effect of Different Key Length on KGR, (b)Impact of Down-Sampling on False Negative Ratio}
  \label{fig:code_length_down-sampling effect} 
  \end{figure}

  B2P can generate about  0.35 keys per second across a variety of key lengths, as shown in Figure \ref{fig: code_length}. The key generation rate decreases slightly when the key length increases, as pairing attempts for longer keys are more likely to fail because of the sensing noise and the difference in sensing mechanisms. B2P usually makes  pairing attempts at  detected change points. So the number of pairing attempts does not change with the key length. 
  For instance, a 1024-bit key can be based on a signal fragment with three respiration cycles, and there can be five CPs in the signal fragment. As a result, the 1024-bit keys generated at these 5 CPs
  will have some overlap.
  Table \ref{tbl: best parameter} shows the parameters and methods
  used to generate the results described above. In the following experiments, we vary the parameters that are being addressed in each experiment and keep the rest of the parameters constant as shown in  Table \ref{tbl: best parameter}.
  \begin{table}[]
    \centering
    \caption{Typical Parameters}
    \begin{tabular}{|l|l|}
    \hline
    \textbf{Parameter} & \textbf{value} \\ \hline
    Key length (bits) & 256 \\ \hline
    BCH parameters (N,K) & (255, 115) \\ \hline
    Number of bits per sample & 2 \\ \hline
    Sync offset (ms) & 250 \\ \hline
    Number of CP offsets & 2 \\ \hline
    CP num per peak-valley & 1 \\ \hline
    Code type & Gray \\ \hline
    CPD method & STD \\ \hline
    Down-sampling & skip \\ \hline
    \end{tabular}
    \label{tbl: best parameter}
    \vspace{-.151in}
    \end{table}
  \subsubsection{\textbf{Impact of Down-Sampling}}
  In this experiment, we compare the two down-sampling methods: skip and average. As shown in Table \ref{devices_spec}, the RIP sensor has a much higher sampling rate than the accelerometers. So down-sampling is applied to respiration signals collected with the RIP sensor.
  
  Table \ref{tbl: down-sampling effect} shows the pairing performance of the two methods. The key generation rates for both methods are very close.  We believe it is because of the filtering step designed to capture respiration dynamics. The low-pass and band-pass filters used in the filtering both have a cut-off frequency much lower than the sampling rates. So the differences between the two methods occurring at the sampling rates, much higher than the cut-off frequency, are largely filtered out.
  \begin{table}[]
    \centering
    \caption{Impact of Down-Sampling on Performance Metrics}
    \begin{tabular}{|l|l|l|}
    \hline
    \textbf{Downsampling} & \textbf{Average} & \textbf{Skip} \\ \hline
    Key generation rate (key/sec) & 0.2385 & 0.2382 \\ \hline
    Entropy & 0.94 & 0.96 \\ \hline
    False positive (\%) & 0.027 & 0.0281 \\ \hline
    False negative (\%) & 0.0008 & 0.0007 \\ \hline
    \end{tabular}
    \label{tbl: down-sampling effect}
    \vspace{-.151in}
    \end{table}
  \subsubsection{\textbf{Impact of Filtering}}
\begin{figure} [htbp]
    \centering
  \subfloat[\label{fig: filtering effect - KeyGenerationRate}]{%
       \includegraphics[width=0.5\linewidth]{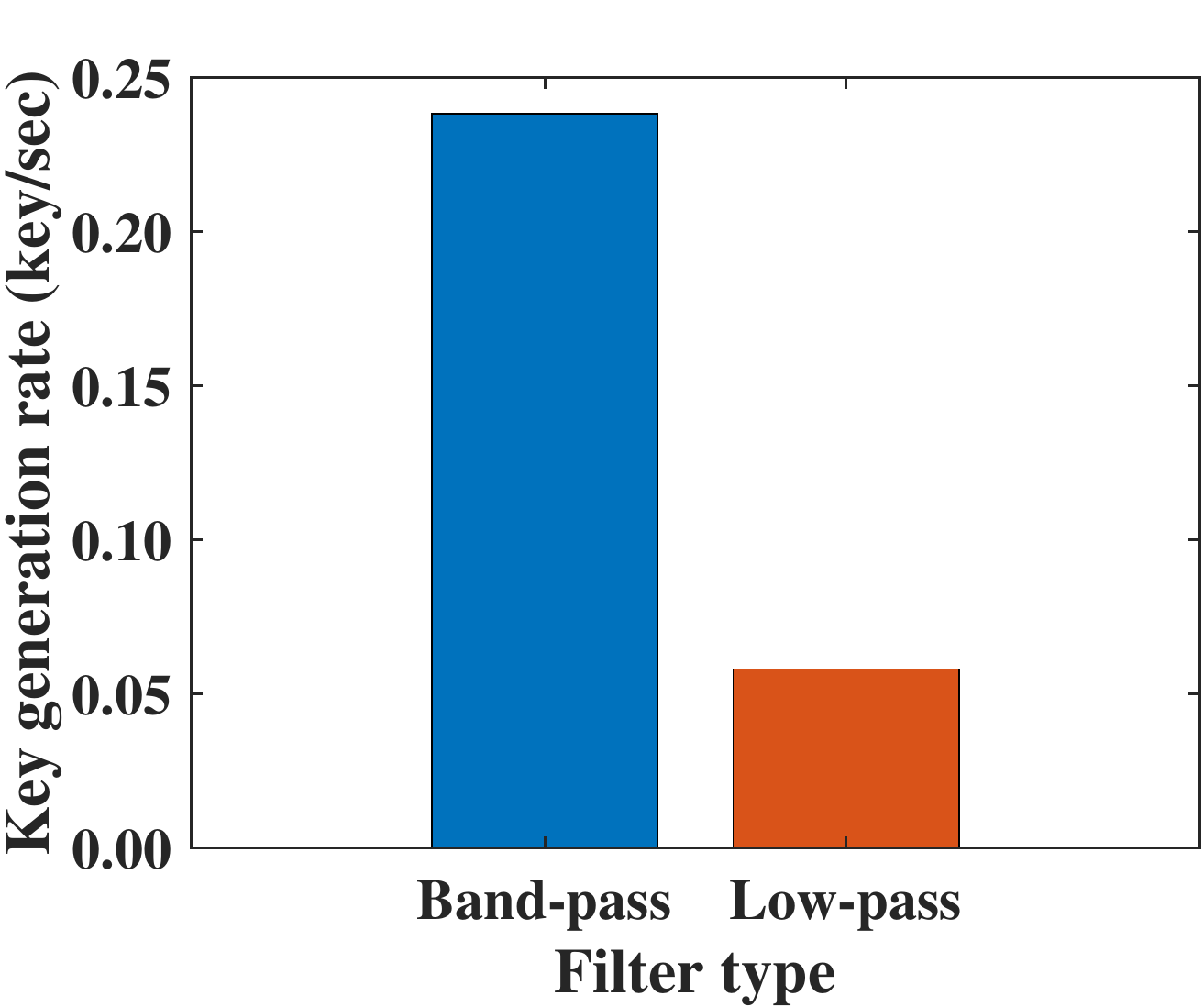}}
    \hfill
  \subfloat[\label{fig: filtering effect - Entropy}]{%
        \includegraphics[width=0.5\linewidth]{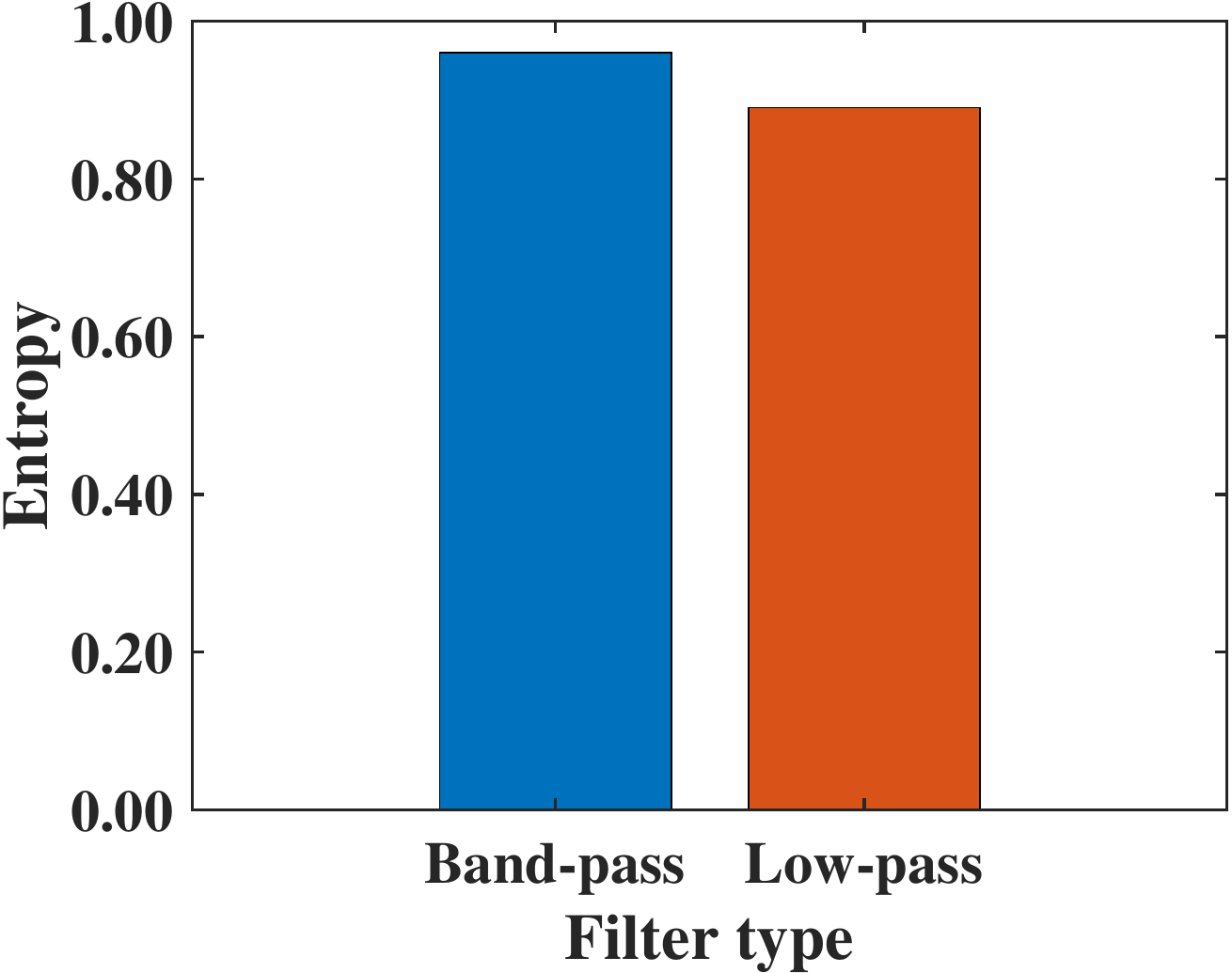}}
        \hfill
  \subfloat[\label{fig: filtering effect - FalsePositive}]{%
        \includegraphics[width=0.5\linewidth]{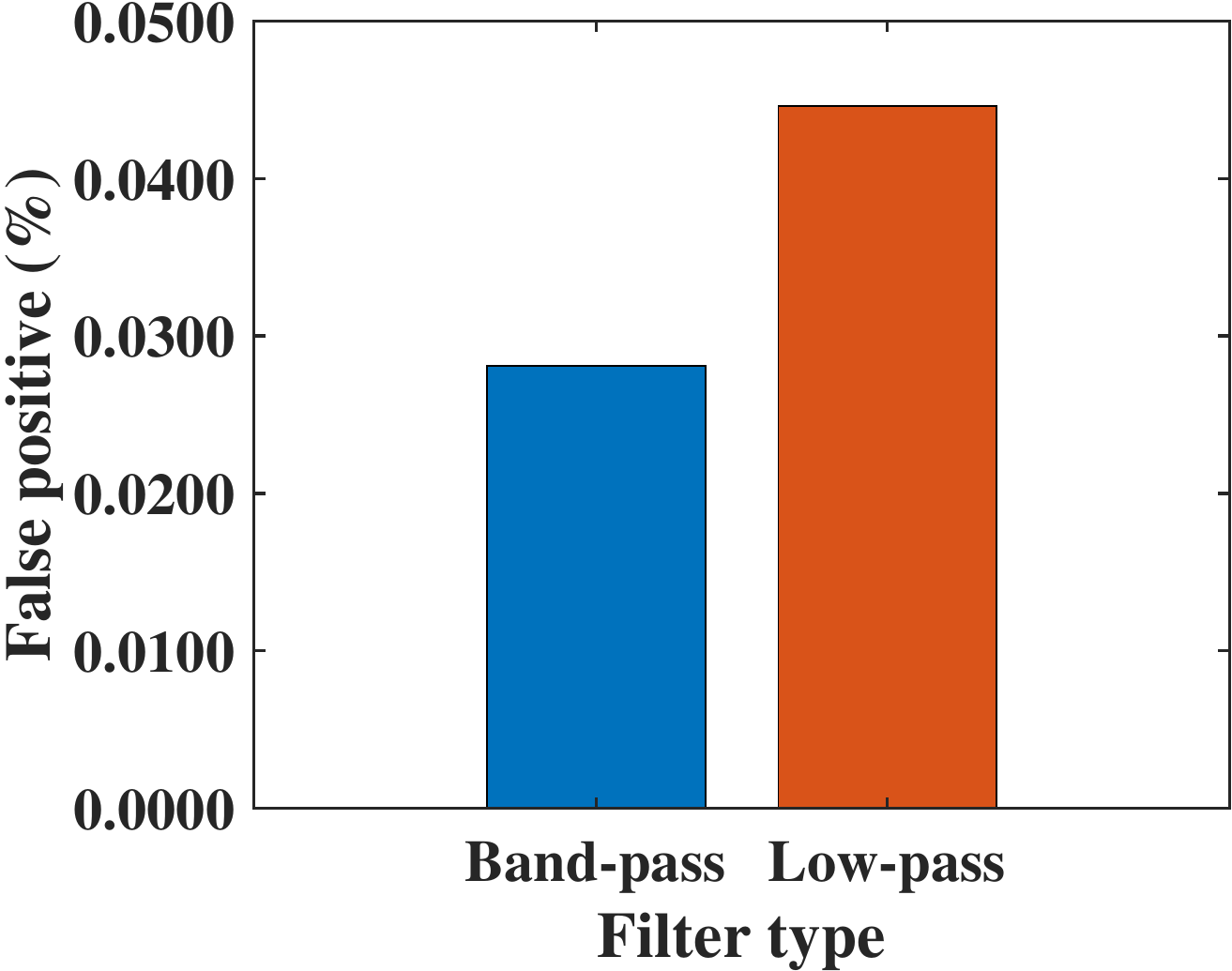}}
        \hfill
  \subfloat[\label{fig: filtering effect - FalseNegative}]{%
        \includegraphics[width=0.5\linewidth]{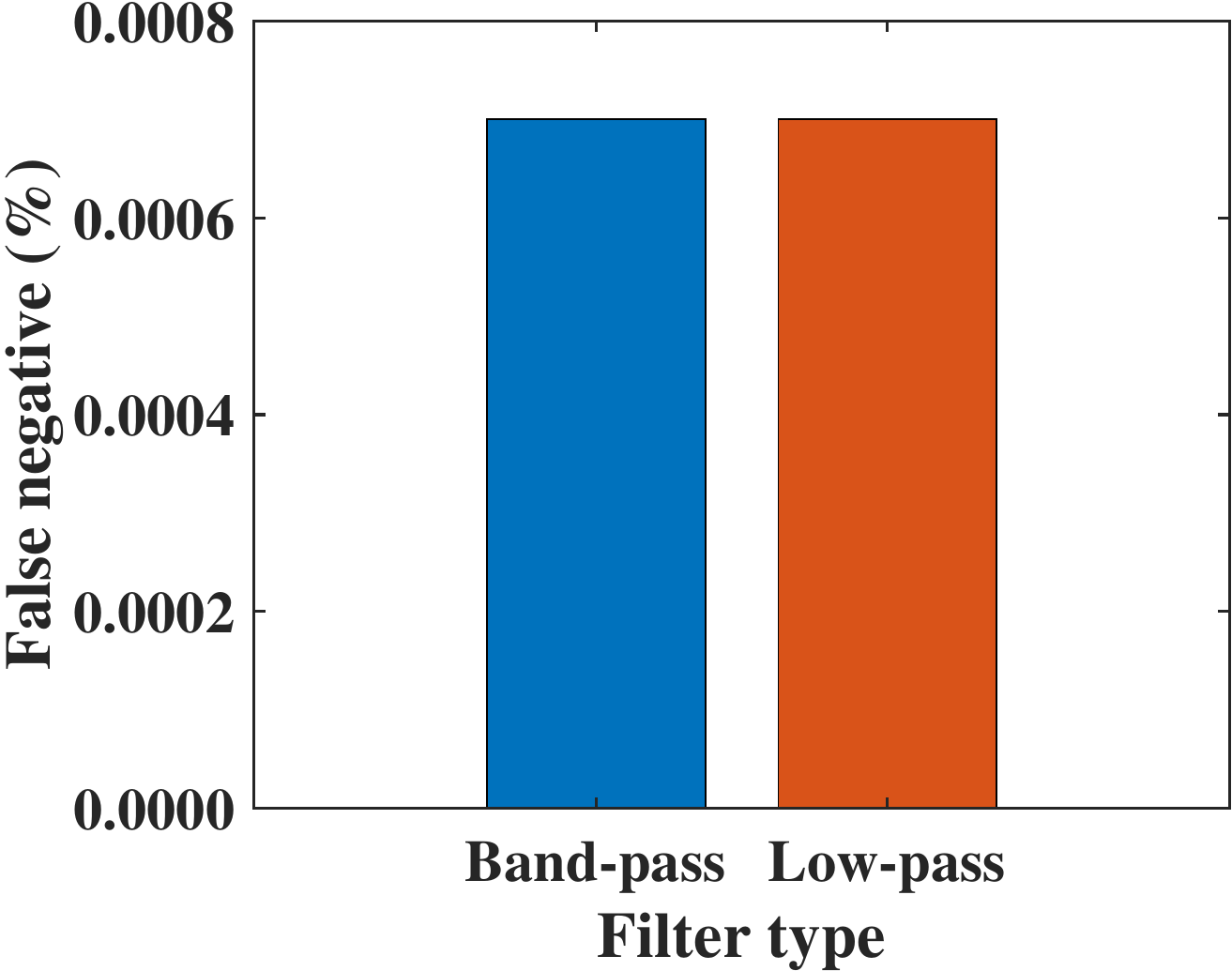}}
  
  \caption{Impact of Filtering on Performance Metrics (a) Key Generation Rate, (b) Entropy, (c) False Positive Ratio, (d) False Negative Ratio.}
  \label{fig: filtering effect} 
  \end{figure}

  Choosing an effective filter can significantly affect the system's performance, as it removes undesired frequencies and noises. Figure~\ref{fig: filtering effect - KeyGenerationRate} shows how, in comparison with a low-pass filter, a band-pass filter with cut-off frequency in the range of (0.1, 0.5) causes a significant improvement of the KGR. The reason is that the accelerometer signal contains a lot of external disturbance that is  due to the user's motion. Since most of the user motion happens at low frequency, the band-pass filter removes most of them while the low-pass filter does not affect the user's low-frequency motions. The most considerable motion noise occurs at the end of the exhalation phase (the valleys in Figure~\ref{res_acc_diff}) because the diaphragm usually has a short stop there \cite{francis2016ultrasonographic}. This stop point affects the detected signal of both devices. The band-pass filter also smoothens these points and leads to an increase in entropy, as shown in Figure~\ref{fig: filtering effect - Entropy}. Figure~\ref{fig: filtering effect - FalsePositive} shows an improvement in FPR that is due to having a smoother signal. However, the FNR does not change by changing the filtering method.
  
  As Table~\ref{tbl: golay filter effect} shows, applying the Savitzky-Golay filter generally improves all the performance metrics because it removes undesired dynamics that the band-pass filter can not. 
  \begin{table}[]
    \centering
    \caption{Impact of Applying the Savitzky-Golay Filter on Performance Metrics}
    \begin{tabular}{|l|p{2cm}|p{2cm}|}
    \hline
    \textbf{} & \textbf{Without Savitzky-Golay filter} & \textbf{With Savitzky-Golay filter} \\ \hline
    \textbf{Key generation (key/sec)} & 0.2382 & 0.2532 \\ \hline
    \textbf{Entropy} & 0.88 & 0.96 \\ \hline
    \textbf{False positive (\%)} & 0.0285 & 0.0281 \\ \hline
    \textbf{False negative (\%)} & 0.0008 & 0.0007 \\ \hline
    \textbf{Bit agreement (\%)} & 59.1 & 58.66 \\ \hline
    \end{tabular}
    \label{tbl: golay filter effect} 
    \vspace{-.171in}
    \end{table}
  \subsubsection{\textbf{Impact of Synchronization}}
  Synchronizing the signals collected from independent devices is always challenging for pairing \cite{DiRienzo2020, shang2020audiokey, jin2015magpairing}. The task is even more challenging for pairing with sensors with different sensing mechanisms. As described in Section~\ref{sec:CPD}, we utilize the change point detection (CPD) method to detect the significant changes in the signal observed by both devices and, we use change points to synchronize the two signals. We experimented with five parameters that affect the detected change points and pairing performances: CPD method, number of change points per peak-valley, CPD threshold, CP offset, and synchronization offset. 
  
\begin{figure} [htbp]
    \centering
  \subfloat[\label{fig: CPDMethod - KeyGenerationRate}]{%
       \includegraphics[width=0.5\linewidth]{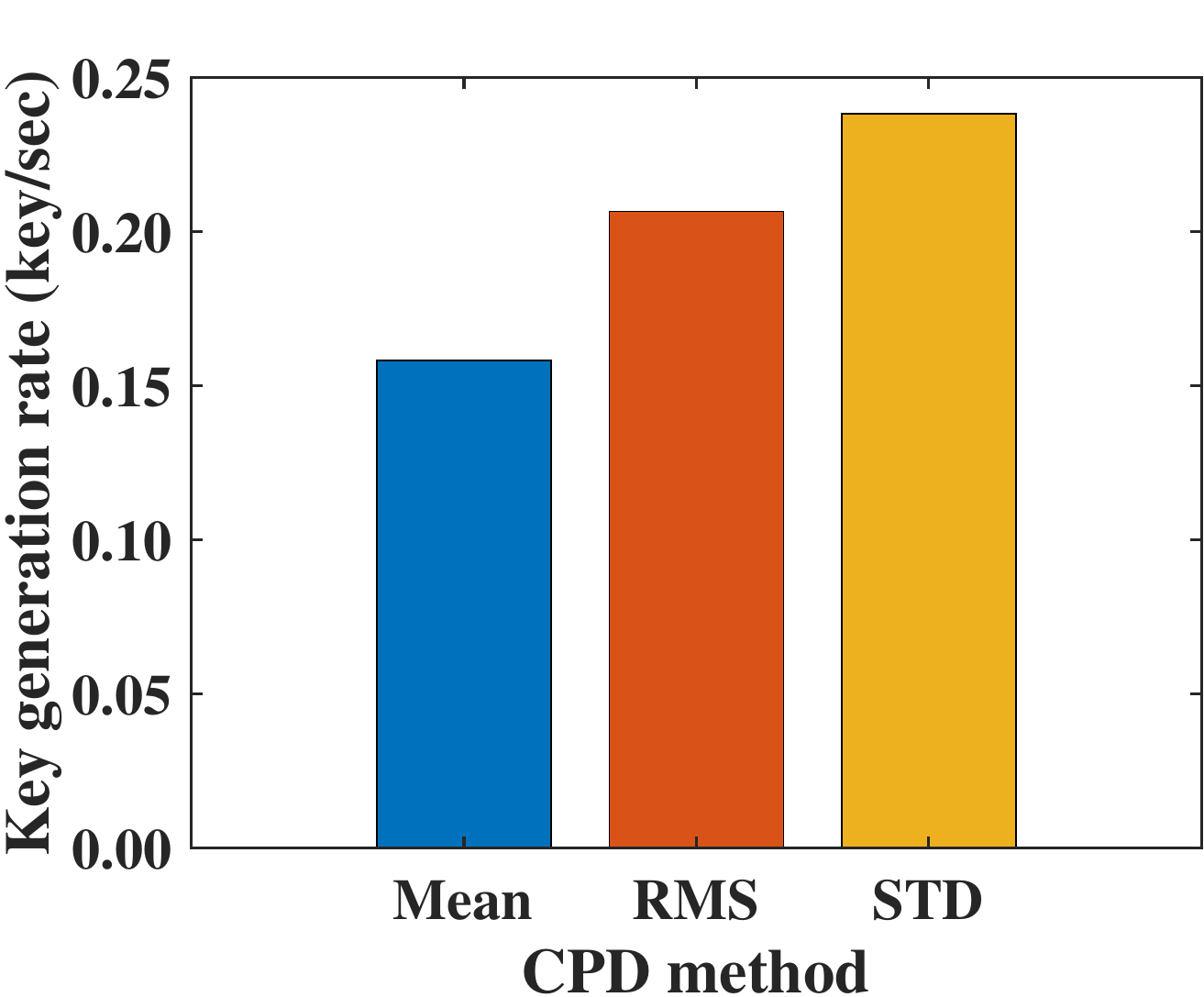}}
    \hfill
  \subfloat[\label{fig: CPDMethod- Entropy}]{%
        \includegraphics[width=0.5\linewidth]{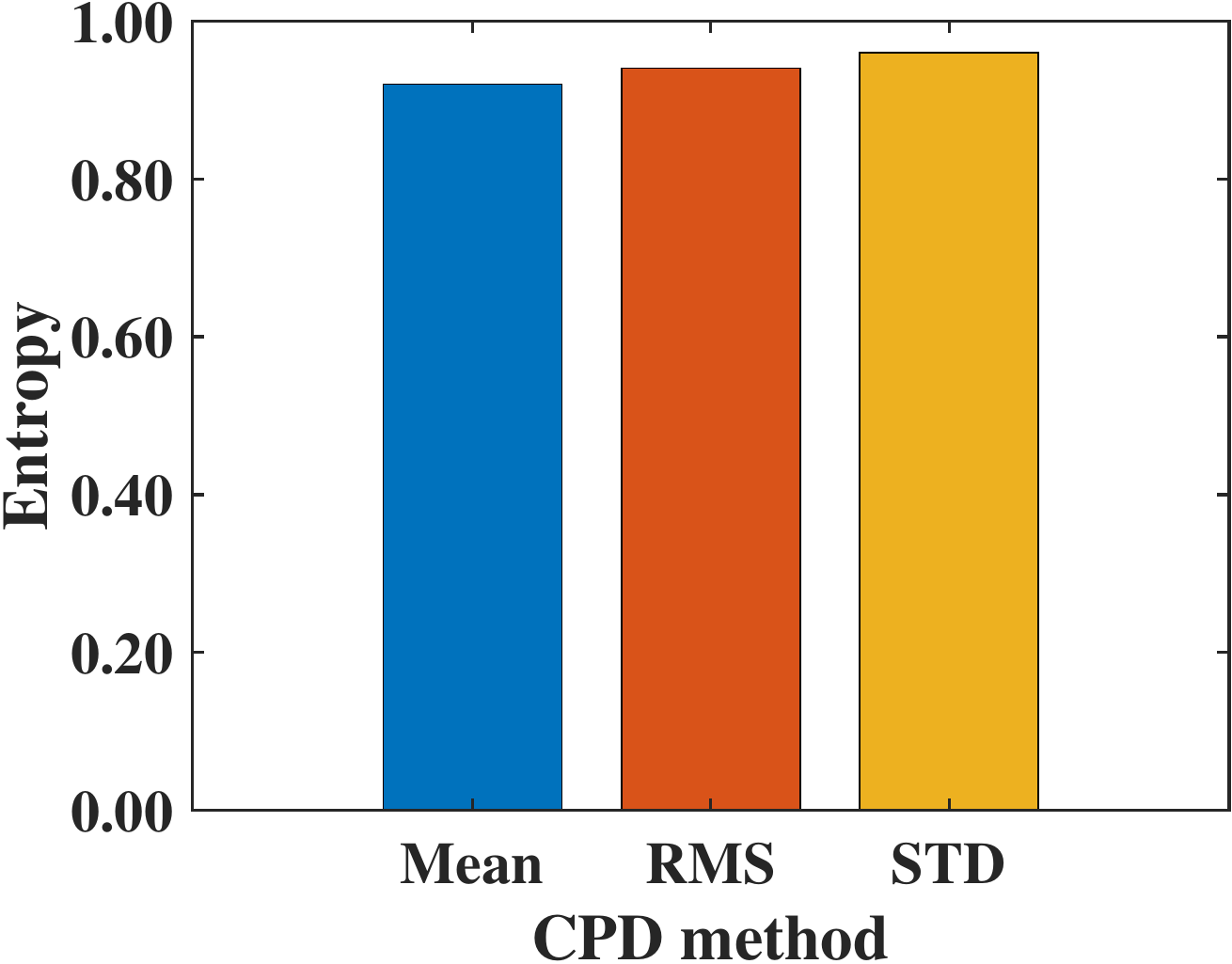}}
        \hfill
  \subfloat[\label{fig: CPNum - KeyGenerationRate}]{%
        \includegraphics[width=0.5\linewidth]{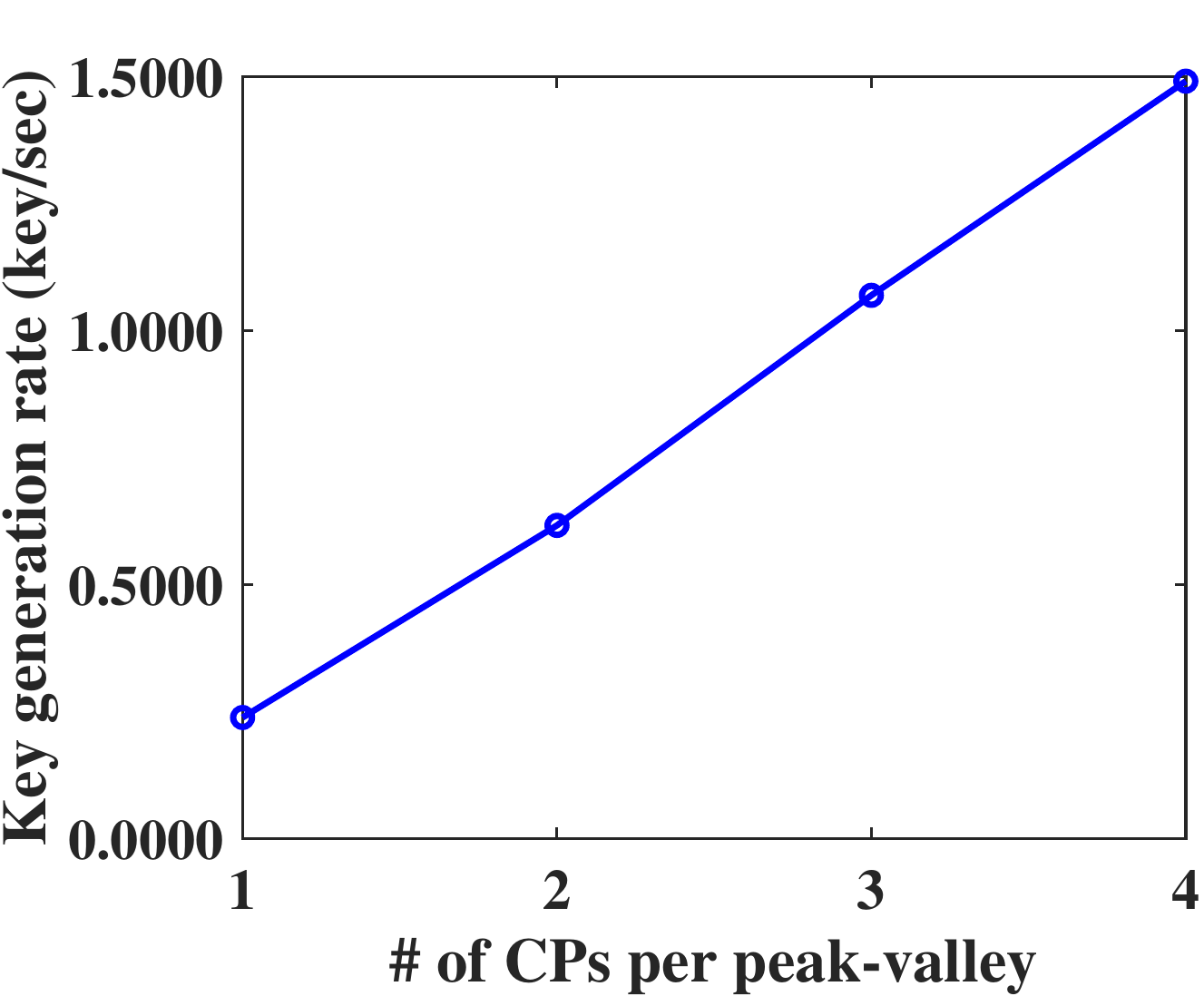}}
        \hfill
  \subfloat[\label{fig: CPNum - Entropy}]{%
        \includegraphics[width=0.5\linewidth]{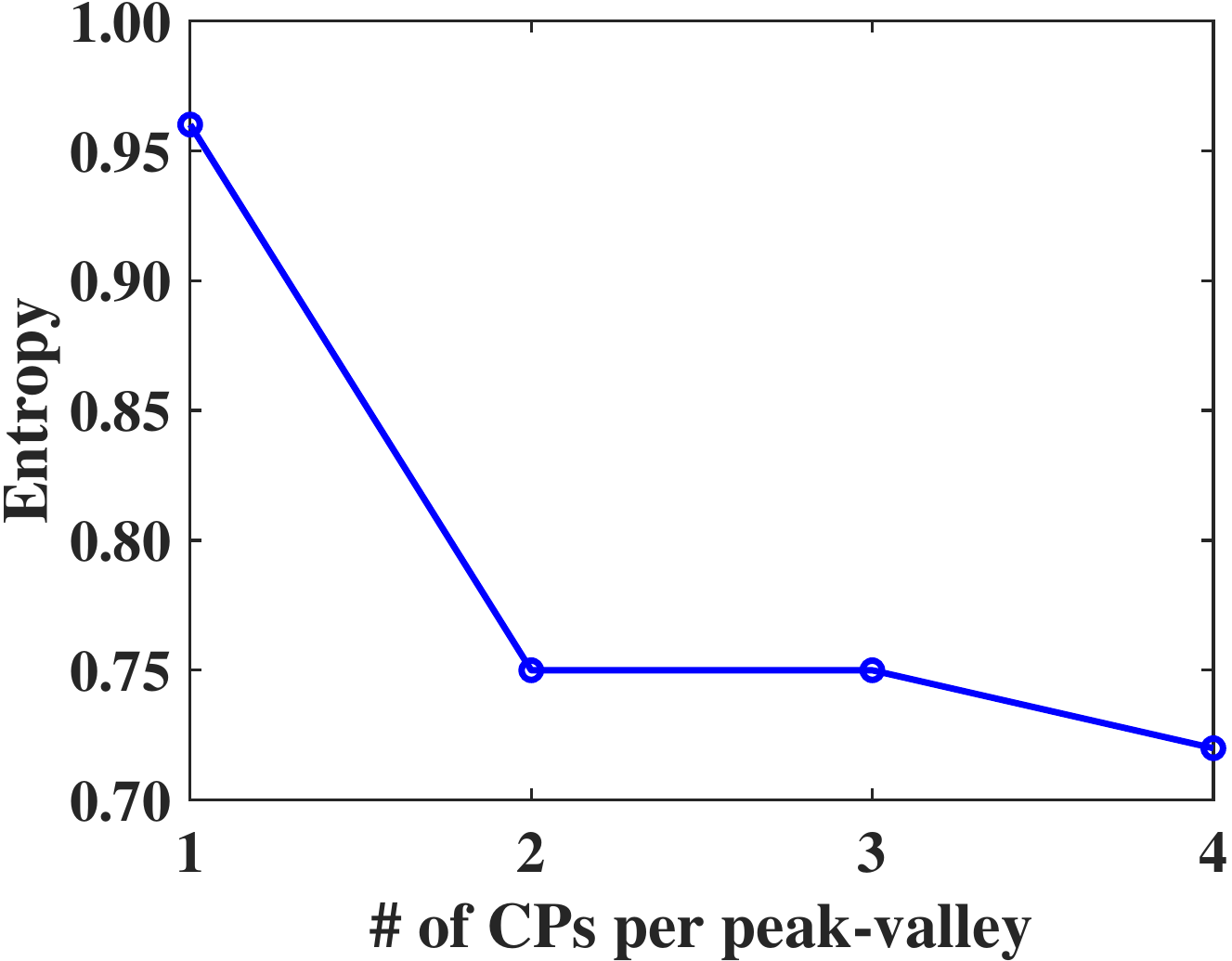}}
        \hfill
  \subfloat[\label{fig: CPDThreshold - KeyGenerationRate}]{%
        \includegraphics[width=0.5\linewidth]{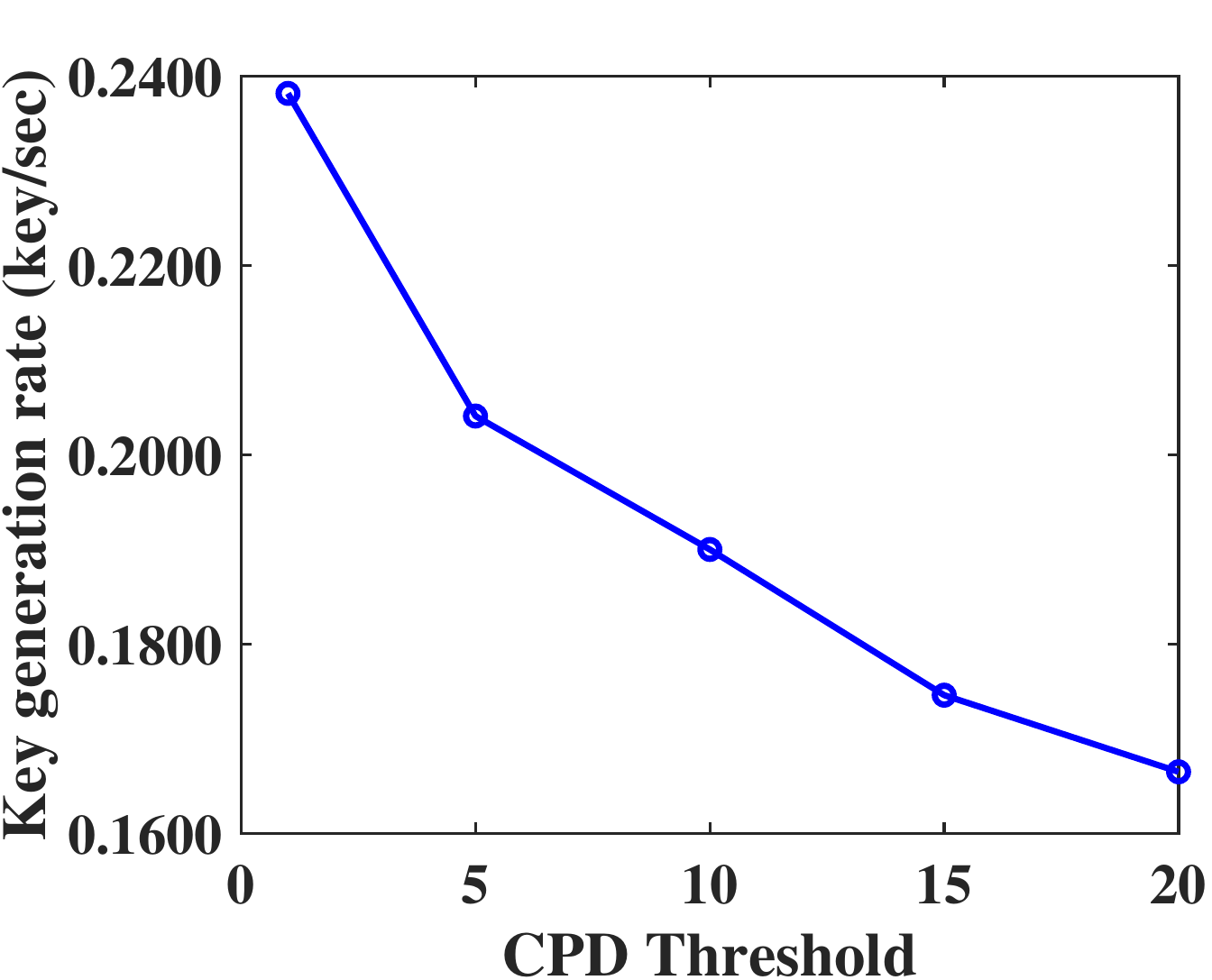}}
        \hfill
  \subfloat[\label{fig: CPOffset - KeyGenerationRate}]{%
        \includegraphics[width=0.5\linewidth]{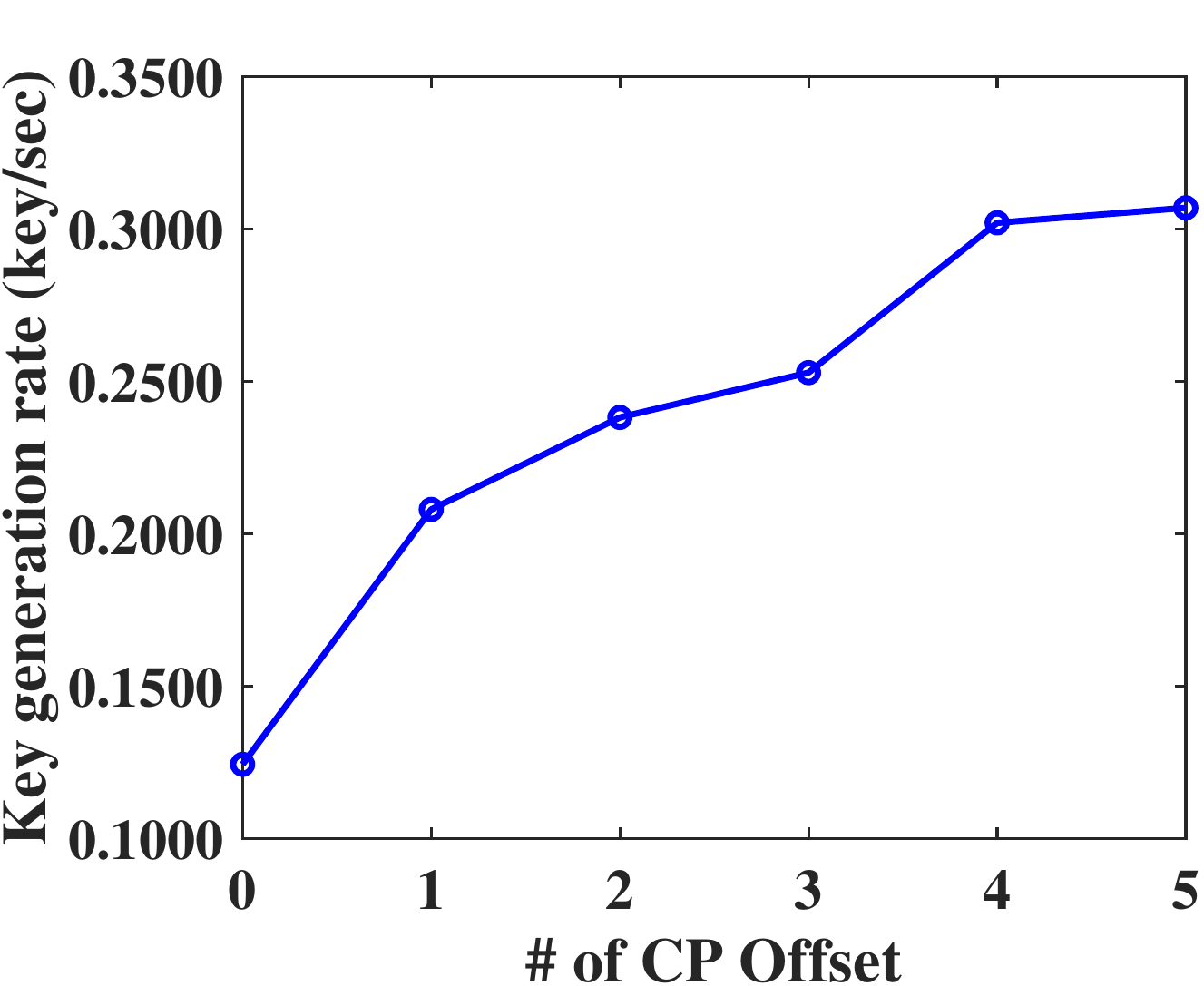}}
        \hfill
  \subfloat[\label{fig: CPOffset - Entropy}]{%
        \includegraphics[width=0.5\linewidth]{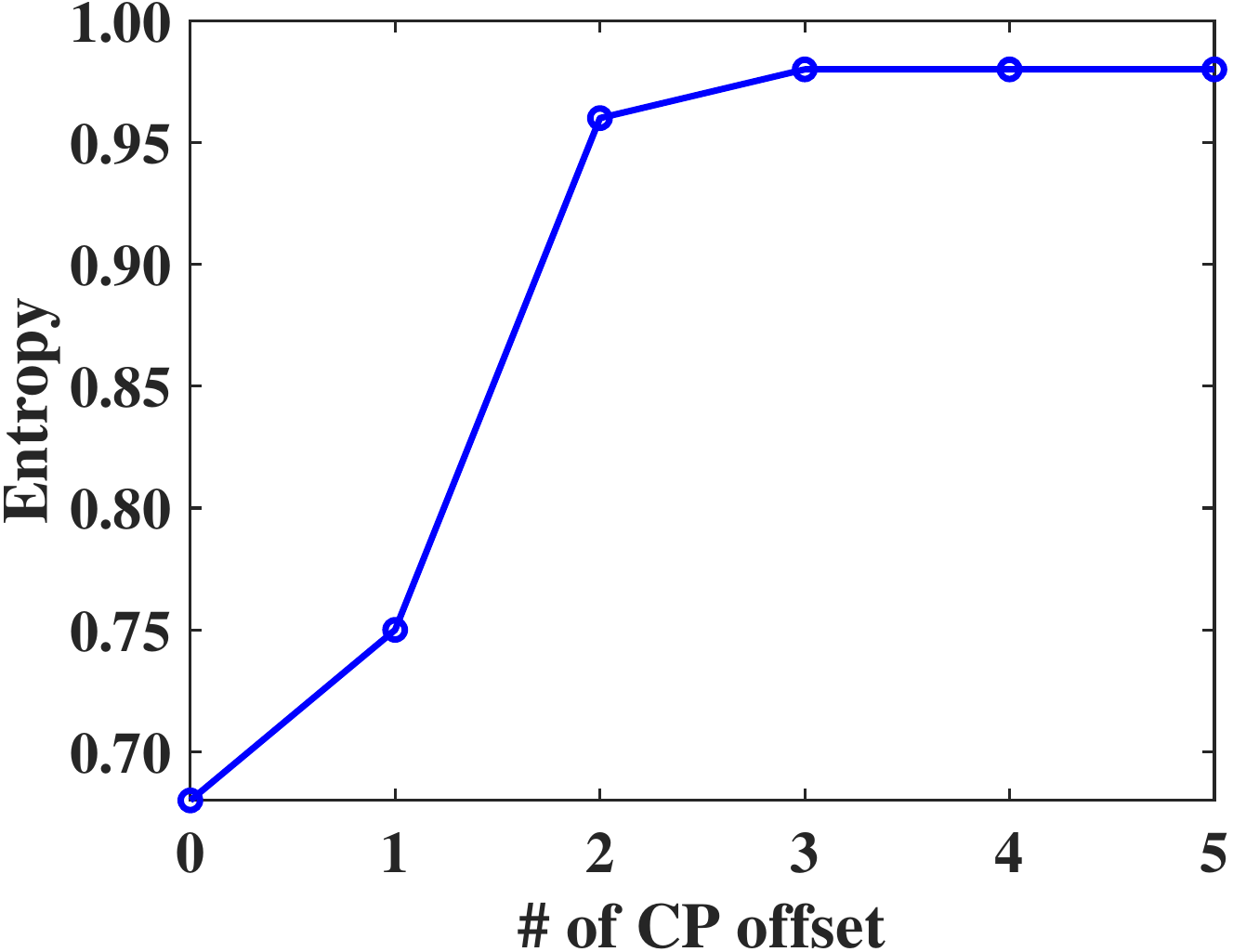}}
        \hfill
  \subfloat[\label{fig: SyncOffset - KeyGenerationRate}]{%
        \includegraphics[width=0.5\linewidth]{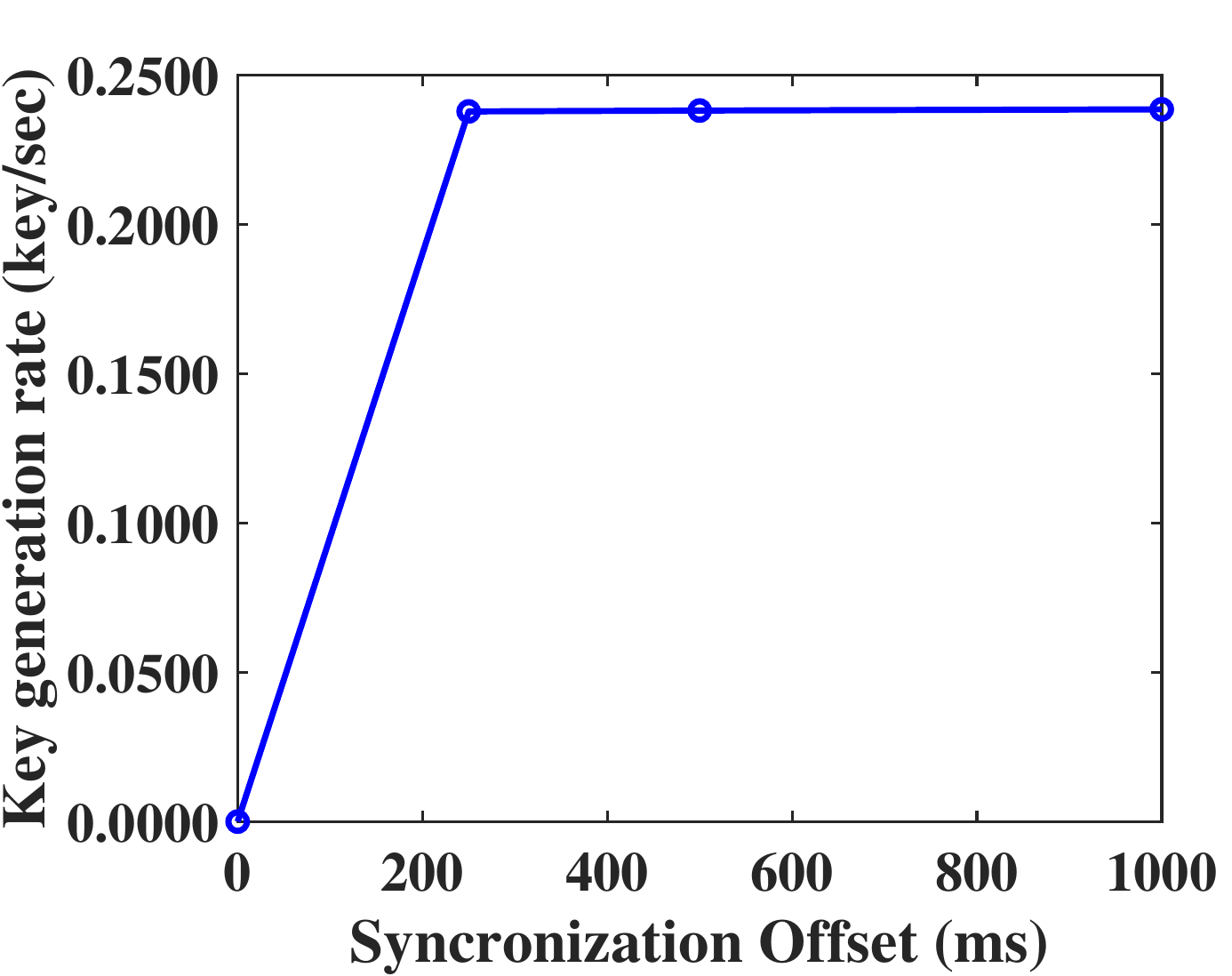}}
        \hfill
  \subfloat[\label{fig: SyncOffset - Entropy}]{%
        \includegraphics[width=0.5\linewidth]{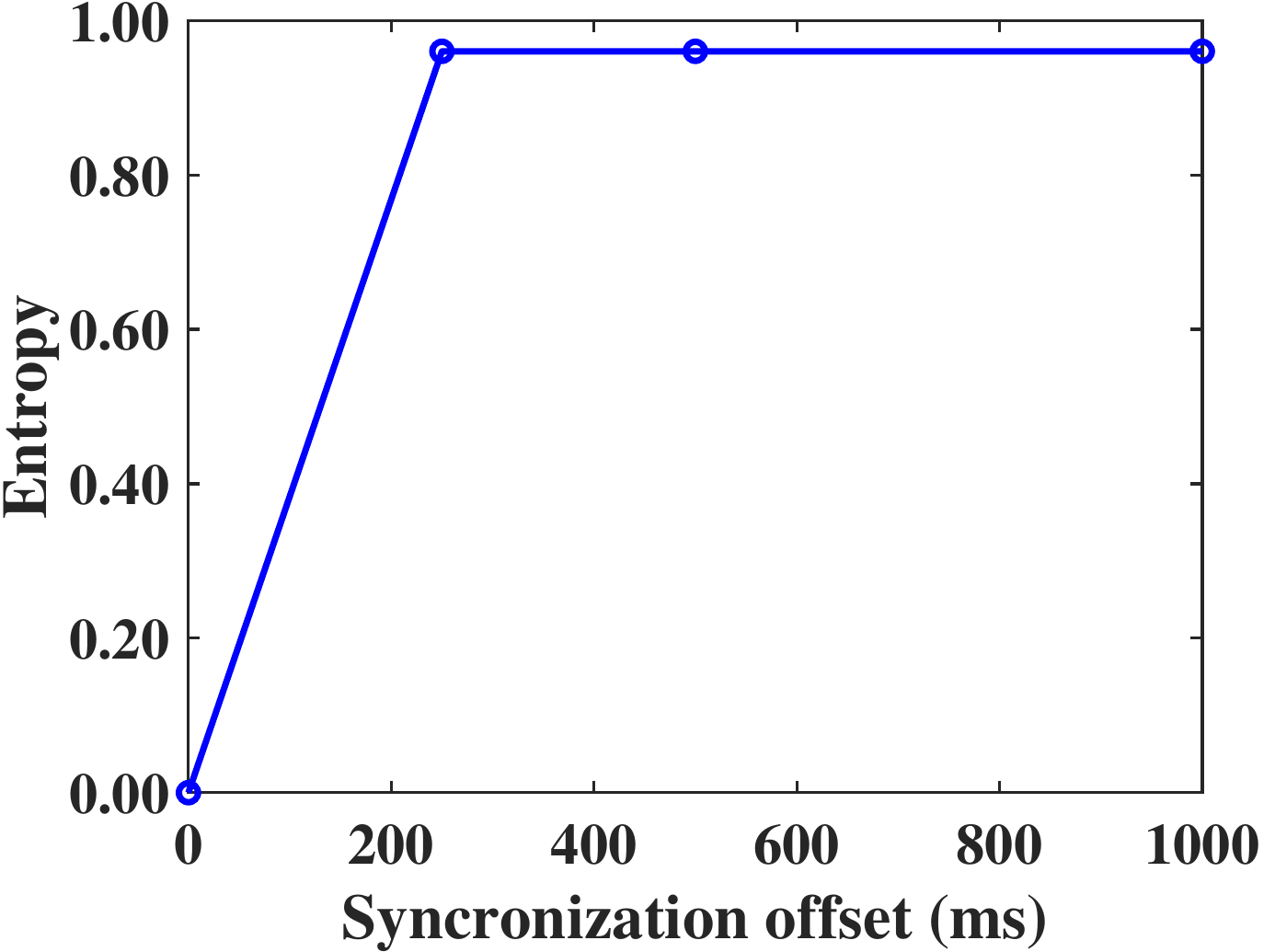}}
  
  \caption{Impact of Synchronization on Performance Metrics. (a) CPD Method {\em vs.} Key Generation Rate, (b) CPD Method {\em vs.} Entropy, (c) Number of CP per Peak-Valley {\em vs.} Key Generation Rate, (d) Number of CP per peak-valley {\em vs.} Entropy, (e) CPD threshold {\em vs.} Key Generation Rate, (f) CP Offset {\em vs.} Key Generation Rate, (g) CP Offset {\em vs.} Entropy, (h) Synchronization Offset {\em vs.} Key Generation Rate, (i) Synchronization Offset {\em vs.} Entropy}
  \label{fig: synchronization effect} 
  \end{figure}

  {\bf CPD Method:} Since the respiration signal is a low-frequency signal, even small changes in its frequency can be detected using the STD-based CPD. Its advantages over other statistical metrics are shown in the Figures~\ref{fig: CPDMethod - KeyGenerationRate} and \ref{fig: CPDMethod- Entropy}. In the rest of the experiments on CPD parameters, we focus on STD-based CPD only. 
  
  {\bf Number of Change Points per Peak-Valley:} To select a change point in a peak-valley duration, the B2P first detects all the change points in that duration, then sorts them and selects the first $n_{CP}$ most significant CPs. Figures~\ref{fig: CPNum - KeyGenerationRate} and \ref{fig: CPNum - Entropy} show that although increasing the number of change points ($n_{CP}$) can increase the KGR significantly, it biases the final key and decreases its entropy in a way that is not reliable as a secure key.
  
  {\bf CPD Threshold Level:} By changing the CPD threshold, we can control the number of change points. Increasing the CPD threshold leads to the loss of some possible synchronization points. Therefore, the KGR will decrease, as shown in Figure~\ref{fig: CPDThreshold - KeyGenerationRate}. However, it reduces the chance of the adversary generating a common key. 
  
  {\bf CP Offset:} Another important parameter for synchronization is the offset of change points, which is the number of CPs adjacent to the selected CP that are allowed to be considered as an alternative to the selected CP. We denote this as $n_{offset}$. A non-zero offset allows for some inconsistency between change point detection on the two devices. For example, if one device misses a change point,  future pairing attempts can be based on change points that are out of sync. The parameter $n_{offset}$ specifies the number of neighboring change points are used for pairing attempts. A value of 1 for $n_{offset}$ means that both the previous and the next change point are used for pairing attempts. Figures~\ref{fig: CPOffset - KeyGenerationRate} and \ref{fig: CPOffset - Entropy} show that increasing the CP offset Number can significantly increase the KGR and entropy.
  
  {\bf Synchronization Offset:} In order to have more control over the synchronization and pairing process, we also parameterize the synchronization offset. This parameter specifies how much lag or lead between two signals is accepted. This can be checked at the confirmation stage by sending the CP's index in which the common key has been generated. Using the generated key, $DevB$ encrypts its CP's index and sends it to $DevA$. $DevA$ decrypts it and checks whether it is within the range of the synchronization offset. If it is, $DevA$ will accept the key as a shared key. Experimental results in the Figures~\ref{fig: SyncOffset - KeyGenerationRate} and \ref{fig: SyncOffset - Entropy} show that increasing this parameter until 250 msec can significantly raise the KGR and entropy. Increasing the synchronization offset beyond 250 msec will only increase the attacker's success rate.
  \subsubsection{\textbf{Impact of Quantization}}
\begin{figure} [htbp]
    \centering
  
  \subfloat[\label{fig: BitPerSample - KeyGenerationRate}]{%
        \includegraphics[width=0.5\linewidth]{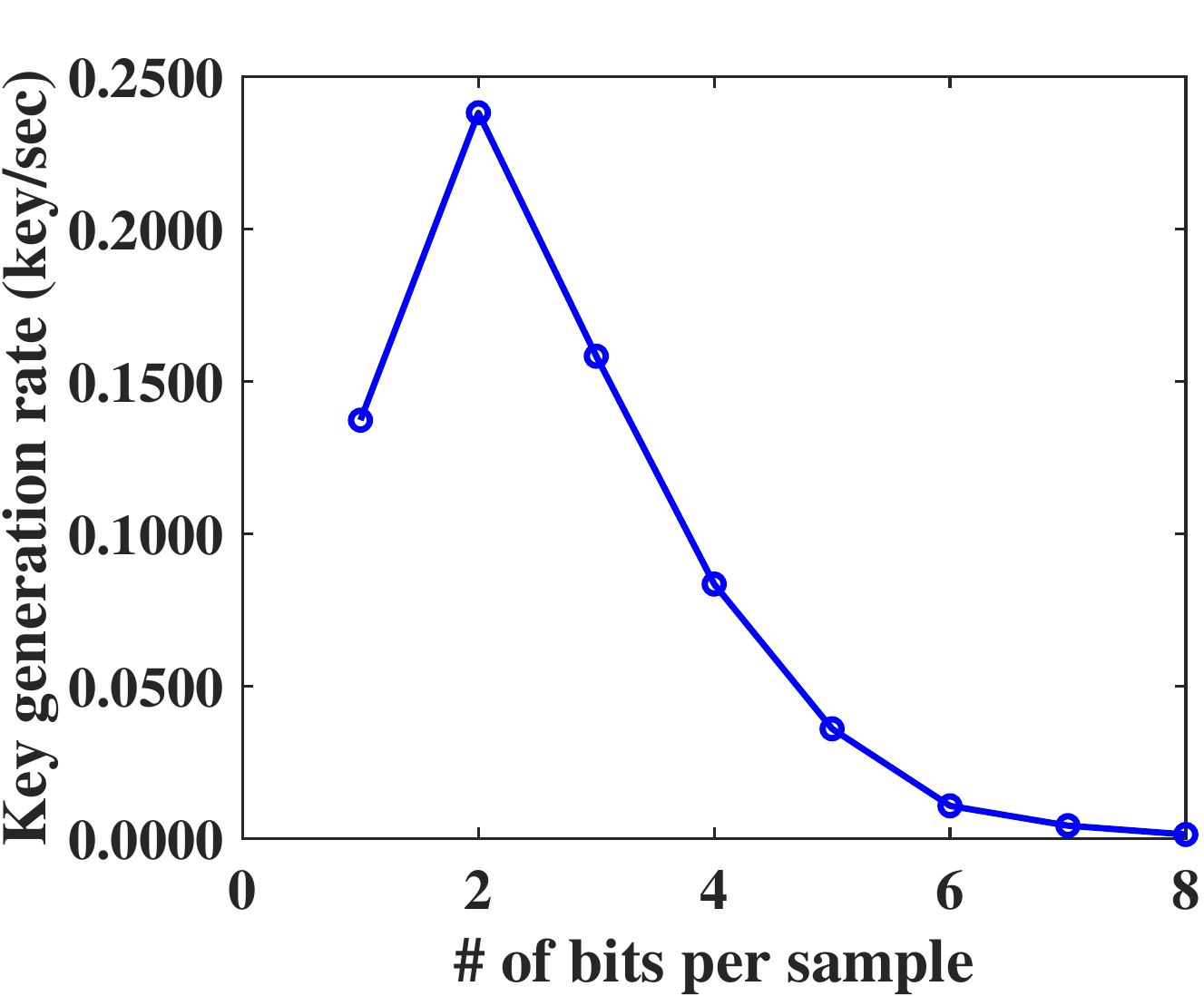}}
        \hfill
  \subfloat[\label{fig: BitPerSample - Entropy}]{%
        \includegraphics[width=0.5\linewidth]{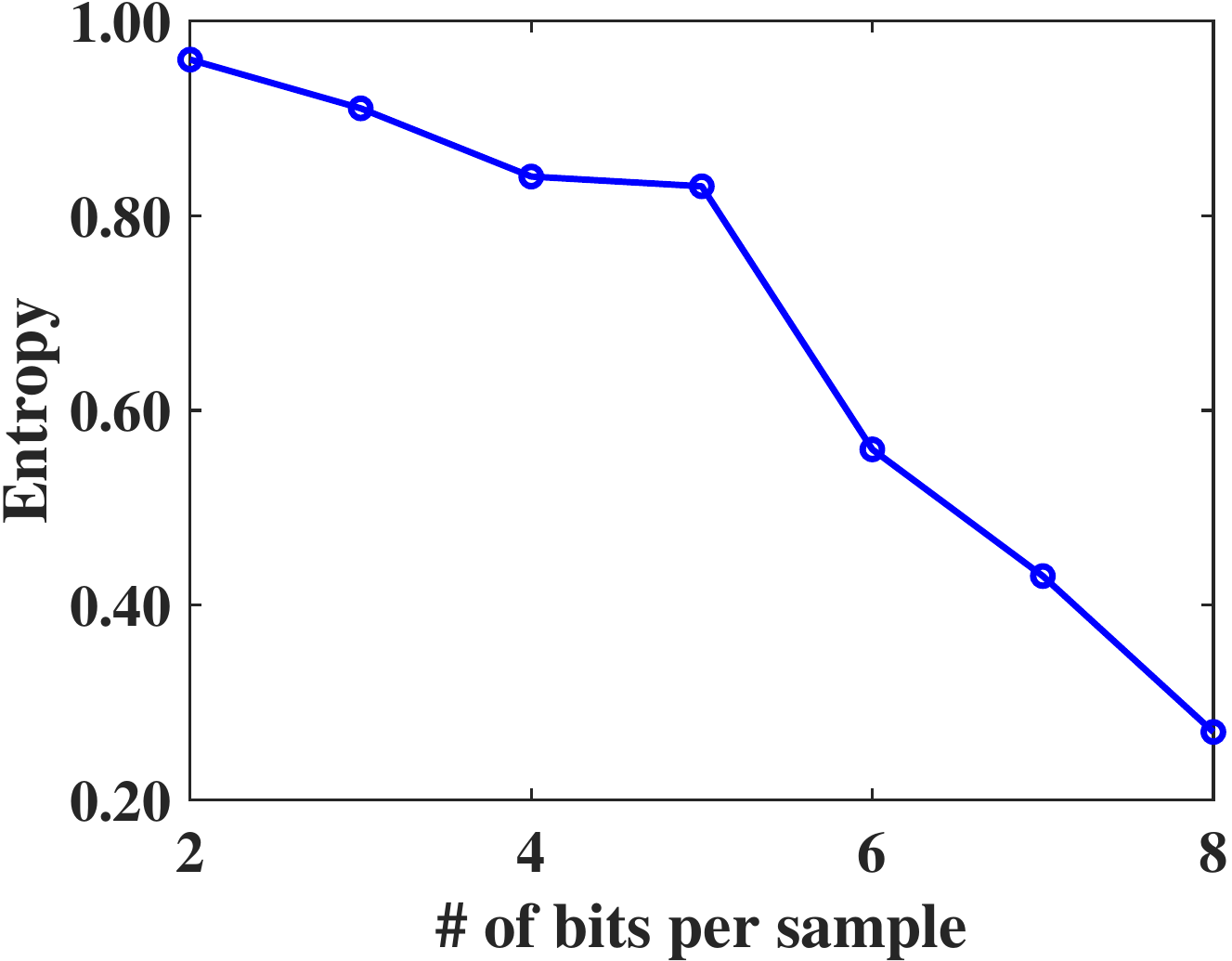}}
        \hfill
  \subfloat[\label{fig: BitPerSample - FalsePositive}]{%
        \includegraphics[width=0.5\linewidth]{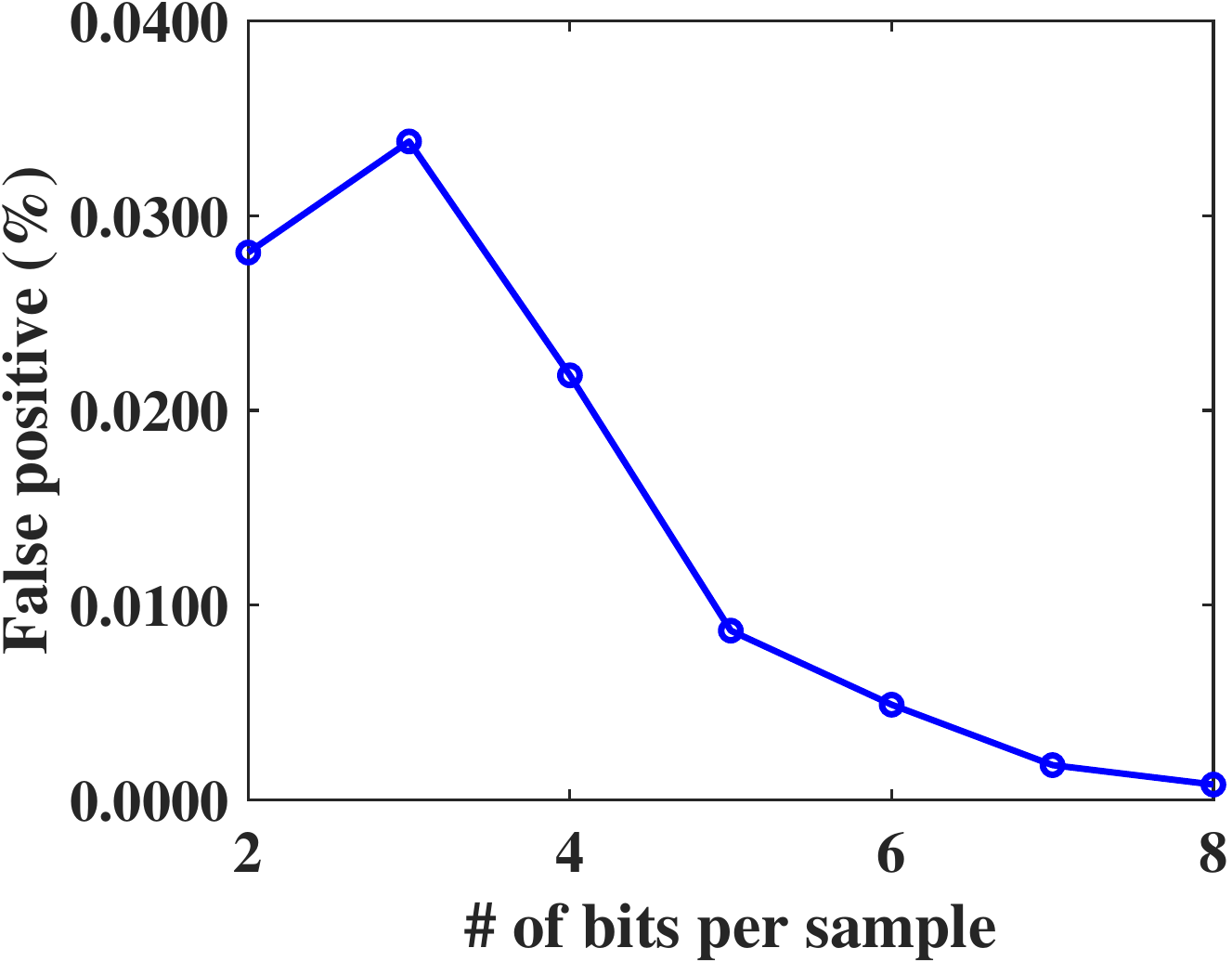}}
        \hfill
  \subfloat[\label{fig: BitPerSample - FalseNegative}]{%
        \includegraphics[width=0.5\linewidth]{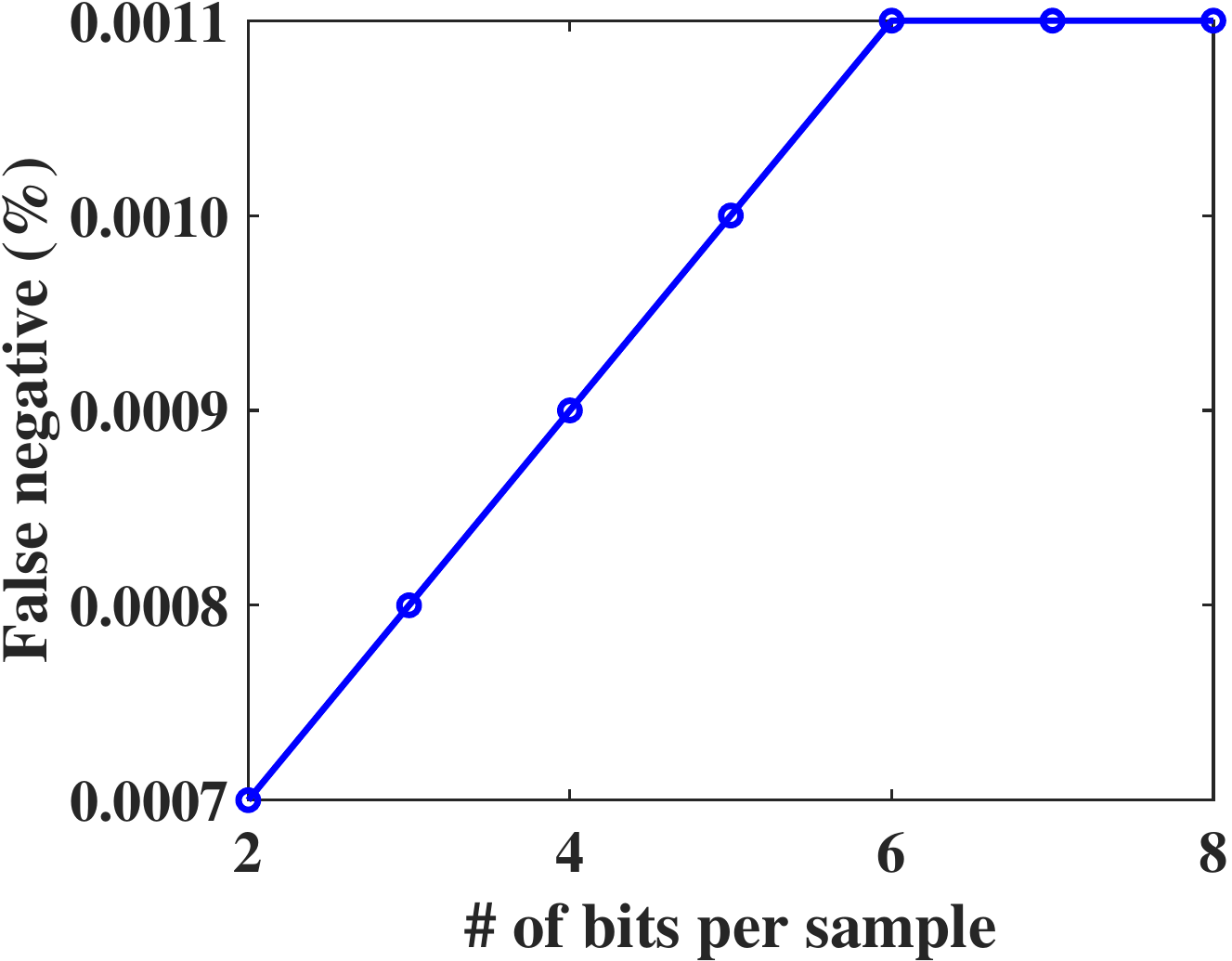}}
        \hfill      
  \subfloat[\label{fig: CodePresentation - KeyGenerationRate}]{%
        \includegraphics[width=0.5\linewidth]{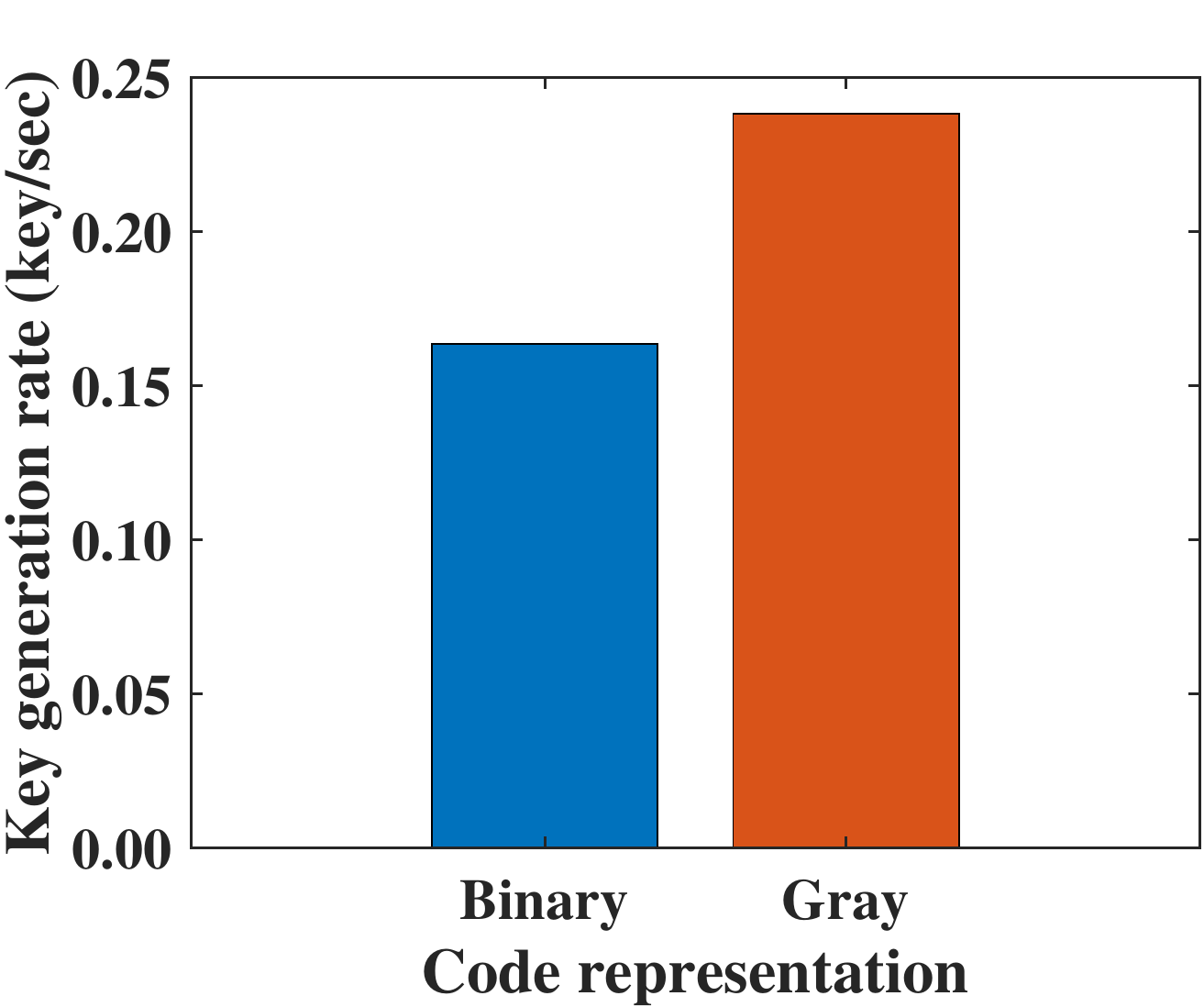}}
        \hfill
  \subfloat[\label{fig: CodePresentation - Entropy}]{%
        \includegraphics[width=0.5\linewidth]{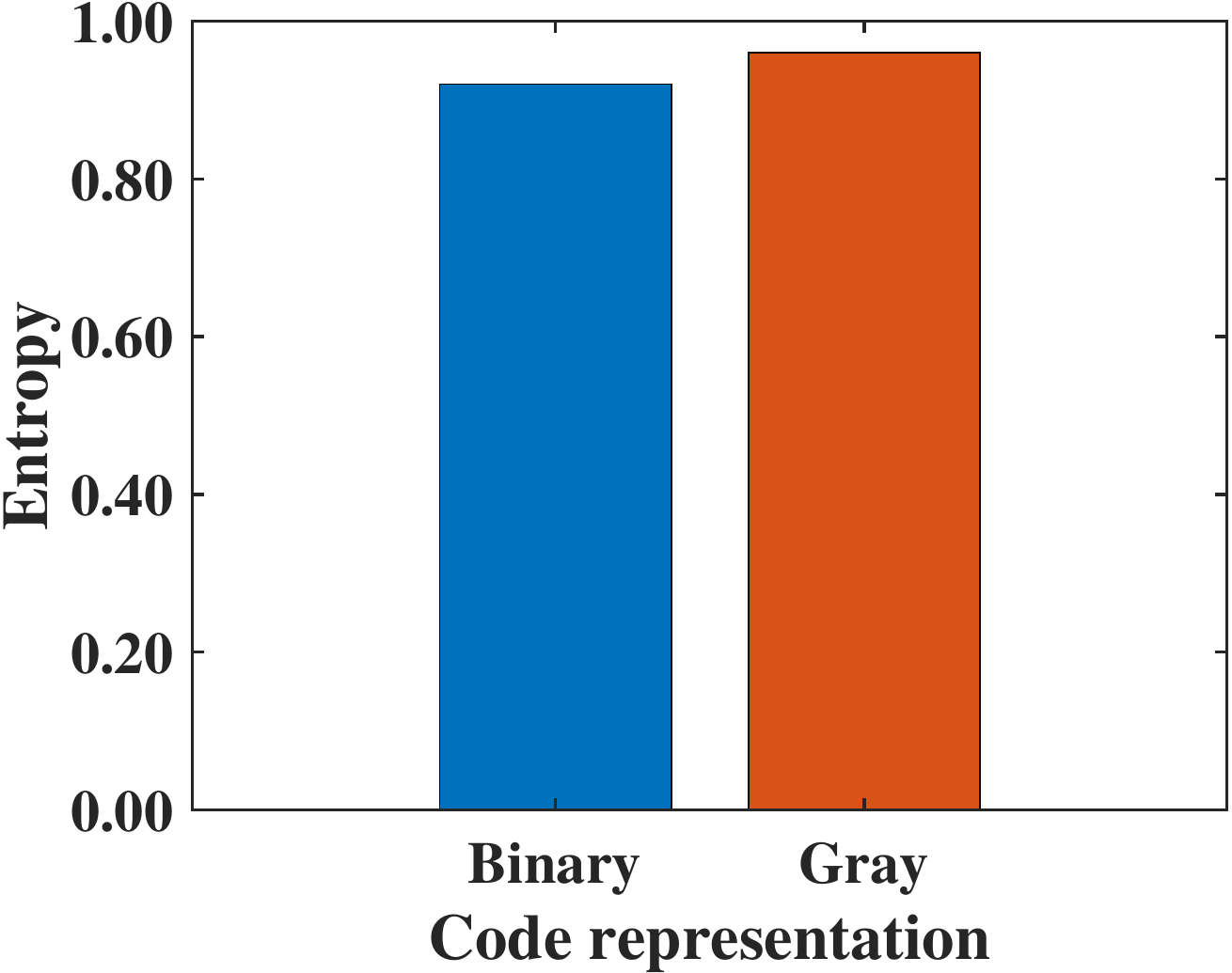}}      
  \caption{Impact of Quantization on Performance Metrics. (a) Number of Bits per Sample {\em vs.} Key Generation Rate, (b) Number of Bits per Sample {\em vs.} Entropy, (c) Number of Bits per Sample {\em vs.} False Positive Ratio, (d) Number of Bits per Sample {\em vs.}  False Negative Ratio, (e) Code Representation Type {\em vs.} Key Generation Rate, (f) Code Representation Type {\em vs.} Entropy}
  \label{fig: quantization effect}
  \vspace{-.251in} 
  \end{figure}

  In this set of experiments, we vary the number of bits per sample and experiment on the bit representation types. Figure \ref{fig: BitPerSample - KeyGenerationRate} shows that the key generation rate decrease when the number of bits per sample increases. The maximal key generation rate is achieved at 2 bits per sample. The result is consistent with our theoretical results presented in Section \ref{sec:quantization}. Figure \ref{fig: BitPerSample - Entropy} shows the decrease in entropy when the number of bits per sample increases. The decrease is because fewer samples are needed to generate a key of a fixed length and  because consecutive samples generally generate key bits with less randomness. Figure \ref{fig: BitPerSample - FalseNegative} shows that the false positive rate increases with the number of bits per sample. More bits per sample lead to more quantization levels and, in turn, to a reduction of the distance between two successive quantization levels. As expected, an increase of the number of bits per sample leads to less tolerance of sensing noise or other differences caused by the different sensing mechanisms in the devices.    The drop in the false positive rate shown in Figure \ref{fig: BitPerSample - FalsePositive} is simply because the fewer samples required to generate a key can increase the chance of false positives. Both Figure \ref{fig: BitPerSample - FalsePositive} and Figure \ref{fig: BitPerSample - FalseNegative} show that false positive and false negative rates are less than 0.04\% and 0.0011\% respectively. 
  
  Figure \ref{fig: CodePresentation - KeyGenerationRate} and Figure \ref{fig: CodePresentation - Entropy} compare the results for Gray coding and binary coding used to represent bits generated by the quantizer. Figure \ref{fig: CodePresentation - KeyGenerationRate} shows that Gray coding can achieve higher key-generation rates than binary coding. The advantage is mainly because of the error correction capabilities of Gray coding \cite{Rostami2013, Lin2019}.
  \subsubsection{\textbf{Impact of Error Correction}}
  A key parameter of the error correction and conformation module is the error correction rate (ECR), defined as the ratio between the number of errors that can be corrected and the length of a codeword in bits. B2P uses BCH mainly to correct mismatches in key bits between two devices to be paired. So the  error correction rate directly affects B2P's key generation rate. For 127-bit codewords and 255-bit codewords, $ECR$s are in the ranges of between 0.79\% and 24.4\% and between 0.39\% and 24.7\%, respectively. 
  
\begin{figure} [htbp]
    \centering
  \subfloat[\label{fig: ErrorCorrection - KeyGenerationRate}]{%
       \includegraphics[width=0.5\linewidth]{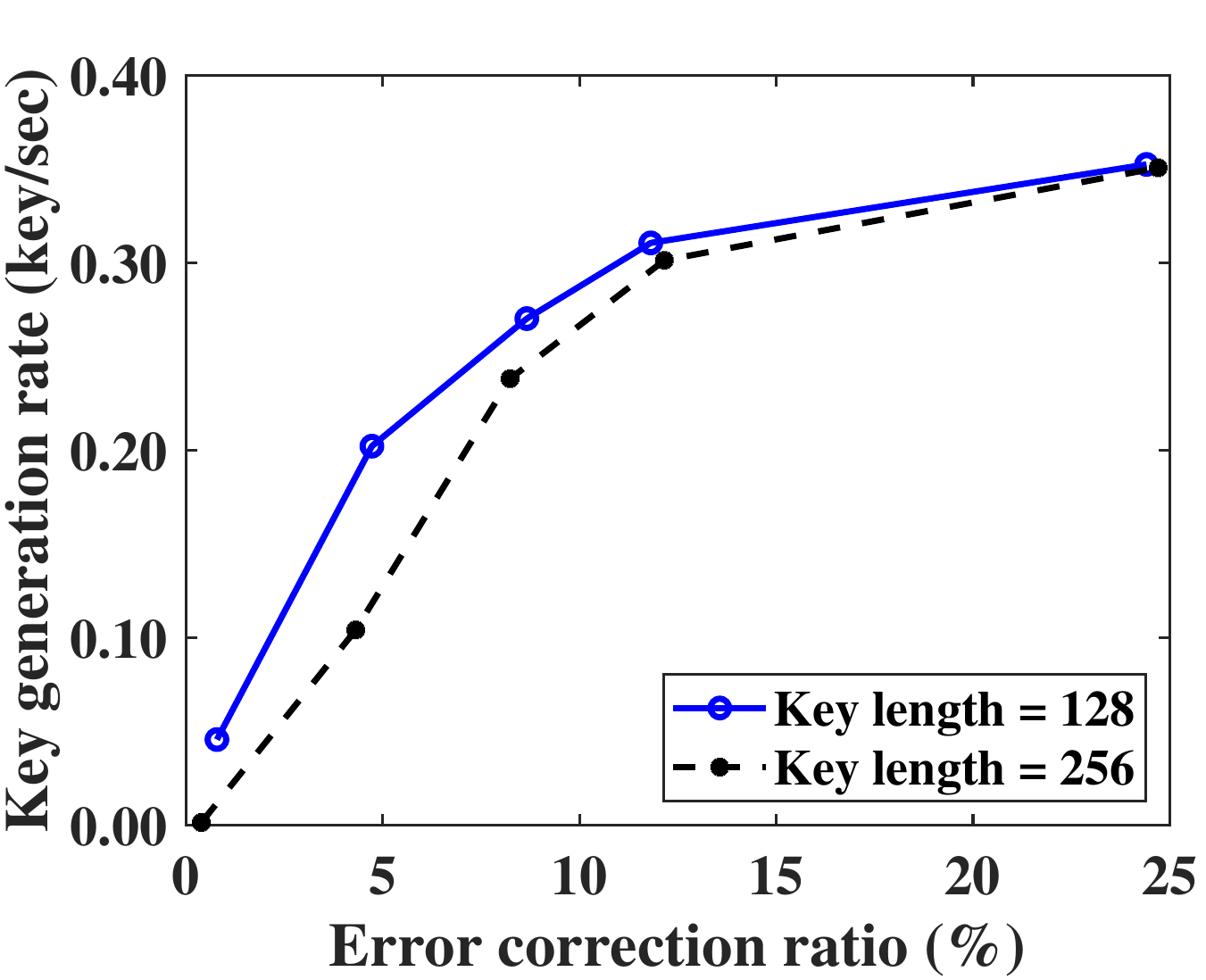}}
    \hfill
  \subfloat[\label{fig: ErrorCorrection- Entropy}]{%
        \includegraphics[width=0.5\linewidth]{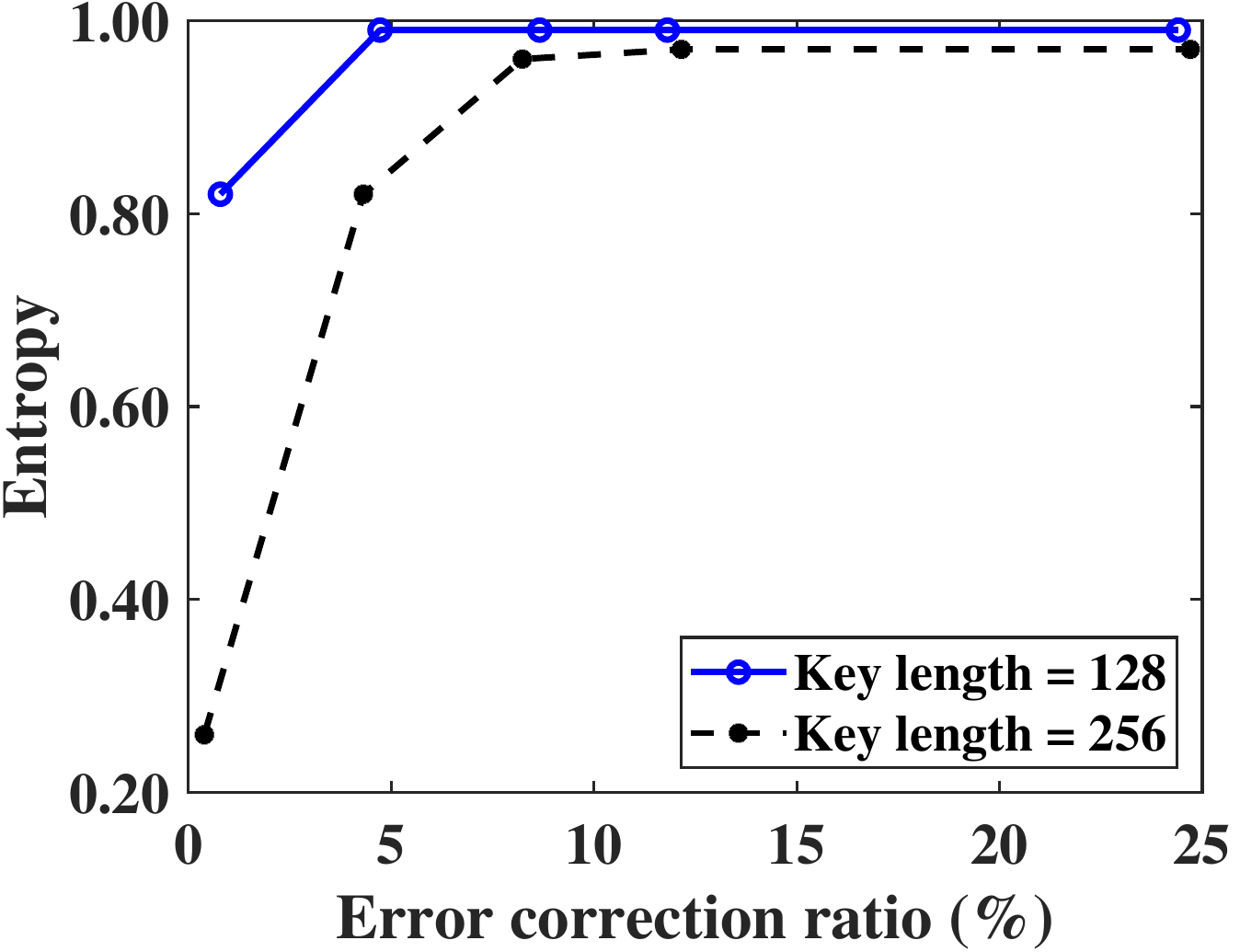}}
        \hfill
  \subfloat[\label{fig: ErrorCorrection - FalsePositive}]{%
        \includegraphics[width=0.5\linewidth]{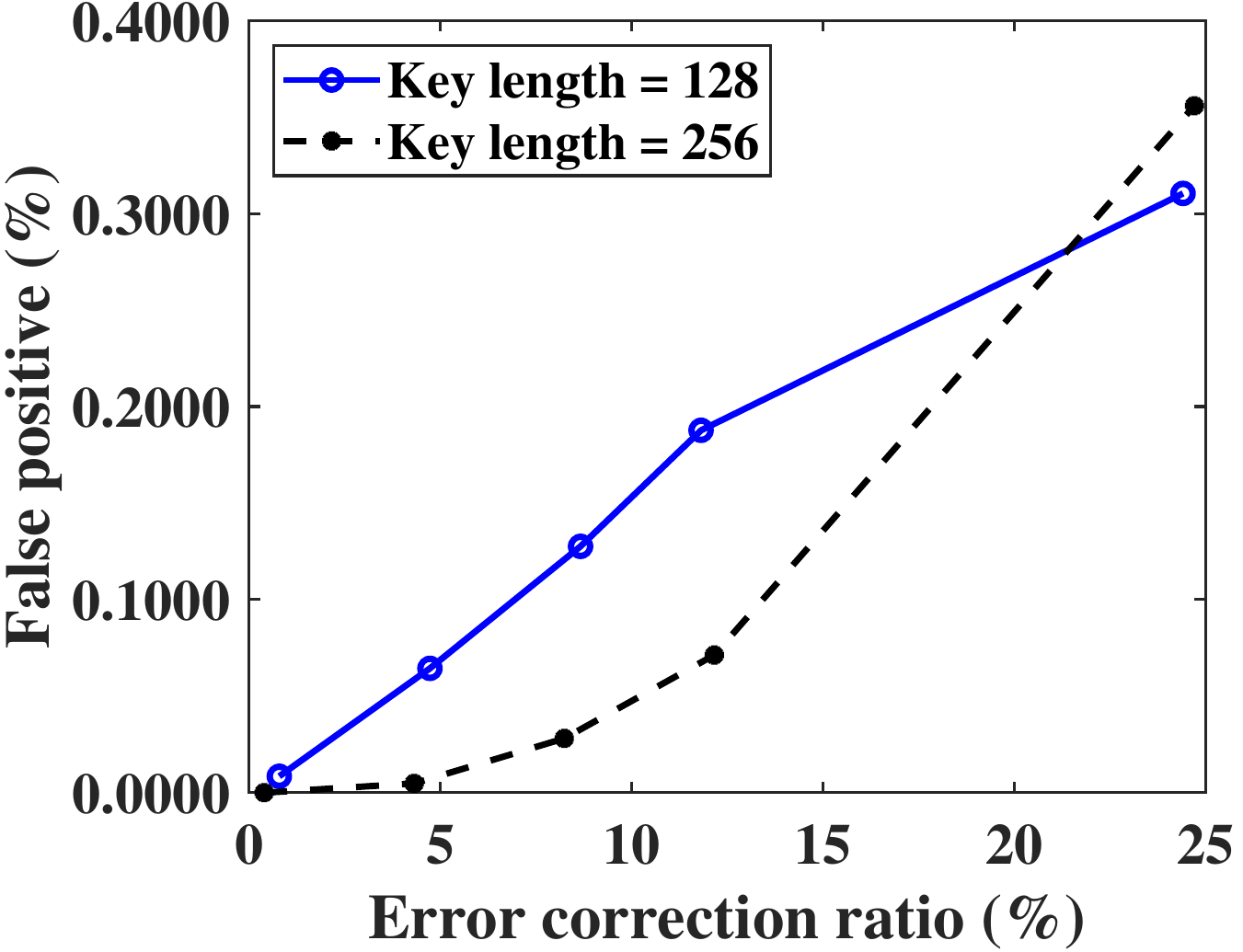}}
        \hfill
  \subfloat[\label{fig: ErrorCorrection - FalseNegative}]{%
        \includegraphics[width=0.5\linewidth]{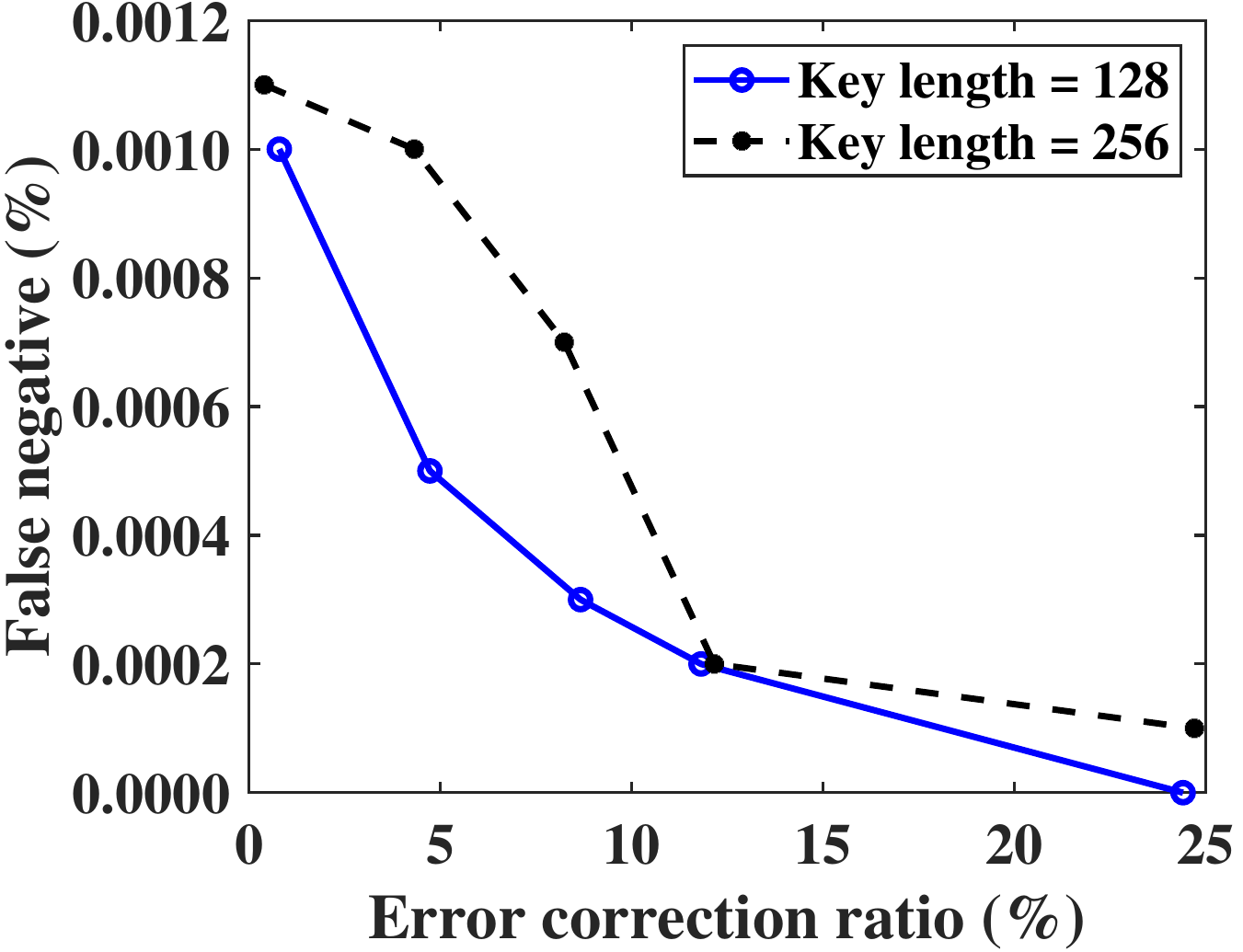}}
        \hfill
  
  \caption{Impact of Error Correction on Performance Metrics. (a) Correction Ratio {\em vs.} Key Generation Rate, (b) Error Correction Ratio {\em vs.} Entropy, (c) Error Correction Ratio {\em vs.} False Positive Ratio, (d) Error Correction Ratio {\em vs.} False Negative Ratio}
  \label{fig: ErrorCorrection effect} 
  \vspace{-.10in}
  \end{figure}

  As we expected, the key generation rates, shown in Figure \ref{fig: ErrorCorrection - KeyGenerationRate}, increase with $ECR$ simply because more bit mismatches can be resolved when $ECR$ increases. The low key-generation rates for both 128-bit and 256-bit keys when $ECR<1\%$ indicate that the mismatches are very unlikely to be removed completely with low-$ECR$ BCH codes. 
  Figure \ref{fig: ErrorCorrection- Entropy} shows that B2P can generate 128-bit keys and 256-bit keys with entropy close to 1, the ideal entropy, when $ECR \geq 5\%$ and $ECR \geq 8\%$,  respectively. 
  Figure \ref{fig: ErrorCorrection - FalseNegative} shows the decrease of false-negative rate when $ECR$ increases. This decrease is consistent with our intuition, as larger $ECR$ can resolve more bit mismatches. The cost of the decrease of false-negative rate is the increase of false-positive rate, shown in Figure \ref{fig: ErrorCorrection - FalsePositive}. In other words, the increase of $ECR$ can also possibly increase the number of false-positive because more bit mismatches can be resolved even for respiration signals not collected from the same body.
  
  We did experiment with other lengths as well. Since bit-string selection is based on the CPs, the KGR relies on the number of CPs: the more CPs in a specific signal duration, the higher the KGR. On the other hand, using band-pass and Savitzky-Golay filters, we have relatively smooth signals with some mismatches that can be corrected in the error correction step. Hence, although increasing the key length can decrease the KGR in principle, the effect in practice is not significant, as exemplified in Figure~\ref{fig: code_length}. Using low-frequency biometric activities brings fundamental trade-offs between key generation rate and key length. For example, for breathing sensors with a sample rate of 64 Hz, a low-frequency respiration signal (around 0.35 Hz), and a quantizer that represents each sample with 2 bits, key lengths lower than 128 bits can not have a high entropy rate. This is because the keys are generated from a small number of samples (there are 128 bits per second, which makes for about one-third of a respiration cycle). Also, key lengths greater than 256 bit may have significant overlap with consecutive keys. For instance, a key length of 1024 bits covers around three respiration cycles, which has at least six CPs.
  \subsection{Resistance to Attacks} \label{subsec:attacks}
    An adversary may try to compromise pairing with the B2P protocol by launching impersonation attacks, which attempt to pair a malicious device with a legitimated device.  We evaluate B2P's resistance to impersonation attacks below.
  %
  
\begin{figure} [htbp]
    \centering
  \subfloat[\label{fig: video attack - face}]{%
       \includegraphics[width=.24\columnwidth]{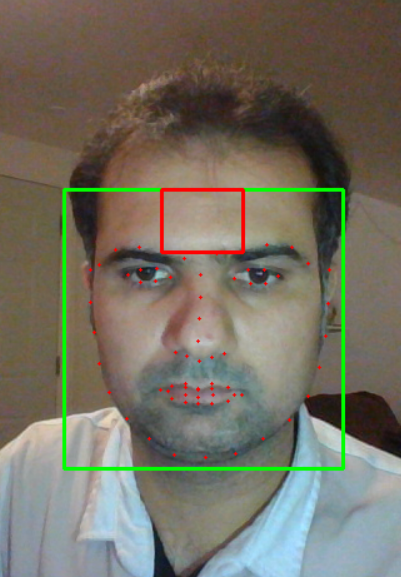}}
    \hfill
  \subfloat[\label{fig: video attack - HSV}]{%
        \includegraphics[width=.24\columnwidth]{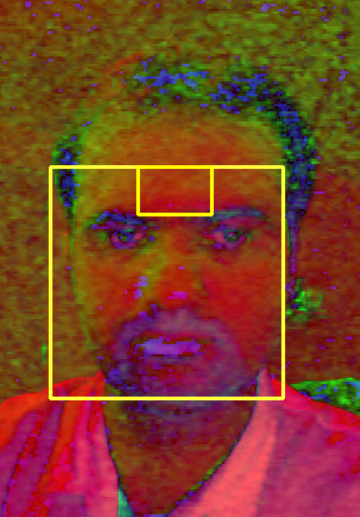}}
        \hfill
  \subfloat[\label{fig: video attack - rr}]{%
              \includegraphics[width=.45\columnwidth]{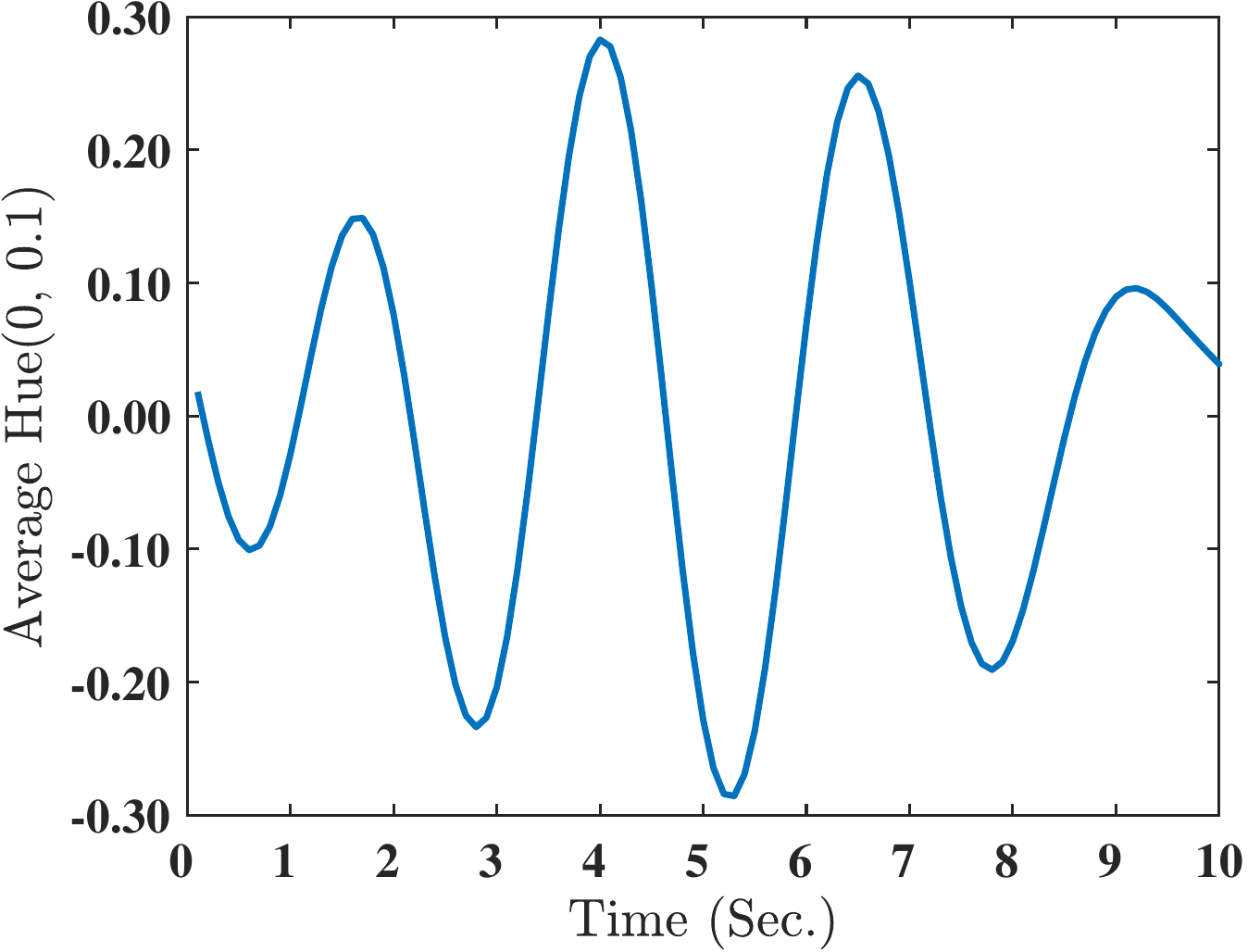}}      
  \caption{Video Attack Procedure (a) Forehead Detection Based on Face Detection and Facial
  Landmarks, (b) HSV Color Space of Detected Face and Forehead, (C) Detected Respiration Signal}
  \label{fig: video attack} 
  \vspace{-0.1in}
  \end{figure}

  We assume that the adversary is able to extract respiration signals with video analysis on the victim's breathing behavior. In \cite{Jafar_HR_RR}, the authors demonstrated that it is possible to monitor the respiration and heart-beat signals from a video. They demonstrated the accuracy of their method by comparing it with the RIP-based method, where the video-based approach achieved a low root mean square error (RMSE). 
   The proposed method, for each video frame, detects the user's face and facial landmarks using the histogram of oriented gradients (HoG) features with the help of a support vector machine (SVM). Then, the method determines the region of interest (ROI), which is a part of the user's forehead, based on the facial landmarks, as shown in Figure~\ref{fig: video attack - face}. Afterward, as shown in Figure~\ref{fig: video attack - HSV}, it converts the frame to hue saturation value (HSV) color space and calculates the average hue of all ROI pixels. The variation in the average values is due to the heart-beat, respiration, and changes in ambient. A band-pass filter with a cut-off frequency in the respiration frequency range $(0.1Hz, 0.5Hz)$ is applied on the array of the average values to extract the respiration signals, as shown in Figure~\ref{fig: video attack - rr}. Based on the extracted respiration signal, the adversary can launch the impersonation attack to pair with a legitimate device attached to the victim.
  
   In our experiments on resistance to the impersonation attack with video analysis, a video camera of resolution $640*480$ pixels and capture rate of 30 RGB frames per second is used to take a video of the victim's face. We use the method described above to extract the victim's respiration signal. Since the method
   extracts the respiration signal in real-time, it loses some frames due
   to the processing time. Our results show that the method can process
   nine frames per second to extract respiration signals. Hence, the extracted signal
   will lose some information of the actual respiration signal. The extracted signals are also affected by ambient light and the tone of
   the user’s skin color, as described 
   in \cite{Jafar_HR_RR}. 
  
\begin{figure} [htbp]
    \centering
    \includegraphics[width=.5\columnwidth]{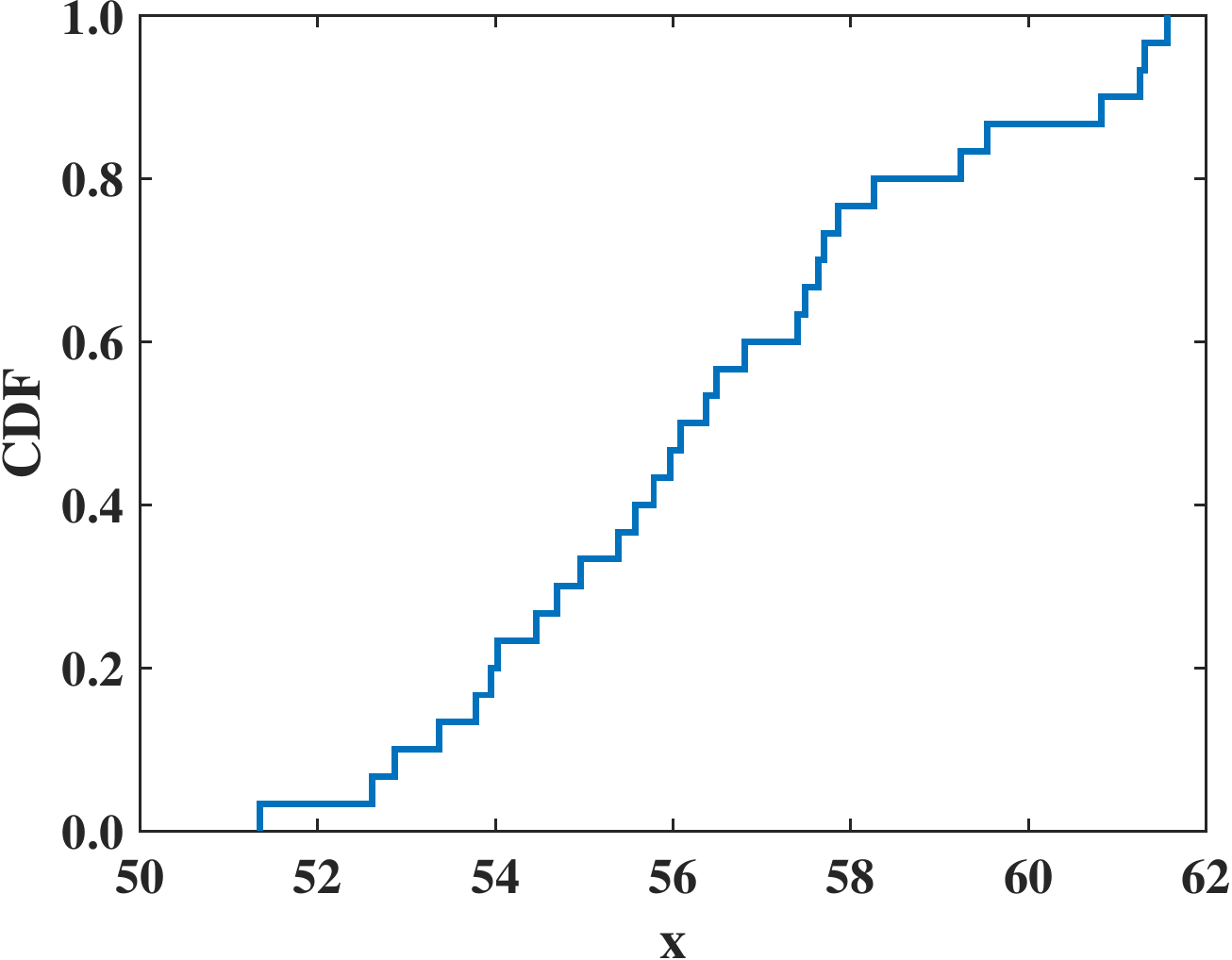}
    \caption{CDF of Bit Agreement Rate}
     \label{Attacker bit agreement}
  \end{figure}

   Our experiment results show that the false-negative rate of this impersonation attack is zero. This means that the adversary  can not generate a shared key. To find out how close the adversary is to a successful pairing with a legitimate device with this impersonation attack,  we measure the bit agreement rate, which represents the percentage of matched key bits. As shown in Figure \ref{Attacker bit agreement}, the maximal bit agreement rate is  $61.56\%$.
   This is because the extracted signal from the user's video cannot preserve the original signal's features, such as change points. Hence, the signal generated for the impersonation attack cannot be synchronized to the victim's signal. The loss of synchronization causes lots of mismatches that the error correction method cannot correct. 
  \section{Discussion}
  To the best of our knowledge, this paper is the first study that uses the respiration signal for pairing. We compare the B2P results with other biometric-based pairing approaches, such as heartbeat-based \cite{jayanth2017wearable,majumder2019energy, brezulianu2019iot} or gait-based \cite{Schurmann2018,xu2016walkie,Sun2017,Groza2012} approaches. As Table~\ref{tbl:result_compare} shows, the B2P protocol can generate a 128-bit key per 3.43 seconds and a 256-bit key at 4.39 seconds while the bit agreement for the attacker tops out at 64\% and 62\%, respectively. In comparison, H2B \cite{Lin2019}, a {\em heart-beat based} pairing protocol based on the interval between two peaks in the heart-beat signal (inter-pulse interval (IPI)) needs more than 40 seconds to generate a 128-bit key, and its bit-agreement rate is 75\%. The authors in \cite{Yuzuguzel2015a} proposed a {\em shaking-based} pairing approach that generates a shared key every 6.58 seconds. The authors did not report on the attacker's bit agreement rate. Another context-based approach in \cite{Rodriguez1989a} introduced a {\em gait-based} pairing method that takes 5 seconds to generate a shared key. The bit-agreement rate for this attack is 60\%. Although the aforementioned studies used signals that have higher frequencies than the respiration signal, their key generation rates are less than our protocol.
  
  Dealing with different types of sensors to generate a shared key is a main contribution of this paper. We showed how it is possible to use RIP and accelerometer sensors to generate a shared secure key in only one breathing cycle. Although RIP sensors are not widely used in common wearable devices at this time, smart shirts such as Hexoskin, Astroskin, Athos Shirt \cite{Athos_Shirt}, and Zephyr \cite{Zephyr} are already available, and they are increasingly being used by professional or amateur sportsmen and women, astronauts, and researchers. With the advancement of healthcare IoT devices, RIP sensors will be embedded in more future wearable devices. Therefore, the B2P protocol can be utilized in these devices to facilitate their secure communication with other IoT devices.
  \begin{table}[]
    \centering
    \caption{B2P Performance in Comparison with Other Pairing Methods}
    \begin{tabular}{|l|l|p{1.2cm}|p{1.2cm}|p{1.5cm}|}
    \hline
    \multicolumn{2}{|l|}{\textbf{Scheme}} & \textbf{Key Length (bit)} & \textbf{User KGR (key/sec)} & \textbf{Attacker bit agreement (\%)} \\ \hline
    H2B & \cite{Lin2019} & 128 & 0.0239 & 75 \\ \hline
    Shake Me & \cite{Yuzuguzel2015a} & 128 & 0.1520 & ? \\ \hline
    Gait-Based & \cite{Rodriguez1989a} & 128 & 0.2000 & 60 \\ \hline
    B2P & This paper & 128 & 0.2913 & 64 \\ \hline
    B2P & This paper & 256 & 0.2278 & 62 \\ \hline
    \end{tabular}
    \label{tbl:result_compare}
    \vspace{-.171in}
    \end{table}
  \section{Related Work}
  The use of biometric activities has become a popular approach for context-based pairing in wearable devices. A number of biometric activities have been studied, such as heart-beat, shaking, and gait, and a variety of sensor technologies have been used to observe (i.e., collect signals of) these activities. These sensors include ECG, PPG, EMG, and accelerometer. Indeed, these signals can provide auxiliary out-of-band (OOB) channels \cite{mayrhofer2012uacap} as a feasible option to facilitate device pairing. In this section, we provide a summary of studies that used the aforementioned signals. As far as we know, this study is the first in using respiration activity for the pairing of wearable devices.
  
  \noindent{\bf ECG and PPG:} 
  The heart-beat is a promising option for wireless body area networks (WBANs) authentication and key generating schemes because its properties are unique, and their features differ from person to person \cite{Thotahewa2014}. Heart-beat signals can be easily collected, and they are hard to copy by other people in comparison to simple pin codes. It is more secure than traditional methods because it requires a user to be available at the time of authentication and pairing process \cite{Sujatha2013, Wang2011}. Heart-beat signals can usually be collected by ECG and PPG sensors. ECG sensors collect the electrical activity of heart muscles through electrodes attached to the body. PPG sensors which can be attached to different parts of the body like the ear and finger, detect the blood level transforms in the microvascular cot of tissue \cite{Rundo2018}. It illuminates the body and measures transforms in light absorption as blood circulates in the body. The heart-beat signal can also be measured by seismocardiogram (SCG), which is the chest movement in response to the heart-beat. Accelerometers and piezo vibration sensors in the wearable devices can measure SCG as well \cite{Lin2019,wang2018unlock,ramos2012heart}. 
  
  Various features extracted from heart-beat signals can be used for authentication and key generating purpose. The most important feature used in WBANs is heart rate variability (HRV) or R-R interval or inter-beat interval (IBI) or Inter-pulse Interval (IPI) \cite{karaa2015biomedical, Sufi2010, Okoh2015} indicates the time interval between consecutive heart-beats \cite{McCraty2015}. Indeed, the fluctuations of heart rate around an average rate are shown by HRV \cite{karaa2015biomedical}. As has been proven by several studies \cite{cherukuri2003biosec,Rostami2013,obrist2012cardiovascular}, HRV is highly random and can be used as a random source to generate keys. Since HRV is a unique characteristic for each person, it can be used as an authentication method to pair devices on the same body.
  
  Rostami \emph{et al.} \cite{Rostami2013} proposed an HRV-based pairing method to authenticate external medical device controllers and programmers to IMDs.
  The authors introduce a touch-to-access policy using a time-varying physiological value (PV) by ECG readings. They utilized statistical characterization of ECG for pairing wearable devices. Another pairing system called H2B is presented by Lin \emph{et al.} \cite{Lin2019}, which utilizes piezo sensors to detect heart-beat signals and generate a secret key.
  
  \noindent{\bf EMG:}
  The EMG or electromyogram signals are the electrical signal generated by contractions of human muscles. According to medical research \cite{merletti2004electromyography,devasahayam2012signals}, the EMG signal is a quasi-random process, i.e., the average value of EMG is correlated to the generated force of the muscle, but it has a random amplitude variation under a given force. In other words, there are stochastic variations of EMG amplitude for a unique gesture and force. Therefore, the EMG signals can be used as a secure source to generate secret keys in physically close contact for some wearable devices like Myo armband \cite{Myo}, Athos gear \cite{Athos}, and Leo smart band \cite{Leo}. Since detecting this kind of signal needs physical contact in proximity, it is extremely difficult for an adversary to perform an eavesdropping attack. EMG-KEY is an EMG-based method proposed by Yang \emph{et al.} \cite{Yang2016} which leverages EMG variation signal to generate a secret key for pairing two wearable devices.
  
  \noindent{\bf Gait:}
  Due to the different properties of an individual’s muscular-skeletal structure, gait patterns are fairly unique among individuals \cite{zhong2014sensor}. Hence, it can be determined if two devices are carried by the same person \cite{Lester2004}. 
  
  Various techniques exploit different features of gait to generate a common key for pairing wearable devices. Sun \emph{et al.} \cite{Sun2017} proposed a method to generate a symmetric key based on the timing information of gait. The authors used the IPI of consecutive gait as a common feature between the two devices. Schürmann \emph{et al.} \cite{Schurmann2018} presented a secure spontaneous authentication scheme that exploits correlation in acceleration sequences from devices worn or carried together by the same person to extract always-fresh secure secrets. In their method, BANDANA, they utilized instantaneous variations in gait sequences with respect to the mean. Walkie-Talkie \cite{xu2016walkie} is another shared secret key generation scheme that allows two legitimate devices to establish a common cryptographic key by exploiting users’ walking characteristics (gait). The authors exploit independent component analysis (ICA) for blind source separation (BSS) to separate accelerometer signals from different body movements such as arm swing and walk. In Gait-Key \cite{xu2017gait} Xu \emph{et al.} extended their method in Walkie-Talkie to examine the effect of multi-level quantization on the pairing success rate. In \cite{xu2019gait} the same authors also proposed using spatial alignment instead of using BSS. A usability analysis of four gait-based device pairing schemes \cite{Schurmann2018,xu2016walkie,Sun2017,Groza2012} are presented in \cite{Brusch2020}.
  
  In complement to the aforementioned studies, our protocol, B2P, is the first one to utilize the breathing signal to enable secure pairing in wearable devices.
  \section{Conclusion}\label{conclude}
  We presented the design, implementation, and evaluation of the Breath to Pair (B2P) protocol, a respiration-based pairing approach for wearable devices. B2P enables generating the same secure keys based on the user's breathing signal detected by different types of sensors, RIP and accelerometer, in independent wearables. We explored several techniques to address the challenges in pairing distinctive wearables that use different types of sensors. B2P utilizes a change point detection technique for synchronizing independent wearables. It uses a Lloyd-Max quantization and an improved BCH error correction method to optimally digitize the respiration signal and correct mismatches between bit-strings. Extensive experiments on 30 participants indicate that B2P can generate a secure 256-bit key every 2.85 seconds, which is about one breathing cycle. We also proposed a sparse segment selection method to prevent the adversary. Experiment results show that B2P is secure against different types of attacks.
  \bibliographystyle{IEEE}
  \bibliography{refs}

\begin{thebibliography}{10}

\bibitem{Guk2019}
Kyeonghye Guk, Gaon Han, Jaewoo Lim, Keunwon Jeong, Taejoon Kang, Eun~Kyung
  Lim, and Juyeon Jung,
\newblock ``{Evolution of wearable devices with real-time disease monitoring
  for personalized healthcare},''
\newblock {\em Nanomaterials}, vol. 9, no. 6, pp. 1--23, 2019.

\bibitem{Vhaduri2019}
Sudip Vhaduri and Christian Poellabauer,
\newblock ``{Multi-modal biometric-based implicit authentication of wearable
  device users},''
\newblock {\em IEEE Transactions on Information Forensics and Security}, vol.
  14, no. 12, pp. 3116--3125, 2019.

\bibitem{chong2014survey}
Ming~Ki Chong, Rene Mayrhofer, and Hans Gellersen,
\newblock ``A survey of user interaction for spontaneous device association,''
\newblock {\em ACM Computing Surveys (CSUR)}, vol. 47, no. 1, pp. 1--40, 2014.

\bibitem{Seneviratne2017}
Suranga Seneviratne, Yining Hu, Tham Nguyen, Guohao Lan, Sara Khalifa, Kanchana
  Thilakarathna, Mahbub Hassan, and Aruna Seneviratne,
\newblock ``{A Survey of Wearable Devices and Challenges},''
\newblock {\em IEEE Communications Surveys and Tutorials}, vol. 19, no. 4, pp.
  2573--2620, 2017.

\bibitem{Bianchi2016}
Andrea Bianchi and Ian Oakley,
\newblock ``{Wearable authentication: Trends and opportunities},''
\newblock {\em it - Information Technology}, vol. 58, no. 5, 2016.

\bibitem{JohnDian2020}
F.~{John Dian}, Reza Vahidnia, and Alireza Rahmati,
\newblock ``{Wearables and the Internet of Things (IoT), Applications,
  Opportunities, and Challenges: A Survey},''
\newblock {\em IEEE Access}, vol. 8, pp. 69200--69211, 2020.

\bibitem{Poongodi2020}
T.~Poongodi, Rajalakshmi Krishnamurthi, R.~Indrakumari, P.~Suresh, and
  Balamurugan Balusamy,
\newblock {\em {Wearable devices and IoT}}, vol. 165,
\newblock Springer International Publishing, 2020.

\bibitem{Pantelopoulos2010}
Alexandros Pantelopoulos and Nikolaos~G. Bourbakis,
\newblock ``{A survey on wearable sensor-based systems for health monitoring
  and prognosis},''
\newblock {\em IEEE Transactions on Systems, Man and Cybernetics Part C:
  Applications and Reviews}, vol. 40, no. 1, pp. 1--12, 2010.

\bibitem{jayanth2017wearable}
S~Jayanth, MB~Poorvi, R~Shreyas, B~Padmaja, and MP~Sunil,
\newblock ``Wearable device to measure heart beat using iot,''
\newblock in {\em 2017 International Conference on Inventive Systems and
  Control (ICISC)}. IEEE, 2017, pp. 1--5.

\bibitem{majumder2019energy}
AKM Majumder, Yosuf~Amr ElSaadany, Roger Young, and Donald~R Ucci,
\newblock ``An energy efficient wearable smart iot system to predict cardiac
  arrest,''
\newblock {\em Advances in Human-Computer Interaction}, vol. 2019, 2019.

\bibitem{brezulianu2019iot}
Adrian Brezulianu, Oana Geman, Marius~Dan Zbancioc, Marius Hagan, Cristian
  Aghion, D~Jude Hemanth, Le~Hoang Son, et~al.,
\newblock ``Iot based heart activity monitoring using inductive sensors,''
\newblock {\em Sensors}, vol. 19, no. 15, pp. 3284, 2019.

\bibitem{milici2016wireless}
Stefano Milici, Javier Lorenzo, Antonio Lazaro, Ramon Villarino, and David
  Girbau,
\newblock ``Wireless breathing sensor based on wearable modulated frequency
  selective surface,''
\newblock {\em IEEE Sensors Journal}, vol. 17, no. 5, pp. 1285--1292, 2016.

\bibitem{shah2019cloud}
Syed Tauhid~Ullah Shah, Faizan Badshah, Faheem Dad, Nouman Amin, and Mian~Ahmad
  Jan,
\newblock ``Cloud-assisted iot-based smart respiratory monitoring system for
  asthma patients,''
\newblock in {\em Applications of Intelligent Technologies in Healthcare}, pp.
  77--86. Springer, 2019.

\bibitem{mahbub2017low}
Ifana Mahbub, Salvatore~Andrea Pullano, Hanfeng Wang, Syed~Kamrul Islam,
  Antonino~S Fiorillo, Gary To, and MR~Mahfouz,
\newblock ``A low-power wireless piezoelectric sensor-based respiration
  monitoring system realized in cmos process,''
\newblock {\em IEEE Sensors Journal}, vol. 17, no. 6, pp. 1858--1864, 2017.

\bibitem{naranjo2018smart}
David Naranjo-Hern{\'a}ndez, Alejandro Talaminos-Barroso, Javier Reina-Tosina,
  Laura~M Roa, Gerardo Barbarov-Rostan, Pilar Cejudo-Ramos, Eduardo
  M{\'a}rquez-Mart{\'\i}n, and Francisco Ortega-Ruiz,
\newblock ``Smart vest for respiratory rate monitoring of copd patients based
  on non-contact capacitive sensing,''
\newblock {\em Sensors}, vol. 18, no. 7, pp. 2144, 2018.

\bibitem{wan2018wearable}
Jie Wan, Munassar~AAH Al-awlaqi, MingSong Li, Michael O’Grady, Xiang Gu, Jin
  Wang, and Ning Cao,
\newblock ``Wearable iot enabled real-time health monitoring system,''
\newblock {\em EURASIP Journal on Wireless Communications and Networking}, vol.
  2018, no. 1, pp. 1--10, 2018.

\bibitem{yoshida2018development}
Shinya Yoshida, Hiroshi Miyaguchi, and Tsutomu Nakamura,
\newblock ``Development of tablet-shaped ingestible core-body thermometer
  powered by gastric acid battery,''
\newblock {\em IEEE Sensors Journal}, vol. 18, no. 23, pp. 9755--9762, 2018.

\bibitem{lamonaca2019overview}
Francesco Lamonaca, Eulalia Balestrieri, Ioan Tudosa, Francesco Picariello,
  Domenico~Luca Carn{\`\i}, Carmelo Scuro, Francesco Bonavolont{\`a}, Vitaliano
  Spagnuolo, Gioconda Grimaldi, and Antonio Colaprico,
\newblock ``An overview on internet of medical things in blood pressure
  monitoring,''
\newblock in {\em 2019 IEEE International Symposium on Medical Measurements and
  Applications (MeMeA)}. IEEE, 2019, pp. 1--6.

\bibitem{murali2018pulse}
Dhanurdhar Murali, Deepthi~R Rao, Swathi~R Rao, and M~Ananda,
\newblock ``Pulse oximetry and iot based cardiac monitoring integrated alert
  system,''
\newblock in {\em 2018 international conference on advances in computing,
  communications and informatics (ICACCI)}. IEEE, 2018, pp. 2237--2243.

\bibitem{sargunam2019iot}
B~Sargunam and S~Anusha,
\newblock ``Iot based mobile medical application for smart insulin
  regulation,''
\newblock in {\em 2019 IEEE International Conference on Electrical, Computer
  and Communication Technologies (ICECCT)}. IEEE, 2019, pp. 1--5.

\bibitem{zeng2017wearia}
Yunze Zeng, Amit Pande, Jindan Zhu, and Prasant Mohapatra,
\newblock ``Wearia: Wearable device implicit authentication based on activity
  information,''
\newblock in {\em 2017 IEEE 18th International Symposium on A World of
  Wireless, Mobile and Multimedia Networks (WoWMoM)}. IEEE, 2017, pp. 1--9.

\bibitem{unar2014review}
JA~Unar, Woo~Chaw Seng, and Almas Abbasi,
\newblock ``A review of biometric technology along with trends and prospects,''
\newblock {\em Pattern recognition}, vol. 47, no. 8, pp. 2673--2688, 2014.

\bibitem{shim2015survey}
Kyung-Ah Shim,
\newblock ``A survey of public-key cryptographic primitives in wireless sensor
  networks,''
\newblock {\em IEEE Communications Surveys \& Tutorials}, vol. 18, no. 1, pp.
  577--601, 2015.

\bibitem{wikiRespiratory}
``Respiratory system, wikipedia, the free encyclopedia,'' .

\bibitem{massaroni2019contact}
Carlo Massaroni, Andrea Nicol{\`o}, Daniela Lo~Presti, Massimo Sacchetti,
  Sergio Silvestri, and Emiliano Schena,
\newblock ``Contact-based methods for measuring respiratory rate,''
\newblock {\em Sensors}, vol. 19, no. 4, pp. 908, 2019.

\bibitem{clarenbach2005monitoring}
Christian~F Clarenbach, Oliver Senn, Thomas Brack, Malcolm Kohler, and Konrad~E
  Bloch,
\newblock ``Monitoring of ventilation during exercise by a portable respiratory
  inductive plethysmograph,''
\newblock {\em Chest}, vol. 128, no. 3, pp. 1282--1290, 2005.

\bibitem{Preejith2017}
S.~P. Preejith, Ahamed Jeelani, Paresh Maniyar, Jayaraj Joseph, and
  Mohanasankar Sivaprakasam,
\newblock ``{Accelerometer based system for continuous respiratory rate
  monitoring},''
\newblock {\em 2017 IEEE International Symposium on Medical Measurements and
  Applications, MeMeA 2017 - Proceedings}, pp. 171--176, 2017.

\bibitem{Vertens2015}
Johan Vertens, Fabian Fischer, Christian Heyde, Fabian Hoeflinger, Rui Zhang,
  Leonhard Reindl, and Albert Gollhofer,
\newblock ``{Measuring respiration and heart rate using two acceleration
  sensors on a fully embedded platform},''
\newblock {\em icSPORTS 2015 - Proceedings of the 3rd International Congress on
  Sport Sciences Research and Technology Support}, , no. icSPORTS, pp. 15--23,
  2015.

\bibitem{JafariTadi2014}
Mojtaba {Jafari Tadi}, Tero Koivisto, Mikko P{\"{a}}nk{\"{a}}{\"{a}}l{\"{a}},
  and Ari Paasio,
\newblock ``{Accelerometer-based method for extracting respiratory and cardiac
  gating information for dual gating during nuclear medicine imaging},''
\newblock {\em International Journal of Biomedical Imaging}, vol. 2014, 2014.

\bibitem{Vanegas2020}
Erik Vanegas, Raul Igual, and Inmaculada Plaza,
\newblock ``{Sensing systems for respiration monitoring: A technical systematic
  review},''
\newblock {\em Sensors (Switzerland)}, vol. 20, no. 18, pp. 1--84, 2020.

\bibitem{Jafar_HR_RR}
Jafar Pourbemany, Essa Almabrok, and Ye~Zhu,
\newblock ``Real time video based heart rate and respiration rate monitoring,''
\newblock {\em IEEE NAICON}, vol. 2, no. 4, 2021.

\bibitem{bentsen2016electromagnetic}
Mariann~HL Bentsen, Morten Eriksen, Merete~S Olsen, Trond Markestad, and Thomas
  Halvorsen,
\newblock ``Electromagnetic inductance plethysmography is well suited to
  measure tidal breathing in infants,''
\newblock {\em ERJ open research}, vol. 2, no. 4, 2016.

\bibitem{sanyal2018algorithms}
Shourjya Sanyal and Koushik~Kumar Nundy,
\newblock ``Algorithms for monitoring heart rate and respiratory rate from the
  video of a user’s face,''
\newblock {\em IEEE Journal of translational engineering in health and
  medicine}, vol. 6, pp. 1--11, 2018.

\bibitem{pmid23201991}
Z.~Zhang, J.~Zheng, H.~Wu, W.~Wang, B.~Wang, and H.~Liu,
\newblock ``{{D}evelopment of a respiratory inductive plethysmography module
  supporting multiple sensors for wearable systems},''
\newblock {\em Sensors (Basel)}, vol. 12, no. 10, pp. 13167--13184, Sep 2012.

\bibitem{lloyd1982least}
Stuart Lloyd,
\newblock ``Least squares quantization in pcm,''
\newblock {\em IEEE transactions on information theory}, vol. 28, no. 2, pp.
  129--137, 1982.

\bibitem{max1960quantizing}
Joel Max,
\newblock ``Quantizing for minimum distortion,''
\newblock {\em IRE Transactions on Information Theory}, vol. 6, no. 1, pp.
  7--12, 1960.

\bibitem{han2010bch}
Yunghsiang~S Han,
\newblock ``Bch codes,''
\newblock {\em Graduate Institute of Communication Engineering, National Taipei
  University Taiwan}, 2010.

\bibitem{Hexoskin}
``Hexoskin smart shirt,'' .

\bibitem{klir2013uncertainty}
George~J Klir and Mark~J Wierman,
\newblock {\em Uncertainty-based information: elements of generalized
  information theory}, vol.~15,
\newblock Physica, 2013.

\bibitem{francis2016ultrasonographic}
Colin~Anthony Francis, Joaqu{\'\i}n~Andr{\'e}s Hoffer, and Steven Reynolds,
\newblock ``Ultrasonographic evaluation of diaphragm thickness during
  mechanical ventilation in intensive care patients,''
\newblock {\em American Journal of Critical Care}, vol. 25, no. 1, pp. e1--e8,
  2016.

\bibitem{DiRienzo2020}
Marco {Di Rienzo}, Giovannibattista Rizzo, Zeynep~Melike Işilay, and Prospero
  Lombardi,
\newblock ``{Seismote: A multi-sensor wireless platform for cardiovascular
  monitoring in laboratory, daily life, and telemedicine},''
\newblock {\em Sensors (Switzerland)}, vol. 20, no. 3, 2020.

\bibitem{shang2020audiokey}
Jiacheng Shang and Jie Wu,
\newblock ``Audiokey: a usable device pairing system using audio signals on
  smartwatches,''
\newblock {\em International Journal of Security and Networks}, vol. 15, no. 1,
  pp. 46--58, 2020.

\bibitem{jin2015magpairing}
Rong Jin, Liu Shi, Kai Zeng, Amit Pande, and Prasant Mohapatra,
\newblock ``Magpairing: Pairing smartphones in close proximity using
  magnetometers,''
\newblock {\em IEEE Transactions on Information Forensics and Security}, vol.
  11, no. 6, pp. 1306--1320, 2015.

\bibitem{Rostami2013}
Masoud Rostami and Ari Juels,
\newblock ``{Authentication for Implanted Medical Devices Categories and
  Subject Descriptors},''
\newblock {\em Ccs}, pp. 1099--1111, 2013.

\bibitem{Lin2019}
Qi~Lin, Weitao Xu, Jun Liu, Abdelwahed Khamis, Wen Hu, Mahbub Hassan, and Aruna
  Seneviratne,
\newblock ``{H2B: Heartbeat-based secret key generation using piezo vibration
  sensors},''
\newblock {\em arXiv}, pp. 265--276, 2019.

\bibitem{Schurmann2018}
Dominik Sch{\"{u}}rmann, Arne Br{\"{u}}sch, Ngu Nguyen, Stephan Sigg, and Lars
  Wolf,
\newblock ``{Moves like Jagger: Exploiting variations in instantaneous gait for
  spontaneous device pairing},''
\newblock {\em Pervasive and Mobile Computing}, vol. 47, no. May 2017, pp.
  1--12, 2018.

\bibitem{xu2016walkie}
Weitao Xu, Girish Revadigar, Chengwen Luo, Neil Bergmann, and Wen Hu,
\newblock ``Walkie-talkie: Motion-assisted automatic key generation for secure
  on-body device communication,''
\newblock in {\em 2016 15th ACM/IEEE International Conference on Information
  Processing in Sensor Networks (IPSN)}. IEEE, 2016, pp. 1--12.

\bibitem{Sun2017}
Yingnan Sun, Charence Wong, Guang~Zhong Yang, and Benny Lo,
\newblock ``{Secure key generation using gait features for Body Sensor
  Networks},''
\newblock {\em 2017 IEEE 14th International Conference on Wearable and
  Implantable Body Sensor Networks, BSN 2017}, pp. 206--210, 2017.

\bibitem{Groza2012}
Bogdan Groza and Rene Mayrhofer,
\newblock ``{SAPHE: Simple accelerometer based wireless pairing with heuristic
  trees},''
\newblock {\em ACM International Conference Proceeding Series}, , no. 2, pp.
  161--168, 2012.

\bibitem{Yuzuguzel2015a}
Hidir Y{\"{u}}zug{\"{u}}zel, Jari Niemi, Serkan Kiranyaz, Moncef Gabbouj, and
  Thomas Heinz,
\newblock ``{ShakeMe: Key generation from shared motion},''
\newblock {\em Proceedings - 15th IEEE International Conference on Computer and
  Information Technology, CIT 2015, 14th IEEE International Conference on
  Ubiquitous Computing and Communications, IUCC 2015, 13th IEEE International
  Conference on Dependable, Autonomic and Se}, pp. 2130--2133, 2015.

\bibitem{Rodriguez1989a}
E.~Rodr{\'{i}}guez, J.~L. Arqu{\'{e}}s, R.~Rodr{\'{i}}guez, M.~Nu{\~{n}}ez,
  M.~Medina, T.~L. Talarico, I.~A. Casas, T.~C. Chung, W.~J. Dobrogosz,
  L.~Axelsson, S.~E. Lindgren, W.~J. Dobrogosz, Leila Kerkeni, Paula Ruano,
  Lismet~Lazo Delgado, Sergio Picco, Liliana Villegas, Franco Tonelli, Mario
  Merlo, Javier Rigau, Dario Diaz, and Martin Masuelli,
\newblock ``{We are IntechOpen , the world ' s leading publisher of Open Access
  books Built by scientists , for scientists TOP 1 {\%}},''
\newblock {\em Intech}, vol. 32, no. tourism, pp. 137--144, 1989.

\bibitem{Athos_Shirt}
``Athos shirt,'' .

\bibitem{Zephyr}
``Zephyr,'' .

\bibitem{mayrhofer2012uacap}
Rene Mayrhofer, J{\"u}rgen Fu{\ss}, and Iulia Ion,
\newblock ``Uacap: A unified auxiliary channel authentication protocol,''
\newblock {\em IEEE Transactions on Mobile Computing}, vol. 12, no. 4, pp.
  710--721, 2012.

\bibitem{Thotahewa2014}
Kasun Maduranga~Silva Thotahewa, Jean~Michel Redout{\'{e}}, and Mehmet~Rasit
  Yuce,
\newblock {\em {Ultra wideband wireless body area networks}}, vol.
  9783319052878,
\newblock 2014.

\bibitem{Sujatha2013}
S.~Sujatha and R.~Govindaraju,
\newblock ``{A Secure Crypto based ECG Data Communication using Modified SPHIT
  and Modified Quasigroup Encryption},''
\newblock {\em International Journal of Computer Applications}, vol. 78, no. 6,
  pp. 27--33, 2013.

\bibitem{Wang2011}
Wei Wang, Honggang Wang, Michael Hempel, Dongming Peng, Hamid Sharif, and
  Hsiao~Hwa Chen,
\newblock ``{Secure stochastic ECG signals based on gaussian mixture model for
  e-healthcare systems},''
\newblock {\em IEEE Systems Journal}, vol. 5, no. 4, pp. 564--573, 2011.

\bibitem{Rundo2018}
Francesco Rundo, Sabrina Conoci, Alessandro Ortis, and Sebastiano Battiato,
\newblock ``{An advanced bio-inspired photoplethysmography (PPG) and ECG
  pattern recognition system for medical assessment},''
\newblock {\em Sensors (Switzerland)}, vol. 18, no. 2, 2018.

\bibitem{wang2018unlock}
Lei Wang, Kang Huang, Ke~Sun, Wei Wang, Chen Tian, Lei Xie, and Qing Gu,
\newblock ``Unlock with your heart: Heartbeat-based authentication on
  commercial mobile phones,''
\newblock {\em Proceedings of the ACM on interactive, mobile, wearable and
  ubiquitous technologies}, vol. 2, no. 3, pp. 1--22, 2018.

\bibitem{ramos2012heart}
Juan Ramos-Castro, J~Moreno, H~Miranda-Vidal, Miguel~A
  Garc{\'\i}a-Gonz{\'a}lez, Mireya Fern{\'a}ndez-Chimeno, Gil Rodas, and
  Ll~Capdevila,
\newblock ``Heart rate variability analysis using a seismocardiogram signal,''
\newblock in {\em 2012 annual international conference of the IEEE engineering
  in medicine and biology society}. IEEE, 2012, pp. 5642--5645.

\bibitem{karaa2015biomedical}
Wahiba Ben~Abdessalem Karaa,
\newblock {\em Biomedical image analysis and mining techniques for improved
  health outcomes},
\newblock IGI Global, 2015.

\bibitem{Sufi2010}
Fahim Sufi, Ibrahim Khalil, and Jiankun Hu,
\newblock ``{ECG-Based Authentication},''
\newblock {\em Handbook of Information and Communication Security}, pp.
  309--331, 2010.

\bibitem{Okoh2015}
Ebenezer Okoh,
\newblock ``{Biometrics Solutions in e-Health Security: A Comprehensive
  Literature Review},''
\newblock {\em Spine}, vol. 19, no. Supplement, pp. 2274S--2278S, 2015.

\bibitem{McCraty2015}
Rollin McCraty and Fred Shaffer,
\newblock ``{Heart rate variability: New perspectives on physiological
  mechanisms, assessment of self-regulatory capacity, and health risk},''
\newblock {\em Global Advances In Health and Medicine}, vol. 4, no. 1, pp.
  46--61, 2015.

\bibitem{cherukuri2003biosec}
Sriram Cherukuri, Krishna~K Venkatasubramanian, and Sandeep~KS Gupta,
\newblock ``Biosec: A biometric based approach for securing communication in
  wireless networks of biosensors implanted in the human body,''
\newblock in {\em 2003 International Conference on Parallel Processing
  Workshops, 2003. Proceedings.} IEEE, 2003, pp. 432--439.

\bibitem{obrist2012cardiovascular}
Paul~A Obrist,
\newblock {\em Cardiovascular psychophysiology: A perspective},
\newblock Springer Science \& Business Media, 2012.

\bibitem{merletti2004electromyography}
Roberto Merletti and Philip~J Parker,
\newblock {\em Electromyography: physiology, engineering, and non-invasive
  applications}, vol.~11,
\newblock John Wiley \& Sons, 2004.

\bibitem{devasahayam2012signals}
Suresh~R Devasahayam,
\newblock {\em Signals and systems in biomedical engineering: signal processing
  and physiological systems modeling},
\newblock Springer Science \& Business Media, 2012.

\bibitem{Myo}
``Myo armband,'' .

\bibitem{Athos}
``Athos gear,'' .

\bibitem{Leo}
``Leo smartband,'' .

\bibitem{Yang2016}
Lin Yang, Wei Wang, and Qian Zhang,
\newblock ``{Secret from muscle: Enabling secure pairing with
  electromyography},''
\newblock {\em Proceedings of the 14th ACM Conference on Embedded Networked
  Sensor Systems, SenSys 2016}, pp. 28--41, 2016.

\bibitem{zhong2014sensor}
Yu~Zhong and Yunbin Deng,
\newblock ``Sensor orientation invariant mobile gait biometrics,''
\newblock in {\em IEEE international joint conference on biometrics}. IEEE,
  2014, pp. 1--8.

\bibitem{Lester2004}
Jonathan Lester, Blake Hannaford, and Gaetano Borriello,
\newblock ``{"Are you with me?" - Using accelerometers to determine if two
  devices are carried by the same person},''
\newblock {\em Lecture Notes in Computer Science (including subseries Lecture
  Notes in Artificial Intelligence and Lecture Notes in Bioinformatics)}, vol.
  3001, pp. 33--50, 2004.

\bibitem{xu2017gait}
Weitao Xu, Chitra Javali, Girish Revadigar, Chengwen Luo, Neil Bergmann, and
  Wen Hu,
\newblock ``Gait-key: A gait-based shared secret key generation protocol for
  wearable devices,''
\newblock {\em ACM Transactions on Sensor Networks (TOSN)}, vol. 13, no. 1, pp.
  1--27, 2017.

\bibitem{xu2019gait}
Weitao Xu and Guohao Lan,
\newblock ``Gait-based smart pairing system for personal wearable devices,''
\newblock in {\em Medical Internet of Things (m-IoT)-Enabling Technologies and
  Emerging Applications}. IntechOpen, 2019.

\bibitem{Brusch2020}
Arne Brusch, Ngu Nguyen, Dominik Schurmann, Stephan Sigg, and Lars Wolf,
\newblock ``{Security Properties of Gait for Mobile Device Pairing},''
\newblock {\em IEEE Transactions on Mobile Computing}, vol. 19, no. 3, pp.
  697--710, 2020.

\end{thebibliography}
  
  \end{document}